\pdfoutput=1

\documentclass[12pt]{iopart}

\usepackage{amscd,amssymb,a4}
\usepackage{graphicx,xspace}

\usepackage[left=3cm,top=3cm,right=3cm,bottom=3cm]{geometry}
\usepackage{color}

\begin{document}

\newcommand{\red}{\color{red}}
\newcommand{\blue}{\color{blue}}
\newcommand{\todo}[1]{\textbf{To do: #1}}
\newcommand{\be}{\begin{equation}}
\newcommand{\ee}{\end{equation}}
\newcommand{\beq}{\begin{equation}}
\newcommand{\eeq}{\end{equation}}
\newcommand{\bea}{\begin{eqnarray}}
\newcommand{\eea}{\end{eqnarray}}
\newcommand{\rar}{\rightarrow}
\newcommand{\lar}{\leftarrow}
\newcommand{\ra}{\right\rangle}
\newcommand{\la}{\left\langle }
\renewcommand{\d}{{\rm d }}
\newcommand{\m}{{\tilde m }}
\newcommand{\p}{\partial}
\newcommand{\nn}{\nonumber }

\newcommand{\fig}[2]{\includegraphics[width=#1]{./figures/#2}}
\newcommand{\Fig}[1]{\includegraphics[width=7cm]{./figures/#1}}
\newlength{\bilderlength}
\newcommand{\bilderscale}{0.35}
\newcommand{\storebilderscale}{\bilderscale}
\newcommand{\bilderskip}{\hspace*{0.8ex}}
\newcommand{\textdiagram}[1]{%
\renewcommand{\bilderscale}{0.2}%
\diagram{#1}\renewcommand{\bilderscale}{\storebilderscale}}
\newcommand{\vardiagram}[2]{%
\newcommand{\bilderscale}{#1}%
\diagram{#2}\renewcommand{\bilderscale}{\storebilderscale}}
\newcommand{\diagram}[1]{%
\settowidth{\bilderlength}{\bilderskip%
\includegraphics[scale=\bilderscale]{./figures/#1}\bilderskip}%
\parbox{\bilderlength}{\bilderskip%
\includegraphics[scale=\bilderscale]{./figures/#1}\bilderskip}}
\newcommand{\Diagram}[1]{%
\settowidth{\bilderlength}{%
\includegraphics[scale=\bilderscale]{./figures/#1}}%
\parbox{\bilderlength}{%
\includegraphics[scale=\bilderscale]{./figures/#1}}}
\bibliographystyle{KAY}

%


%








\newcommand{\atanh}
{\operatorname{atanh}}

\newcommand{\ArcTan}
{\operatorname{ArcTan}}

\newcommand{\ArcCoth}
{\operatorname{ArcCoth}}

\newcommand{\Erf}
{\operatorname{Erf}}

\newcommand{\Erfi}
{\operatorname{Erfi}}

\newcommand{\Ei}
{\operatorname{Ei}}

\newcommand{\sgn}{{\mathrm{sgn}}}
\def\be{\begin{equation}}
\def\ee{\end{equation}}

\def\bea{\begin{eqnarray}}
\def\eea{\end{eqnarray}}

\def\e{\epsilon}
\def\l{\lambda}
\def\d{\delta}
\def\o{\omega}
\def\cb{\bar{c}}
\def\Li{{\rm Li}}

\title[The KPZ equation with flat initial condition and the DP with one free end]{The KPZ equation with flat initial condition and the directed polymer with one free end}

\author{Pierre Le Doussal}

\address{CNRS-Laboratoire de Physique Th\'eorique de l'Ecole Normale Sup\'erieure\\
24 rue Lhomond, 75231 Paris
Cedex-France
}

\author{Pasquale Calabrese}

\address{Dipartimento di Fisica dell Universit\`a di Pisa and INFN, 56127 Pisa Italy}

%



\date{\today}

\begin{abstract}
We study the directed polymer (DP) of length $t$ in a random potential in dimension $1+1$ in the continuum
limit, with one end fixed and one end free. This maps onto the Kardar-Parisi-Zhang growth
equation in time $t$, with flat initial conditions. We use the Bethe Ansatz solution for the replicated problem which is an
attractive bosonic model. The problem is more difficult than the previous solution of the
fixed endpoint problem as it requires regularization of the spatial integrals over the Bethe eigenfunctions.
We use either a large fixed system length or a small finite slope KPZ initial conditions (wedge). The latter allows
to take properly into account non-trivial contributions, which appear as deformed strings in the former. By considering a half-space model 
in a proper limit we obtain an expression for the generating function of all positive integer moments 
$\overline{Z^n}$ of the directed polymer partition function. We obtain the generating function 
of the moments of the DP partition sum as a Fredholm Pfaffian. 
At large time, this Fredholm Pfaffian, valid for all time $t$, exhibits convergence 
of the free energy (i.e. KPZ height) distribution to the GOE Tracy Widom distribution 
\end{abstract}

\maketitle

\section{Introduction}

\subsection{Overview}

The continuum Kardar-Parisi-Zhang (KPZ) equation describes the non-equilibrium
growth in time $t$ of an interface of height $h(x,t)$ in the presence
of noise \cite{KPZ}. It defines a universality class believed to encompass 
numerous models and physical systems \cite{kpzreviews}.
The KPZ problem maps to the equilibrium statistical mechanics of a directed polymer (DP) in a random
potential, the simplest example of a glass \cite{directedpoly} with applications to 
vortex lines \cite{vortex}, domain walls \cite{lemerle}, biophysics \cite{hwa} and
Burgers turbulence \cite{Burgers}. Via the exact Cole-Hopf transformation, the DP polymer with two fixed endpoints 
maps onto the KPZ growth starting from a ``droplet'' initial condition, while the DP with one
fixed and one free endpoint maps to KPZ growth from a flat initial interface. 
The third important case is the stationary growth in a finite geometry, equivalent to a DP on a cylinder. 
In the continuum, the DP and KPZ problems completely identify, and the free energy of the polymer 
is simply proportional to the KPZ height while its length $t$ is proportional to the KPZ time.

An important property of the growing KPZ interface is that it becomes, at large $t$, statistically self-affine 
with universal scaling exponents.
In $d=1$, its width is predicted  to grow as $\delta h \sim t^{1/3}$ \cite{exponent}, as observed in experiments 
\cite{exp1,exp2,exp3}. Furthermore, in a remarkable recent experiment on liquid crystals 
\cite{exp4} it was possible to measure with great accuracy, not only the scaling exponents, but also the full
probability distribution of the height field for various initial conditions. It was observed that this distribution
depends on whether the interface starts from a droplet or from a flat configuration.

Since the KPZ equation has resisted direct analytical solutions, progress came indirectly, i.e. from solving models
believed to be in the same universality class. Analytical progress on the KPZ universality class in $d=1$ came from
exact solutions of a lattice DP model at zero temperature \cite{Johansson2000},
discrete growth models such as the PNG model \cite{spohn2000,png}, 
asymmetric exclusion models \cite{spohnTASEP} and vicious 
walkers \cite{greg}. An analog of the
height field $h(x,t)$ was identified and, in the large size limit, its one-point (scaled) probability
distribution was shown to equal the (scaled) distribution of the smallest eigenvalue of a 
random matrix drawn from the famous Gaussian ensembles, the so-called
Tracy Widom (TW) distribution \cite{TW1994}, which also appears in many
other contexts \cite{othersTW}.
It was found \cite{spohn2000,ferrari1} that depending on 
the boundary condition, e.g. droplet versus flat initial condition for the interface, 
one gets either the TW distribution $F_2(s)$ of the Gaussian unitary ensemble (GUE)  or $F_1(s)$
of the Gaussian orthogonal  ensemble (GOE) for droplet and flat initial conditions respectively. 
The corresponding many point distributions were identified as 
determinantal space-time processes, the Airy process, $Ai_2$ for droplet, and $Ai_1$ for flat which are both
naturally expressed as Fredholm determinants \cite{spohnreview}.

These advances gave valuable, but only indirect information on the {\it continuum}
KPZ equation, i.e. a conjecture for its infinite $t$ limit (termed the KPZ renormalization fixed
point \cite{corwinRG}), but no information about the approach to this limit, that is 
of great importance for experiments \cite{exp1,exp2,exp3,exp4}, which always probe a large but finite time window.
This long time approach to the asymptotic value is particularly relevant in the recent experiments \cite{exp4} which 
seem to confirm that the height fluctuations are indeed described by the GUE and GOE TW distributions at large time
depending on the type of initial condition. 

Recently, we found \cite{we,dotsenko} an exact solution for the continuum directed polymer  
with both endpoints fixed, i.e. the KPZ equation with droplet initial conditions, 
by exploiting the exact Bethe ansatz solvability of the {\it replicated}
problem (initiated in Refs. \cite{kardareplica,bo-90,bb-00}). A similar result for 
the droplet initial condition was derived within different approaches 
\cite{spohnKPZEdge,corwinDP}. This solution shows that the 
generating function (called $g(s)$ below) of the free-energy distribution 
is a Fredholm determinant, not only in the infinite time limit (the TW distribution
is naturally written as a Fredholm determinant), but also for 
finite time. It thus provides an exact solution for the universal crossover in time in the continuum KPZ equation.
It was also shown \cite{we} that this universal distribution (depending on a single parameter $t$) 
describes, in the DP framework, the high temperature regime \cite{we}
that has remarkable universal features \cite{highT}. In the growth problem,
this corresponds to a universal large diffusivity-weak noise limit, at fixed 
correlation length of the noise.

Several authors then exploited the powerful replica Bethe Ansatz approach to the KPZ equation and 
the directed polymer, and a number of other quantities and initial conditions 
have been considered \cite{ps-2point,ps-npoint,ps-more,ps-more2,we-flat,it-11,it-12,dot2,dot3}.
In particular, very recently, the two remaining important classes of initial conditions have been
solved using replica, the flat initial condition \cite{we-flat} and the stationary class 
\cite{it-12}. Other remarkable recent results concern the distribution of the
position of the DP endpoint \cite{greg2,quastelendpoint} and have been obtained by completely
different methods. 
Some recent more mathematical achievements  have been reported in Refs. \cite{variousmath} (see also Ref. \cite{reviewCorwin} for a
review). 

\subsection{Quantities of interest, strategy of the calculation and main results} 

The aim of this paper is to provide a detailed account of our recent solution of the KPZ equation with 
flat initial condition \cite{we-flat}, i.e. the directed polymer with one free end and one fixed. The 
continuum model has natural space and time units and we define
\bea \label{lambdadef}
\lambda = \frac{1}{2} (\bar c^2 t/T^5)^{1/3} ,
\eea 
the single dimensionless parameter which describes the crossover from the diffusive regime
at small time $t$ to the glass fixed point regime at large $t$. Here $T$ is the temperature,
$\bar c$ the disorder strength and $t$ the length of the DP. In terms of
KPZ parameters this is $T=2 \nu$ and $\bar c=D \lambda_0^2$. All parameters are
defined below. Note that (\ref{lambdadef}) is exactly the same 
parameter $\lambda$ as in our previous work \cite{we} (and in \cite{dotsenko}) on the fixed endpoints problem.

Our aim here is to characterize the probability distribution (PDF) of the free energy of the DP
\bea
&& F = - T \ln Z = T \lambda f\,,
\eea
equivalently, of the height field of the KPZ equation
\bea \label{defxi}
&& \frac{\lambda_0}{2 \nu} h = \ln Z = v_0 t + \lambda \xi_t \,.
\eea 
To achieve this, we first calculate by means of the Bethe ansatz the disorder average (denoted here by overbars) of the integer 
powers of partition function $\overline{Z^n}$ as function of time. 
We then avoid the explicit $n\to 0$ limit by introducing the generating function
\be
 g_\lambda(s) = \overline{e^{- e^{- \lambda s} Z}} = 1 + \sum_{n=1}^\infty \frac{(- e^{- \lambda s})^n}{n!} \overline{Z^n} = 
\overline{ \exp( - e^{- \lambda (s + f)} ) }\,, 
\ee
where the subscript $\lambda$ is often omitted below. Once $g_\lambda(s)$ is known, the PDF of the rescaled free energy $P(f)$ 
at large time (i.e. $\lambda  \to \infty$) is immediately extracted as
\be \label{conv}
 \lim_{\lambda \to \infty} g_\lambda(s) = \overline{ \theta(f+s) } = {\rm Prob}(f > -s) = \int_{-s}^{+\infty} df P(f) \,. 
\ee
and it can also be extracted at finite time (e.g. via the procedure described in \cite{we}). In this paper $\theta(x)$ denotes the Heaviside function. The same
generating function was introduced and computed in \cite{we} for the fixed endpoint problem when
it takes the form of a Fredholm determinant (FD) at any time $t$ (any $\lambda$). Here the
calculation is much more involved but we finally find that $g_\lambda(s)$ takes the form of a
Fredholm Pfaffian (i.e. the square root of a Fredholm determinant) for any $\lambda$. 
We then show that at large time this Fredholm Pfaffian, giving the free energy (i.e. KPZ height) 
distribution, converges to the GOE Tracy Widom distribution, i.e. the scaled random variable $\xi_t=-f$ defined in (\ref{defxi}) 
obeys
\bea
g_\infty(s) = \lim_{t \to \infty} {\rm Prob}(\xi_t < s) = F_1(s) \,.
\eea 

As we will see, the calculations are far more complicated than for the DP with fixed endpoint. 
The first step, i.e. obtaining explicit expressions for the integer moments $\overline{Z^n}$ is much more involved.
In the fixed endpoint calculation one could directly consider the infinite space $L=+\infty$ and perform a
summation over the string states, which have an easy structure in that limit. Now however we
need to calculate the spatial integrals over the Bethe eigenfunctions, and these are not nicely convergent on the infinite space
(one easily gets either zero or infinite result!). The first approach we try, which we call {\it direct approach} is 
to consider (i) a periodic system of finite size $L$, for large $L$. 
The correct treatment however requires the use of the Bethe equations
which enforce periodicity. This approach quickly becomes intractable, and we have been able to calculate only small
integer values of $n$ (we explicitly report only up to $n=4$), or to sum for all $n$ but at fixed number of strings $n_s=1,2$.
A structure in the contributions however starts to appear, confirmed later by other methods. The second approach is to consider
(ii) a small finite slope $w>0$ in the KPZ initial conditions, i.e. a wedge initial condition. There the
limit $L=+\infty$ can be taken from the start and the desired integrals can be obtained by considering
poles in $1/w$. The {\it symmetric wedge} allows to make some progress, in particular we identify a contribution
which we missed at first in the {\it direct approach}, which can be identified as a {\it deformed
string} in the large $L$ limit. Hence, both methods taken together give reliable results for small values of $n$ and/or $n_s$. The true
progress however comes from considering the {\it half-space wedge}. By taking a combination of
limits (a) large distance from the wedge (b) small $w$, we are able to compute systematically
all contributing poles in $1/w$ and derive their Pfaffian structure. The rest of the calculation
relies on further tricks to perform the summation over all string states in the form
of a Fredholm Pfaffian (FP). This FP takes various forms, some of them can be
useful for numerical evaluations, while one is particularly suited to take the large time
limit and to show convergence to the GOE TW distribution.

\subsection{Organization of the manuscript}

The paper is organized as follows. 
In Sec. \ref{model} we introduce the DP and KPZ models, the various boundary conditions, and the notations 
employed in this paper.
In Sec. \ref{BAsec} we recall the needed details about replica and the Bethe ansatz solution for the attractive Bose gas.
The starting formula for the moments and the string partition sums are given in Sec. \ref{sec:starting} and \ref{sec:starting2}, and the
Airy trick which allows to perform divergent sums is recalled. 
In the following sections we start our new computations. 
As a first step, in Sec. \ref{DIRECT}, we attempt to calculate $\overline{Z^n}$ directly for the flat initial condition.
Although it does show a nice structure emerging, the actual calculation is very cumbersome and only results for $n=2,3,4$ and $n_s=1,2$ are reported. The exact result for $n_s=1$ already reproduces the correct right tail of $F_1(s)$ (i.e. left tail
for $P(f)$), as discussed in Sec. \ref{sec:ns1}. 
In Sec. \ref{HSsec} we move to the wedge initial condition. The full-space or symmetric wedge is discussed only
in the \ref{FSsec} but as we show it allows to reproduce flat or narrow initial conditions, respectively. We thus proceed with the 
half-space wedge and analyze its pole structure in Sec. \ref{sec:poles} by taking proper limits. This leads to the full
analytical formula for $Z(n_s,s)$ which is presented in Section \ref{sec:fullgenerating}, and checked to agree with all previous
small $n,n_s$ results by the two other methods, as well as with a non-trivial sum rule at $t=0$. 
The reader uninterested in the technicalities involved in deriving the expression of
$Z(n_s,s)$ can skip sections \ref{DIRECT} and \ref{FSsec}. In Sec. \ref{Pfsec} we proceed with the Airy trick 
to derive the final expression for $g(s)$ as a Fredholm Pfaffian. It involves on the way some new Pfaffian identities
which allow to perform the summations over the number of particles in the strings. The explicit expressions for the
kernels are given in Sections \ref{sec:Airy} to \ref{sec:FP}. The large time limit is then studied 
in Section \ref{sec:larget} and the convergence to the GOE Tracy Widom distribution is proved. 
Section \ref{conclusion} contains concluding remarks. In a series of appendices we report the most technical parts of the calculations.

\section{Models and boundary conditions}
\label{model}

\subsection{Directed polymer in the continuum}

Here we study the continuum model for the directed polymer in $d=1+1$ dimension in a gaussian random potential,
defined by the partition sum $Z(x,t|y,0)$ of paths $x(\tau) \in R$ starting at $x(0)=y$ and ending at $x(t)=x$ 
\be \label{zdef} 
Z(x,t|y,0) = \int_{x(0)=y}^{x(t)=x}  Dx e^{- \frac{1}{T} \int_0^t d\tau [ \frac{1}{2}  (\frac{d x}{d\tau})^2  + V(x(\tau),\tau) ]}\, ,
\ee
which is the solution of the Feynman-Kac equation
\be \label{diffeq}
\partial_t Z = - H_d Z = \frac{T}{2} \nabla_x^2 Z - \frac{1}{T} V(x,t) Z\,,
\ee
with initial condition $Z(x,t=0|y,0)=\delta(x-y)$. 
In the $\delta$-correlated continuum model, the random potential $V(x,t)$ is furthermore chosen centered Gaussian with correlator 
\be \label{delta}
\overline{V(x,t) V(x',t)} = \bar c \delta(t-t') \delta(x-x'). 
\ee
By a change of units, i.e. by a rescaling $x \to 
T^3 x$ and $t \to 2 T^5 t$ one can eliminate the temperature $T$: below we work in these units hence set $T=1$ everywhere. 
In addition it is also possible to also set $\bar c=1$ by a further change of units 
$x \to x/\bar c$ and $t \to t/\bar c^2$, a freedom that we sometimes use below (when explicitly indicated). 

As discussed in details in \cite{highT,we}, the $\delta$-correlated continuum model (\ref{zdef}) 
describes the universal high $T$ limit of (i) the DP on a lattice (ii) a continuum DP model, but
where the disorder has a finite correlation length $r_f$ along $x$ (by contrast $r_f=0$ in (\ref{delta})). 
In both cases the description by the $\delta$-correlated model becomes exact 
when the large scale which appears at high temperature is much larger than
the small scale cutoff of the model, e.g. $T^3/\bar c \gg r_f$ in case (ii). 

\subsection{KPZ equation in the continuum}

An important motivation to study the DP problem comes from its relation to the KPZ growth equation for the 
height field $h(x,t)$
\be \label{kpzeq}
\partial_t h = \nu \nabla^2 h + \frac{1}{2} \lambda_0 (\nabla h)^2 + \xi(x,t),
\ee
in presence of a noise $\xi(x,t)$, where $\nu$ is diffusivity and the strength $\lambda_0$ of the non-linear term
has no relation to the parameter $\lambda$. The Cole-Hopf mapping 
\be
Z(x,t)=e^{\frac{\lambda_0}{2 \nu} h(x,t)}, \qquad  T=2 \nu, \qquad V(x,t)=- \lambda_0 \xi(x,t),
\ee 
maps (\ref{diffeq}) into (\ref{kpzeq}). More precisely, the exact relation in terms of the initial condition $h(x,t=0)$ for the KPZ interface is
\be
e^{\frac{\lambda_0}{2 \nu} h(x,t)} = \int dy Z(x,t|y,0) e^{\frac{\lambda_0}{2 \nu} h(y,t=0)}.
\ee
where $Z(x,t|y,0)$ is the DP partition function given by (\ref{zdef}). At this stage, these relations are
valid for arbitrary $V(x,t)$ and $\xi(x,t)$ (and also in any dimension but here we study $d=1$). In the standard
KPZ problem the noise is also assumed to be a centered Gaussian white noise with
\bea
\overline{\xi(x,t) \xi(x',t')}=D \delta(x-x') \delta(t-t')\,,
\eea
and so it corresponds to the $\delta$-correlated continuum DP (\ref{delta}), upon 
the further identification $\bar c = D \lambda_{0}^2$. We mwntion that 
there are some mathematical difficulties in defining KPZ equation with white noise \cite{corwinDP,reviewCorwin}, but that
they are usually surmounted by defining it precisely as the Cole-Hopf transform of
the DP problem, which does not suffer from such problems. 

By correspondence with the DP (see above), the white noise KPZ equation describes
a universal limit of growth models. For instance, if we consider a continuum KPZ equation
where the noise has finite correlation lengths (along $x$ and $t$) it will be described by
the white noise limit whenever the characteristic time $t^* = 2 (2 \nu)^5/D^2 \lambda_0^4$ 
and space $x^*=\sqrt{\nu t^*}$ scales are much larger than these correlation lengths
\footnote{in that case $D$ is defined as the space time integral of the two point correlator,
see Refs. \cite{highT,we} for the same statement on the DP}. Hence it corresponds 
to the high diffusivity $\nu$ or weak noise $D$ limit.
Similar statements can be made for discrete growth models although we will
not elaborate here. 

\subsection{Boundary conditions} 

It is well known that the critical exponents are independent of the chosen initial condition \cite{reviewCorwin} 
and so they have an high degree 
of universality. By contrast, the distribution of the fluctuations of the height field $h(x,t)$ (i.e. the probability density distribution
of the free energy of the directed polymer) depends on the initial conditions $h(x,t=0)$ (or at least some
features of them) even for large time and so the initial condition 
specifies different universality classes (see \cite{reviewCorwin} p. 16 for a detailed classification). 
Three main categories of initial conditions have been identified that are (i) the sharp wedge (also called droplet) initial condition, 
(ii) the flat one and (iii) the random one (which includes the stationary one). 
For large time, the probability distribution functions are expected to crossover to distinct universal distributions, some related to 
the Tracy-Widom distributions for the largest eigenvalues of various random matrix ensembles. In addition to the infinite time behaviour, for the
continuum KPZ equation with white noise (as defined above) each main class 
exhibits also a one parameter family of universal finite time crossover distributions (indexed by $\lambda$)
and here we will mainly study the one associated to the flat initial condition.

A slightly more general, and particularly useful initial condition that we also study here is the so-called {\it wedge} initial condition
\be
\frac{\lambda_0}{2 \nu} h_{\rm wedge}(x,t=0)= - w |x|,
\label{wedgeic}
\ee
that has the important property to interpolate (for any $t$) between the infinitely narrow wedge for $w \to \infty$ (sharp initial condition) and the 
{\it flat interface} initial condition for $w\to 0^+$, thus having as limiting cases two of the three important random matrix ensembles. In addition,
as discussed below, for fixed $w$ but infinite $t$ it converges to the fixed endpoints (droplet) class, as physically reasonable. 
 
In our previous work \cite{we} we studied the sharp wedge initial condition for the KPZ interface, that corresponds to the 
DP with the two ends fixed (say at $x_0=0$), i.e. $Z=Z(0,t|0,0)$, and obtained the distribution of $\ln Z$ for any time. 
In this paper we are interested in the {\it flat interface} initial conditions of the KPZ growth problem, i.e. $h(x,t=0)=0$. 
This corresponds to a DP where one end has been fixed and the other is instead free, so that the partition function of interest is
\be
Z_{\rm flat}(x,t) = \int dy Z(x,t|y,0), 
\label{Zflatdef}
\ee
i.e. the sum of all directed paths which join a given point $x(t)=x$ to any point $y$ on the line at $t=0$. 
As already stated, in terms of the wedge initial condition, this corresponds to the opposite limit $w \to 0^+$. 
We will also discuss here the general case, i.e. for any $w\geq 0$, which interpolates between the two problems, but we are still 
unable to provide a full solution. 


In order to achieve the solution of the flat initial condition, we first attempt to solve directly the case $w=0$. 
Although the integral (\ref{Zflatdef}) is perfectly convergent, due to the fast decay of the diffusion kernel 
(stretching energy of the DP), divergences appear in the Fourier representation of the Bethe states 
in the infinite space problem, and force us to start from a finite (periodic) system of length $L$ along $y$, and 
take the limit $L\to\infty$ only after the calculations of the moments of the partition function $Z(x,t)$ in (\ref{Zflatdef}). 
This way of proceeding is plagued by further technical difficulties that allow us to get only partial results.
For this reason we move to the full-space wedge initial condition (\ref{wedgeic}) with DP partition sum
\be
 Z_{\rm fs,w}(x,t) = \int_{-\infty}^{\infty} dy e^{-w |y|} Z(x,t|y,0)\,,
\ee
with $Z(x,t=0)= e^{-w |x|}$. In this case all divergences are canceled by the exponential decay of the initial condition and it is 
possible to work directly in the thermodynamic limit $L=+\infty$. 
We reproduce all the results found in the previous approach and extend them.
However, some other technical problems make difficult a full solution. 
While it does not seem impossible to overcome these technical issues, we have found much easier to consider the 
wedge initial condition for  
(left) half-space problem with partition sum
\be \label{hsdef} 
Z_{\rm hs,w}(x,t) = \int_{-\infty}^{0} dy e^{w y} Z(x,t|y,0)\,,
\ee
with $Z(x,t=0)=\theta(-x) e^{w x}$. 
As the physical intuition suggests, this half-space problem has the important property that at fixed $w=0^+$ it interpolates between 
(i) the narrow wedge initial condition for $x \to +\infty$, since the polymer is then stretched and 
(ii) the flat, full-space initial condition (\ref{Zflatdef}) for $x \to -\infty$, because very far from the origin the polymer does not feel the presence of 
a boundary.
It is easy to show this from the statistical tilt symmetry (STS) of the problem (see
\ref{sts}). 
This implies the relation for the moments of the partition function for the half-space problems
\bea \label{stsrel}
\overline{Z^n_w(x,t)} = e^{-\frac{n x^2}{4 t}} \overline{Z^n_{w+\frac{x}{2 t}}(0,t)} \,.
\eea
Thus changing the endpoint is the same as changing $w$, up to a simple additive piece
in the free energy. The problem with $w=0$ and $x>0$ is the same
as $w=\frac{x}{2 t}$ and endpoint at $0$, and in the limit $x \to + \infty$ it thus identify with the
sharp edge problem. Eq. (\ref{stsrel}) also implies that $\overline{Z^n_w(x,t)} = e^{n w (x + w t) } \overline{Z^n_0(x+ 2 w t,t)}$.
Hence if $x \to - \infty$ faster than $2 w t$ tends to $+\infty$ the problem becomes a full space one
with an edge far away, as claimed above. The only difference is a uniform shift $2 w t$ in space 
and $w(x+w t)$ in the average free energy. 
 
Thus the final solution for the flat initial condition will be achieved by considering the limit
\be 
\lim_{x \to - \infty, w \to 0} Z_{\rm hs}(x,t)  \equiv Z_{\rm flat}(x,t).
\ee
The limit being taken as discussed above. 
Note that this limit still has $x$ dependence in a given disorder realization.

\section{Quantum mechanics}
\label{BAsec}

The calculation of the $n$-th integer moment of a DP partition sum can be expressed as a quantum mechanical problem 
(see e.g.  \cite{kardareplica,bb-00}).
Upon replication of (\ref{zdef}), averaging over  disorder  and using the Feynman-Kac formula, one finds that the moments 
\be
{\cal Z}_n := \overline{Z(x_1 t|y_1,0)..Z(x_n t|y_n,0)}\,,
\ee 
satisfy \be
\partial_t {\cal Z}_n = - H_n {\cal Z}_n\,,
\ee 
where $H_n$ is the (attractive) Lieb-Liniger (LL) Hamiltonian \cite{ll} for $n$ particles with interaction parameter $c=-\bar c$
\be
H_n = -\sum_{j=1}^n \frac{\partial^2}{\partial {x_j^2}}  + 2 c \sum_{1 \leq  i<j \leq n} \delta(x_i - x_j).
\label{LL}
\ee
In other words the ${\cal Z}_n$ can be written as (imaginary time) quantum mechanical expectations:
\bea \label{expect}
{\cal Z}_n = \langle x_1,..x_n | e^{- t H_n} |y_1,..y_n \rangle\,.
\eea
At this stage, and to compute ${\cal Z}_n$ for arbitrary endpoints one would need all
eigenstates of $H_n$, not only the symmetric ones (bosons). As explained below however,
for the observables computed here we will need only the bosonic ones.

This many body quantum mechanical model is solvable by means of Bethe ansatz \cite{ll}.
This consists in an educated guess for the symmetric eigenstates $|\mu \rangle$ of the many-body wave-functions 
$\Psi_\mu(x_1,..x_n)= \langle x_1,..x_n |\mu \rangle$ that
are superpositions of plane waves \cite{ll}  of the form
\be \label{def1}\fl
\Psi_\mu(x_1,..x_n) =  \sum_P A_P \prod_{j=1}^n e^{i \sum_{\alpha=1}^n \l_{P_\alpha} x_\alpha} \, , \quad 
A_P=\prod_{n \geq \beta > \alpha \geq 1} \Big(1- \frac{i c ~\sgn(x_\beta - x_\alpha)}{\lambda_{P_\beta} - \lambda_{P_\alpha}}\Big)\,.
\ee
The sum runs over all $n!$ permutations $P$ of the rapidities $\l_j$.
As written, these Bethe wave functions are not normalized and satisfy 
\be 
\langle x \cdots x |\mu\rangle =  \Psi_\mu(x,..x) = n! e^{ i x \sum_\alpha \lambda_\alpha} \,.
\label{Psixeq}
\ee 
The rapidities must be all different, i.e. $\lambda_\alpha \neq \lambda_\beta$ for $\alpha \neq \beta$ (the wavefunctions are antisymmetric in exchange of 2 rapidities, 
hence vanish if 2 are equal). 
When working with periodic boundary condition, also $ \Psi_\mu(x_1,..x_n)$ must be periodic in any of the variables and 
this forces the set of rapidities $\{ \l \}$ to be solution to the  equations 
\begin{eqnarray}
e^{i \lambda_{\alpha} L} = \prod_{\beta \neq \alpha} \frac{\lambda_{\alpha} - \lambda_{\beta} + i c}
{\lambda_{\alpha} - \lambda_{\beta} - i c},
\label{BE}
\end{eqnarray}
that are the {\it Bethe equations} for the Lieb-Liniger model.
In particular these imply $e^{i \sum_{\alpha=1}^n \lambda_\alpha}=1$ and hence the total moment 
$K=\sum_{j=\alpha}^n \lambda_\alpha=2 \pi p/L$ is quantized in terms of integer numbers as it must be.

The solution for the attractive case has been described soon after the Lieb-Liniger repulsive solution by McGuire \cite{m-65}.
We report here the main ingredients to understand the structure of the wave functions.
For $\bar{c}=-c > 0$ the Bethe equations are
\begin{eqnarray}
e^{i \lambda_{\alpha} L} = \prod_{\beta \neq \alpha} \frac{\lambda_{\alpha} - \lambda_{\beta} - i\bar{c}}
{\lambda_{\alpha} - \lambda_{\beta} + i\bar{c}}.
\label{BEa}
\end{eqnarray}
Consider now a complex rapidity $\lambda_{\alpha} = \lambda + i \eta$.  
The Bethe equation for this rapidity is
\begin{eqnarray}
e^{i\lambda_{\alpha} L} = e^{i \lambda L - \eta L} = 
\prod_{\beta \neq \alpha} \frac{\lambda_{\alpha} - \lambda_{\beta} - i\bar{c}}
{\lambda_{\alpha} - \lambda_{\beta} + i\bar{c}}.
\end{eqnarray}
We consider $n$ to be finite  and $L \rightarrow \infty$.  
If $\eta > 0$, we have $e^{-\eta L} \rightarrow 0$ on the left-hand side.  
Looking at the finite product on the right-hand side, 
we conclude that there must thus be a rapidity $\lambda_{\alpha'}$ such that 
$\lambda_{\alpha'} = \lambda_{\alpha} - i \bar{c} + \mbox{O}(e^{-\eta L})$.
On the other hand, if $\eta < 0$, we have $e^{-\eta L} \rightarrow \infty$  on the left-hand side,
and there must thus be a rapidity $\lambda_{\alpha'}$ such that 
$\lambda_{\alpha'} = \lambda_{\alpha} + i \bar{c} + \mbox{O}(e^{-|\eta| L})$.
Furthermore if $\lambda=\lambda_\alpha$ is a solution also $\lambda^*$ (if $\lambda\neq \lambda^*$) must be for some 
$ \beta\neq\alpha$ as evident by taking the complex conjugate of the Bethe equations (\ref{BEa}). 

Thus the rapidities solution to the Bethe equations for $c<0$ and $L\to\infty$ arrange in object formed by an arbitrary number of 
particles $m\leq n$ with the same real part and imaginary parts that differ by integer multiples of $i c$ and that are symmetric 
with respect to the real axis. These objects are called {\it strings}.
A general eigenstate is built by partitioning the $n$ particles into a set of $n_s \leq n$ 
strings formed by $m_j \geq 1$ particles with $n=\sum_{j=1}^{n_s} m_j$. 
The rapidities associated to these states are written as
\be\label{stringsol}
\l^{j, a}=k_j +\frac{i\cb}2(m_j+1-2a)+i\d^{j,a}.
\ee 
Here, $a = 1,...,m_j$ labels the rapidities within the string $j=1,\dots n_s$.  
$\d^{j,a}$ are deviations which fall off exponentially with system size $L$.
As it should be clear, perfect strings (i.e. with $\d=0$) are exact eigenstates in the limit $L\to\infty$ for arbitrary $n$.
The (unnormalized) wave function for a string of $m$ particles (which in the following will be simply called an {\it m-string}) is
\begin{eqnarray}
\Psi(x_1,..x_m) = n! 
\exp \Big({- \frac{\bar c}{2} \sum_{1\leq i <j\leq m} |x_i-x_j|}\Big)\,,
\end{eqnarray}
and so each string represents a bound states of $m_j$ particles with its own energy and momentum given by
\be
K_j= m_j k_j,\qquad E_j= m_j k_j^2-\frac{\cb^2}{12} m_j(m_j^2-1).
\ee 
The fact that all deviations are exponentially small in $L$ implies that we can obtain 
a much simpler set of equations involving only the string momenta $K_j$ reported in Ref. \cite{cc-07}, that in the 
limit of $L=\infty$ gives vanishing coupling between strings. 
Thus for all our aims, the only effect of the (attractive) interaction is to produce a set of bound states.
When instead interested in the directed polymer in a finite geometry, other methods to tackle 
exactly the Bethe equations should be used, as done e.g. in Ref. \cite{bb-00}.

The general eigenstate consisting of partitioning the $n$ particles into a set of $n_s$ 
strings formed by $m_j \geq 1$ particles with $n=\sum_{j=1}^{n_s} m_j$ have momentum  and energy 
\be
K_\mu=\sum_{j=1}^{n_s} m_j k_j,\qquad E_\mu=\sum_{j=1}^{n_s} \left(m_j k_j^2-\frac{\cb^2}{12} m_j(m_j^2-1)\right).
\label{enmom}
\ee 
Clearly, by minimization of the energy, the ground-state is the single string built with $n$ particles and 
having zero total momentum, while excited states are obtained by either giving momentum to the ground state, 
or by partitioning it into smaller strings to which individual momenta can be given.

\subsection{The replica method for the flat and wedge initial condition}
\label{sec:replica}

In this paper we are interested in the moments of the partition function for the flat initial condition in 
Eq. (\ref{Zflatdef}). By definition in the replica approach these are
\be
\overline{Z_{\rm flat}(x,t)^n} =   \int_{-\infty}^\infty \Big(\prod_{j=1}^n dy_j\Big) \langle y_1 \dots y_n |e^{- t H_n} |x \dots x \rangle \,.
\ee
As compared to (\ref{expect}) we have, for future convenience, exchanged the starting and the end points of the polymer, using that
$\overline{Z_{\rm flat}(x,t)^n}$ is real and $H_n$ hermitian. Assuming completeness of the string states, we can insert a resolution of the identity over the Bethe states $|\mu\rangle$ (representing all the possible allowed rapidities) and obtain
\bea
\overline{Z_{\rm flat}(x,t)^n}&=& \sum_\mu   \int_{-\infty}^\infty \Big(\prod_{j=1}^n dy_j\Big)
\frac{ \langle y_1 \dots y_n |\mu\rangle \langle\mu| x \dots x \rangle}{||\mu ||^2} e^{-t E_\mu}\nn \\&=&
\sum_\mu \frac{\Psi_\mu^*(x,..x)}{||\mu ||^2} e^{-t E_\mu}  
\int_{-\infty}^\infty \Big(\prod_{j=1}^n dy_j\Big) \Psi_\mu(y_1\dots y_n)\,.
 \label{sumf}
\eea
Notice that it has been possible to use the solution of the bosonic problem only because the state $|x\dots x\rangle$
is symmetric for exchange of particles (replicas). Thus it has vanishing overlap with anti-symmetric states representing 
fermionic wave functions which are not included in the above Bethe ansatz solution. 

In Eq. (\ref{sumf}), the energy $E_\mu$ is given in Eq. (\ref{enmom}) and the equal points wave function  $\Psi_\mu^*(x,..x)$ 
in Eq. (\ref{Psixeq}).
The norms of the Bethe states within our normalization have been calculated in Ref. \cite{cc-07} by adapting the
famous Gaudin-Korepin formula \cite{gk} for norms of Bethe states to the case of strings. 
The final result can be written as  \cite{cc-07}
\be
\fl \frac{1}{||\mu ||^2} = \frac{\bar{c}^{n}}{n!  (L \bar{c})^{n_s} } \Phi[k,m]  \prod_{j=1}^{n_s} \frac1{m_j^{2}}  \, , \quad  \Phi[k,m]=
\prod_{1\leq i<j\leq n_s} 
\frac{4(k_i-k_j)^2 +(m_i-m_j)^2 c^2}{4(k_i-k_j)^2 +(m_i+m_j)^2 c^2} \label{norm}\,.
\ee
It is then evident that the only ingredient missing in Eq. (\ref{sumf}) is the integral of Bethe wave function for string states 
that will be one of the main tasks of the following.

It is straightforward to write down the corresponding formulas for wedge initial condition both in full- and half-space problem.  
We have simply 
\be
  \overline{Z_w(x,t)^n} =   
 \sum_\mu \frac{ \Psi_\mu^*(x,..x)}{||\mu ||^2} e^{-t E_\mu}  \int^w \Psi_\mu\,,
\ee
where we defined  
\bea
\fl \int^w \Psi_\mu := \int_{-\infty}^\infty  \Big(\prod_{j=1}^n dy_j\Big) e^{-w \sum_{i=1}^n |y_i|}  \Psi_\mu(y_1,..y_n)\,,  
  &\quad & \textrm{for the full-space problem}, \\
\fl \int^w \Psi_\mu := \int_{-\infty}^0 \Big(\prod_{j=1}^n dy_j\Big) e^{w \sum_{i=1}^n y_i}  \Psi_\mu(y_1,..y_n)\,,  
 \fl & &\textrm{for the half-space problem}. \label{int1}
\eea
Note that in the limit $w \to + \infty$ one should recover the DP with both endpoints fixed (one at $x$ the other at $0$) i.e. the droplet initial
condition, by the simple replacement $e^{-w |y|} \to \frac{2}{w} \delta(y)$ for the
full space and $e^{w y} \theta(-y) \to \frac{1}{w} \delta(y)$ for the half-space model. The factors $2/w$ and $1/w$ lead to an
additive normalization $\sim \ln w$ in the free energy. 

\subsection{Starting formula for moments} \label{sec:starting} 

To summarize we have the starting expression for the moments
\bea  \label{partsum}
\fl \overline{Z(x,t)^n} = \sum_{n_s=1}^n \frac{\bar c^n}{n_s! (\bar c L)^{n_s}} \sum_{(m_1,\dots m_{n_s})_n} 
\prod_{j=1}^{n_s} \frac{1}{m_j^2} \nn\\\fl\qquad\qquad\qquad\qquad\qquad \times
\sum_{k_j} (\int  \Psi_\mu)  \Phi[k,m] 
 \prod_{j=1}^{n_s}e^{(m_j^3 - m_j) \frac{\cb^2 t}{12}- m_j k_j^2 t - i x m_j k_j} \,,
 \label{zn} 
\eea
where $\int  \Psi_\mu$ stands for the various cases, and is a function of $[k,m]$. Here 
$(m_1,\dots m_{n_s})_n$ stands for all the partitioning of $n$ such that $\sum_{j=1}^{n_s} m_j=n$
with $m_j \geq 1$.  In the limit $L\to\infty$ the momenta sums become
continuous and one can use that the string
momenta $m_j k_j$ correspond to free particles i.e. 
$\sum_{k_j} \to m_j L \int \frac{dk_j}{2 \pi}$. We will however also often make use of
the discrete sum (\ref{partsum}) below whenever the momenta are constrained
to special values, e.g. to vanish. 

\subsection{Starting formula for the generating function} \label{sec:starting2} 

Let us recall that the generating function 
\begin{eqnarray} \label{defg}
g(s) = 1+ \sum_{n} \frac{\overline{Z(x,t)^n}}{n!} (-1)^n e^{- \lambda n s} =  1 +  \sum_{n_s=1}^\infty \frac1{n_s!}  Z(n_s,s) \,,
\end{eqnarray}
admits an expansion in the number of strings
\begin{eqnarray}  
&&  \fl Z(n_s,s) =  \sum_{m_1,\dots m_{n_s}=1}^\infty  \frac{(- \bar c)^{\sum_j m_j}  }{(\bar c L)^{n_s} (\sum_j m_j)! }
 \prod_{j=1}^{n_s} \frac{1}{m_j^2} \nn\\ && \fl \qquad \qquad \qquad\qquad\qquad\times
 \sum_{k_j} (\int  \Psi_{\mu}) \Phi[k,m]
 \prod_{j=1}^{n_s}e^{(m_j^3 - m_j) \frac{\cb^2 t}{12}- m_j k_j^2 t - i x m_j k_j - \lambda m_j s}, 
  \label{Zns}
\end{eqnarray}
obtained by permuting the sum over the number of particles and strings, which allows to free
the constraint on the summations over the number of particles, as in a grand canonical ensemble
where $\lambda s$ acts as a chemical potential. 

As discussed in \cite{we,dotsenko}, while the formula (\ref{partsum}) is perfectly well defined, because of the exponential cubic divergence of the series, the formula 
(\ref{Zns})  should be taken in some analytical continuation sense, i.e. it is valid
as a formal series in $t$. This continuation is performed in the following using the famous  {\it Airy trick} 
which for ${\rm Re}(w)>0$ ensures
\be  \label{airytrick} 
 \int_{-\infty}^\infty dy Ai(y) e^{y w} = e^{w^3/3}\,.
\ee
the summations can then be later carried at fixed $y$.

\section{Direct approach to the flat initial condition}
\label{DIRECT}

From Eq. (\ref{sumf}) we need to evaluate the integral
\be
\int_{-\infty}^\infty \Big(\prod_{j=1}^n dy_j\Big) \Psi_\mu(y_1\dots y_n)\,,
\ee
over all possible string wave functions. 
In order to do so let us start by considering the auxiliary integral over the principal domain
$PD=\{\{y_j\}| 0<y_1<y_2<..<y_n<L\}$ 
\be
\fl B_n(\lambda_1,..\lambda_n) = \int_{PD} dy_1.. dy_n e^{i \lambda_1 y_1 + .. + i \lambda_n y_n}  =  
i^n \big(  \sum_{j=1}^{n} (-1)^j e^{i L \sum_{i=n-j+1}^n \lambda_i} \frac{1}{a_j b_j} + \frac{1}{a_{0} b_{0}} \big), 
\ee
where we introduced
\be
a_j = \prod_{p=1}^j \sum_{i=n-j+1}^{n-p+1} \lambda_i \quad , \quad b_j = \prod_{p=j}^{n-1} \sum_{i=n-p}^{n-j} \lambda_i\,.
\ee
Clearly this formula is true if and only if the set $\{ \Lambda_{ij}|_{1 \leq i \leq j \leq n} \}$ with
\begin{eqnarray}
&& \Lambda_{i,j} = \sum_{p=i}^{p=j} \lambda_i \,,
\end{eqnarray}
does {\it not} contain any zero, else the corresponding term in the sum diverges.
Using the symmetry under permutation of the $y_j$ we have for a general Bethe state (\ref{def1}) 
\be
 \int d^n y  \Psi_\mu(y)  = n! \sum_P B_n(\lambda_{P_1},..\lambda_{P_n}) 
\prod_{n \geq \ell > k \geq 1} (1- \frac{i c}{\lambda_{P_\ell} - \lambda_{P_k}})\,.
\ee
Fixing the values of $\lambda_j$ by using the Bethe equations (\ref{BE}), it is possible (but cumbersome) to show that this 
combination always vanishes:
\be
 \int d^n y  \Psi_\mu(y)  = 0\,.
\ee

This result shows that in order to have a non-vanishing contribution of the Bethe wave function, we need to require that at least one of the 
$\Lambda_{ij}$ should be zero (up to permutation). To obtain the result in that case one needs to (i) enumerate all possibilities for
a subset of $\Lambda_{ij}$ to vanish (ii) perform limits in the above expression. Below we give some partial results, with few details,
being assumed that in each reported case we have searched all such possibilities and performed the necessary limits (which is
tedious). 

In order to search for the general structure of the states that contribute non-trivially to this problem, we first consider the
calculation at fixed $n$, i.e. the average of the integer moments of the partition function of the directed polymer. 
Next, we also obtain some results for fixed number of strings $n_s=1,2$. For actual calculations 
a useful formula is:
\begin{eqnarray} \label{laplaceL}
&& \int_0^\infty dL e^{-s  L} B_n(\lambda_1,..\lambda_n) = \frac{1}{s} \prod_{i=1}^n \frac{1}{s - i \mu_i} \quad , \quad \mu_i=\sum_{j=i}^n \lambda_j.
\end{eqnarray}
Upon summing over permutations, Laplace inversion and using the Bethe
equations (BE) we obtain the results discussed below. One can see on this formula that multiple poles
in $1/s$ lead to powers of $L$.

In the rest of this section, to have a light notation, we remove the subscript ``flat'' from the partition function and we  fix 
(without lost of generalities) $x=0$. 
Thus everywhere in this section  $\overline{Z^n}$ stands for $\overline{Z_{\rm flat}(x,t)^n}$ in Eq. (\ref{sumf}).

\subsection{Fixed $n$ results}

We show how to obtain from the previous expression $\overline{Z^n}$ for $n=2,3,4$.
An important check that will be a guide to check the correctness of our manipulations and to understand if we 
considered all the possible Bethe states is that for $t=0$ all the results should just give the trivial expectation value
of the flat condition $h(x,t=0)=0$, i.e.  $\overline{Z_{\rm flat}(x,t=0)^n}=1$.

\subsubsection{$n=2$.}
 
This is the simplest possible case. Clearly there are no Bethe states  with $\lambda_j=0$, thus the only possibility is
$\lambda_1=-\lambda_2$. For these states, simple algebra leads to
\be \label{n2}
\int d^2 y  \Psi(y)  = - 2 \bar c L \frac{1}{\lambda_1^2}\,.
\ee
There are two possible string states satisfying $\lambda_1=-\lambda_2$, i.e.
\begin{itemize}
\item one string of two particles (2-string state) with zero momentum ($n_s=1$); 
\item states with two particles (1-string) with opposite real momentum $\lambda_1=-\lambda_2=k$ ($n_s=2$). 
\end{itemize}
We can now write the contribution of each of these states to $\overline{Z^2}$ using the general formula
(\ref{zn}). Summing up these contributions and integrating over $k$ we obtain the exact result for the
second moment
\begin{eqnarray} \label{z2}
 \overline{Z^2} &=&   2 e^{\frac{c^2}{12} 6 t}  +  \frac{\bar c^2}{2! (\bar c L)^{2}} L \int \frac{dk_1}{2 \pi} 
 (- 2 \bar c L \frac{1}{k_1^2}) \frac{4 k_1^2}{4 k_1^2 + c^2} e^{- 2 k_1^2 t} \\
& =&   2  e^{\frac{c^2}{2}  t} - 4  \bar c  \int \frac{dk_1}{2 \pi} 
   \frac{1}{4 k_1^2 + c^2} e^{- 2 k_1^2 t}  =
   e^{2 \lambda^3} (1+ {\rm erf}(\lambda^{3/2} \sqrt{2}) )\,,
\end{eqnarray}
with ${\rm erf}(z)=\pi^{-1/2} \int_0^z dt e^{-t^2}$ and $\lambda^3 = c^2 t/4$. 
Notice the $1/2!$ factor in front of the integral comes from the $1/n_s!$ in (\ref{zn}) and 
avoids double counting of the states. For $t=0$ (i.e. $\lambda=0$) 
we recover $ \overline{Z^2}=1$ as a non-trivial check of the calculation.

\subsubsection{$n=3$.}

Only a subset of the states with some $\Lambda_{ij}=0$ provide a non-vanishing contribution to $\overline{Z^3}$. 
We report here only the calculation for the states with a non-zero contribution that are
\begin{itemize}
\item A 3-string state of zero momentum ($n_s=1$)
for which $\lambda_1^2=\lambda_3^2=-c^2$ and $\lambda_2=0$. The integral is 
\be
 \int d^3 y  \Psi(y)  = \frac{36 L}{c^2}\,.
\ee
\item One 2-string with $k=0$, i.e. $\lambda_2^2=-c^2/4$ and one particle with $k=0$ ($n_s=2$), with integral
\be
 \int d^3 y  \Psi(y)  = 72 \frac{L^2}{c}\,.
\ee
\item Three 1-strings with  $(\lambda_1,\lambda_2,\lambda_3)=(k_1,k_2,k_3)=(0,k,-k)$ ($n_s=3$), with integral
\be
 \int d^3 y  \Psi(y)  = \frac{6 c L^2 (c^2 + k^2)}{k^4}\,.
\ee
\end{itemize}
We now sum up these three contributions to $\overline{Z^3}$ using the general formula
(\ref{zn}) and displaying all factors for clarity
\begin{eqnarray} \label{z3res} 
  \overline{Z^3} &=&   
4 e^{2 c^2 t}  +\frac{\bar c^3}{2! (\bar c L)^{2}} 2 \frac{1}{4} 
(- 72 \frac{L^2}{\bar c}) \frac{c^2/4}{9 c^2/4} e^{\frac{1}{2} c^2 t}  
\\
&& + 3 \frac{\bar c^3}{3! (\bar c L)^{3}} L \int \frac{dk}{2 \pi}  \frac{6 c L^2 (c^2 + k^2)}{k^4} (\frac{k^2}{k^2 + c^2})^2 \frac{4 k^2}{4 k^2 + c^2} e^{-2 k^2 t}  \nn \\
& =& 
 4 e^{2 c^2 t} - 2  e^{\frac{1}{2} c^2 t}  - 3 \int \frac{dk}{2 \pi} \frac{4 \bar c k^2}{(k^2 + c^2)(4 k^2 + c^2)} e^{-2 k^2 t}\,,\nn
\end{eqnarray}
where the factor $3$ in front of the integral in the last line is the symmetry factor for the number of distinct triplets $(0,k,-k)$.
Integrating over $k$ we obtain the exact formula for the third moment:
\begin{eqnarray}
&& \overline{Z^3} =  4 e^{8 \lambda^3 } - 2 e^{2 \lambda^3 }  -  2 e^{8 \lambda^3 } {\rm erfc}(\lambda^{3/2} 2 \sqrt{2}) + e^{2 \lambda^3  } {\rm erfc}(\lambda^{3/2} \sqrt{2}) )\,,
\end{eqnarray}
with ${\rm erfc}(x)=1-{\rm erf}(x)$. For $t=0$ it reproduces correctly $ \overline{Z^3}=1$

\subsubsection{$n=4$.}

Let us consider now $n=4$ and report only the states giving a non-zero contribution. We will give only the final result
while the detailed contributions and the integrals are given in \ref{app:n4}. The contributing states are:
\begin{itemize}
\item $n_s=4$, four 1-strings with  $k_2=-k_1$ and $k_4=-k_3$. 
\item  $n_s=3$, two 1-strings of opposite momenta $k$ and one 2-string of zero momentum  (i.e. total state 
$(i \frac{c}{2}, -i \frac{c}{2},k,-k)$). 
\item $n_s=2$ with two 2-strings, of opposite momenta, i.e. $(k+i c/2, k-i c/2, - k+ic/2,-k-i c/2)$.
\item $n_s=2$ one 3-string and one 1-string both of zero momentum, i.e. $(i c, 0,-i c, 0)$.
Note that from very general arguments  this state {\it should not exist}, because two rapidities have the same value. 
However, this state exists and is the infinite $L$ manifestation of a deformed string (see for a discussion of the topic Ref. \cite{hc-07}).
In the specific case, the deformed string originating this state is $(i c+\delta_1, \epsilon_1,-i c+\delta_1, \epsilon_2)$,
where $\epsilon_i$ are the string deviations discussed previously. For any finite system $\epsilon_1\neq\epsilon_2$ and so the state
is allowed and becomes  pathological only in the thermodynamic limit. 
As discussed in Ref. \cite{hc-07} these deformed strings are essential to ensure the completeness of the string states. Here we missed this term at first, but found it
later by considering the wedge initial condition, which allows to treat directly $L=+\infty$, in the limit
$w=0^+$. This is detailed in \ref{app:n4}. 

\item $n_s=1$, i.e. a single 4-string of zero momentum. 
\end{itemize}
Summing up all the non-vanishing contributions we have
\begin{eqnarray}
\fl \overline{Z^4} &=&   8 e^{20 \lambda^3} - 8 e^{8 \lambda^3}
-4  e^{8 \lambda^3}  (-2 {\rm erfc}(2 \lambda^{3/2}) + e^{12 \lambda^3} {\rm erfc}(4 \lambda^{3/2}))
\nn \\ \fl&& - 4  e^{18 \lambda^3} {\rm erfc}(3 \sqrt{2} \lambda^{3/2}) e^{2 \lambda^3}
+  48  \int_0^\infty dx 
 \frac{\left(\frac{2 x+1}{\sqrt{4 x+1}}-\frac{\sqrt{2 x+1}}{x+1}\right) e^{-8 \lambda^3 x}}{4 \pi  (4 x (x+3)+5)} \nn\\
 \fl & =& 1 + 12 \sqrt{\frac{2}{\pi}} \lambda^{3/2}  + 51.6394 \lambda^3 + \dots\,,
 \end{eqnarray}
that for $\lambda=0$ gives $1$ as it should.
The contribution of the deformed string is essential in order to recover $\overline{Z^4}=1$ at $t=0$ as a 
reflection of its importance for the completeness.

\subsubsection{First conclusions.}

Although some rules seem to emerge from the study of $n=2,3,4$ as to which are the contributing states, we will not pursue further
here the direct approach to arbitrary $n$. It becomes indeed very tedious for higher $n$ since the BE must be used to show that
some states will not contribute (in an a priori non-obvious way). At this stage we found that contributing states contain
either strings of zero momenta or pairs of strings of opposite momenta. This rule will appear much more clearly and
easily in the wedge initial condition studied in the next Section. More precisely it will appear that the
contributing states are obtained by splitting the $n_s=2 N+M$ strings into $N$ pairs of strings of opposite momenta 
and $M$ single strings of zero momentum. It would be nice to prove this rule in the direct approach, but we have not
attempted it. Of course, once the rule is known it may be possible to pursue the calculation in the direct
approach, but we leave this task for the future.

To close the study at fixed but large $L$ we now show that some interesting but only partial results can also
be obtained for arbitrary $n$ but fixed $n_s$.

\subsection{Summation at fixed number of strings $n_s$}

We can also perform summation over $n$ at fixed number of strings $n_s$ and
obtain the first terms in the generating function (\ref{defg},\ref{Zns}). 

\subsubsection{One string contribution $n_s=1$.} \label{sec:ns1} 

It should be clear from the previous section that for $n_s=1$ the only non-zero
contributions come from the  $n$-string wavefunctions with $k=0$, i.e. the ground state, that is 
\be
 \Psi_\mu(x_1,..x_n) = n! e^{- \frac{\bar c}{2} \sum_{i <j} |x_i-x_j|}\,, 
\ee
whose integral is calculated by standard methods
\be
 \int d^n y  \Psi_\mu(y)  = \frac{2^{n-1} n^2 L}{\bar c^{n-1}}  \,.
\ee
Using Eq. (\ref{zn}), this gives the contribution of the ground state to the moments for arbitrary $n$
\be \label{moments1}
\overline{Z^n}|_{n_s=1} =  2^{n-1} e^{\frac{c^2}{12} (n^3-n) t } \,.
\ee
It is convenient to perform the usual shift in the free energy i.e. define
$Z = \tilde Z e^{-c^2 t/12}$, hence $g(s) = \tilde g(s+ \frac{\lambda^2}{3})$ and to
study from now on $\tilde g(s)$. This is equivalent to set
\bea \label{defv0}
v_0 t = - \frac{\lambda^3}{3}\,,
\eea 
in Eq. (\ref{defxi}). Inserting the above expression in Eq. (\ref{Zns})
we obtain the very simple result
\be
 Z(1,s)  =  \sum_{m=1}^\infty  \frac{(-1)^{m} }{m! }
    2^{m-1}  e^{m^3  \frac{\cb^2 t}{12} - \lambda m s } \,.
\ee
Performing now the Airy trick we have  
\be    \label{firstla}
\fl Z(1,s)   = \int dy Ai(y) \sum_{m=1}^\infty  \frac{(-1)^{m} }{m! } 2^{m-1}  e^{\lambda m y  - \lambda m s }  = \frac{1}{2} \int dy Ai(y+s) (e^{- 2 e^{\lambda y}}-1) \,,
\ee
that for $t\to\infty$ is
\be  \label{first} 
 \lim_{\lambda \to + \infty} Z(1,s) = - \frac{1}{2} \int_{y>0} dy Ai(y+s) = - \frac{1}{2} \int_{y>-s} dy Ai(y) \,.
\ee


\paragraph{Analysis of $Z(1,s)$ at infinite $\lambda$:} 
although quite simple, it is remarkable that $Z(1,s)$ at infinite $\lambda$ already reproduces the first term in the 
asymptotic expansion of the GOE Tracy-Widom distribution $F_1(s)$. 
Indeed, as explained in more details in Section  \ref{sec:larget}, $F_1(s) = {\rm Det}[I - B_s]$ is a Fredholm determinant
of the kernel $B_s(x,y) = \theta(x) Ai(x+y+s) \theta(y)$ and as such it can be formally expanded in
powers of $B_s$, i.e. $F_1(s)= 1 - {\rm Tr} B_s + O(B_s^2)$. 
The one string contribution (i.e. the ground state for any $n$) in Eq.  (\ref{first}) 
is exactly $- {\rm Tr} B_s$. 
It turns out that this expansion is organized exactly in the same way as the expansion for 
large negative $s$, which gives $P(f)$ for large negative $f$ (deep states), i.e one finds, from the
asymptotics of the Airy function at large positive argument:
\bea
&& Ai(y) \approx f_{Ai}(y) e^{- \frac{2}{3} y^{3/2}} \quad , \quad  f_{Ai}(y) = \frac{1}{2 \sqrt{\pi} y^{1/4}} \sum_{k=0}^\infty \frac{(-1)^k}{(\frac{2}{3} y^{3/2})^k} Ai_k \,,\\
&&   Ai_n = \frac{(6n-1)(6 n-5)}{72 n} Ai_{n-1} \quad , \quad Ai_0=1\,,
\end{eqnarray}
i.e. $Ai_1=5/72, Ai_2=385/10368,..$, that (for infinite $\lambda$):
\bea
&& \fl {\rm Prob}(f > -s) \approx {\rm Prob}_1(f > -s) = 1 + Z(1,s) \approx 1 + f_1(-s) e^{- \frac{2}{3} (-s)^{3/2} }  \,,
\eea
with $f_1(y)=- \frac{1}{4 \sqrt{\pi}} \sum_{k=0}^\infty C_k y^{- \frac{3}{4}- \frac{3}{2} k}$ with $C_0=1$ and 
$C_k + \frac{3}{2} (k- \frac{1}{2}) C_{k-1} = (-1)^k Ai_k (2/3)^{-k}$ for $k \geq 1$. We recall the discussion
in Ref. \cite{we} showing that one can construct an independent string approximation which gives a 
bona-fide cumulative probability $g_{ind}(s) = e^{Z(1,s)}$ with the same asymptotics as the
exact one for large $s$. 

\paragraph{Analysis $Z(1,s)$ at finite $\lambda$:} 

For any $\lambda$, the moments (\ref{moments1}) can be formally written as
\be
 \overline{Z^n}|_{1string} = \int_{-\infty}^{+\infty} df P_1(f) e^{- \lambda f n} \quad , \quad  P_1(f) = \frac{1}{2} Ai(- f + \frac{\ln2}{\lambda}) .
\ee
Clearly $P_1(f)$ is not a positive function and cannot be a probability distribution:
it should not be indeed, since it contains only contributions of $n_s=1$. However, for large negative $f$
we expect from the above considerations, at least at large enough $\lambda$, 
that it gives the correct tail
\bea
P(f ) \approx_{f \to - \infty} P_1(f) .
\eea 

\subsubsection{Contribution of two strings, $n_s=2$.}  \label{sec:ns2} 
In this case we find that there are only two kinds of contributions: (i) those with the two strings with $m_1=m_2=m$ with non-zero opposite 
momenta (ii) those with $m_1\neq m_2$ both forced to have zero momentum. 
Thus we write $Z(2,s)=Z_a(2,s)+Z_b(2,s)$ that we now calculate. 

$Z_a(2,s)$ is generated only by the states with $n_s=2 m$ even that must be divided in two strings of the same size $m$.
Denoting with $k$ the opposite momentum variable of the two strings (which have total momenta $\pm m k$), we find that 
the integral can be written as 
\be
 \int \Psi = \frac{2 m^3 (2 m-1)! }{\prod_{j=0}^{m-1} (k^2 + c^2 j^2/4) } \bar c (-1)^m L\,, \quad , \quad \Phi=\frac{k^2}{k^2 + m^2 c^2/4},
\ee
after some heuristic manipulations reported in \ref{app:ns2}. We have also written the norm factor $\Phi$ corresponding
to two equal size strings. Inserting this expression in Eq. (\ref{Zns}) we get
\bea
\fl Z_a(2,s)&=& \sum_{m=1}^\infty \frac{(-1)^m}{m} \bar c^{2m-1} \int \frac{dk}{2 \pi} \prod_{j=1}^m \frac{1}{k^2 +c^2  j^2/4} e^{- 2 m k^2 t + 2 \frac{\lambda^3}{3} m^3 - 2 m \lambda s}\nn \\
\fl & =& \sum_{m=1}^\infty \frac{(-1)^m}{m} 2^{2m} \int \frac{dk}{2 \pi} \prod_{j=1}^m \frac{1}{4 k^2 + j^2} e^{- 8 m  \lambda^3 k^2 + 2 \frac{\lambda^3}{3} m^3 - 2 m \lambda s}\,.
\label{Z2a}
\eea
where we have changed $k^2 \to c^2 k^2$ and used $c^2 t=4 \lambda^3$. 

For $m_1\neq m_2$ the calculation is sketched in the \ref{app:ns2} . The final result is
\be \label{ns2Psi}
\fl \int \Psi = \Big(  \frac{2}{\bar c}\Big)^{m_1+m_2-2}\, \frac{m_1 m_2  (m_1+m_2)^2 
  \Gamma(m_1+m_2)}{|m_1-m_2| \Gamma (m_1) \Gamma(m_2)  }   (-1)^{\min(m_1,m_2)} L^2\,.
\ee
and the norm factor has $\Phi = \frac{(m_1-m_2)^2}{(m_1+m_2)^2}$. Inserting into Eq. (\ref{Zns}) we find
\be \label{Z2direct}
\fl Z_b(2,s)= \frac{1}{4} \sum_{m_1 \neq m_2} \frac{(-2)^{m_1+m_2}}{m_1! m_2!} 
\frac{|m_1-m_2|}{m_1+m_2} e^{ \frac{\lambda^3}{3} (m_1^3+m_2^3) - \lambda s (m_1+m_2)}\,.
\label{Z2b}
\ee
An important remark is in order. As already discussed in the previous section for the case $n=4$, one would generally expect that different rapidities cannot have the same values and so one would conclude that $m_1$ and $m_2$ must be of different parities. Indeed this is
the condition which we used in the \ref{app:ns2} for the derivation of (\ref{ns2Psi}) by the direct method. However, as discussed
there, this restriction would lead to missing states. In fact the states with two strings with $m_1\neq m_2$ and with the same parity 
do exist because they are the infinite $L$ limit of deformed strings that at finite $L$ have different rapidities. We will discover it more
easily using the wedge initial condition in Section \ref{sec:poles} and recover exactly (\ref{Z2direct}), see Eq.(\ref{zunpaired}), where the sum is indeed over all pairs $(m_1,m_2)$. We have not attempted a direct derivation of the contribution of these deformed strings.

It is possible to show that starting from the above expression for $Z(2,s)$, using the Airy trick, summing over $m_j$ and taking the limit $t\to\infty$ one recovers the second order in the expansion of $F_1(s)$. Since it is shown for general $n_s$ in the following, we refer to
section \ref{sec:larget} for this calculation.

This concludes the study of the direct method using fixed size $L$ and the Bethe equations. We now turn to
the wedge initial condition which, after a few non trivial steps, will allow to nicely circumvent the 
difficulties of the direct method.

\section{The wedge initial condition in half-space}
\label{HSsec}

In this section we report the calculation of the generating function for the wedge initial condition. 
For the symmetric wedge we have only partial results which are given in \ref{FSsec}. 
Conversely, the half-space wedge allows for a full solution that we report in the following. 

\subsection{Spatial integral of the Bethe eigenfunctions}

Let us consider the (left) half-space wedge model defined by the partition sum (\ref{hsdef}) with $w>0$. 
To compute its moments (\ref{zn}) and their generating function (\ref{Zns}) we need the spatial integral (\ref{int1}) 
over the Bethe states. 
By performing explicit Integration in Eq. (\ref{def1}) over the negative half-space we find
\begin{eqnarray}
&&\fl  \int^w  \Psi_\mu  = n! \sum_P G^w_{\lambda_{P_1},..\lambda_{P_n}} 
\prod_{n \geq \ell > k \geq 1} (1+ \frac{i \bar c}{\lambda_{P_\ell} - \lambda_{P_k}})\,, \nn \\
&&\fl  G^w_{\lambda_1,..\lambda_n} = \int_{-\infty< x_1<x_2< ..<x_n < 0} 
dx_1.. dx_n e^{(w+ i \lambda_1) x_1 + .. + (w+ i \lambda_n) x_n }  = \prod_{j=1}^{n} \frac{1}{ j w + i \lambda_1 + .. + i \lambda_j }\,,
 \nonumber 
\end{eqnarray}
where the dependence on $w$ only occurs as an (imaginary) shift of the rapidities. 

It turns out that the sum over permutation can be performed explicitly and can be expressed, for any $n$ and any 
set of rapidities, as a product over pairs of rapidities, a manifestation of the remarkable properties of the BA
\be
\int^w \Psi_\mu   =  \frac{n!}{i^n \prod_{\alpha=1}^n (\lambda_\alpha - i w)}  \prod_{1 \leq \alpha < \beta \leq n} \frac{i + \lambda_\alpha + \lambda_\beta - 2 i w}{ \lambda_\alpha + \lambda_\beta - 2 i w} .
\ee
In this formula and below we set $\bar c=1$. This formula has been also checked with
mathematica up to large $n$.

If we now inject the string solution $\lambda_{j,a}$ with $a=1,..m_j$ as in Eq. (\ref{stringsol}), 
the above can be rewritten as a product of intra-string contributions and inter-string pairwise contributions in the form
\bea  \label{inths}
&& \int^w \Psi_\mu  = n! (-2)^{n} \prod_{i=1}^{n_s} S^w_{m_i,k_i}
 \prod_{1 \leq i < j \leq n_s} D^w_{m_i,k_i,m_j,k_j}\,, 
 \eea
where
\bea
&& D^w_{m_i,k_i,m_j,k_j} =  \prod_{1 \leq a \leq m_i} \prod_{1 \leq b \leq m_j} 
 \frac{i + \lambda_{i,a} + \lambda_{j,b} - 2 i w}{ \lambda_{i,a} + \lambda_{j,b} - 2 i w}, \\
 && S^w_{m_i,k_i} = \frac{1}{(-2 i)^{m_i} \prod_{a=1}^{m_i} (\lambda_{i,a} - i w)} \prod_{1 \leq a < b \leq m_i} \frac{i + \lambda_{i,a} + \lambda_{i,b} - 2 i w}{ \lambda_{i,a} + \lambda_{i,b} - 2 i w} .
\eea 
The intra-string factor $S^w_{m_i,k_i}$ can be evaluated for a single string (for convenience
we choose here the string parameterization $\lambda_{j,a} =- \frac{i}{2} (m_j+1 - 2 a) + k_j$, 
obtained from (\ref{stringsol}) with the replacement $a \to m_j+1-a$)
\bea
&& \fl S^w_{m,k} =  
  \frac{1}{(- i)^m} \prod_{a=1}^{m}  \frac{1}{f(2 a)} \prod_{1 \leq a < b \leq m} \frac{f(a+b+1)}{f(a+b)}
 = \frac{1}{(- i)^m} \prod_{a=1}^{m}  \frac{1}{f(2 a)} \prod_{b=1}^m \frac{f(2 b)}{f(1+b)} \\
 && = \frac{1}{(- i)^m}  \prod_{b=1}^{m_i} \frac1{- i (m  -  b) + 2 k - 2 i w},
 \eea 
with $f(x)=-i  (m+1 -  x) + 2 k - 2 i w$. Hence we obtain
\bea \label{sw}
&& \fl S^w_{m,k} = \frac{\Gamma (1-z-m)}{\Gamma (1-z)} = \frac{(-1)^m \Gamma(z)}{\Gamma(z + m)}  = \frac{(-1)^m}{z(z+1)..(z+m-1)} \quad , \quad z= 2 i k + 2 w.
\eea

The inter-string factor $D^w_{m_1,k_1,m_2,k_2}$ can be computed similarly
\bea
&& \fl D^w_{m_1,k_1,m_2,k_2} 
 = \prod_{1 \leq a \leq m_1} \prod_{1 \leq b \leq m_2} 
  \frac{f(a+b+1)}{f(a+b)}  = \prod_{b=1}^{m_2} \frac{f(1+b+m_1)}{f(1+b)} \\
  && =
 \prod_{b=1}^{m_2} \frac{- i (\frac{m_2-m_1}{2} - b) + k_1 + k_2 - 2 i w}{- i (\frac{m_1+m_2}{2}  -   b)  +  k_1+k_2 - 2 i w}, \nn
\eea
with $f(x)=-i  (\frac{m_1+m_2}{2} +1 -  x) + k_1+k_2 - 2 i w$. This leads to two equivalent forms
\bea \label{dw}
\fl  D^w_{m_1,k_1,m_2,k_2}& =&
\frac{\Gamma \left(1-z -\frac{m_1+m_2}{2}\right) \Gamma \left(1-z +\frac{m_1+m_2}{2}\right) }{
\Gamma \left(1-z +\frac{m_1-m_2}{2}\right) \Gamma \left(1-z -\frac{m_1-m_2}{2}\right)}
\\ \fl
&  =& (-1)^{m_2} \frac{\Gamma(1- z + \frac{m_1+m_2}{2}) \Gamma(z + \frac{m_1-m_2}{2}) }{
\Gamma(1- z + \frac{m_1-m_2}{2}) \Gamma(z + \frac{m_1+m_2}{2})} \quad , \quad z=  i k_1+i k_2 + 2 w\,. \label{dw2}
 \eea
This factor is in general complicated, even for $w=0$, although in that case we found that
its norm simplifies to
\bea \label{remarquable} 
\fl \quad |D^{w=0}_{m_1,k_1,m_2,k_2} |^2 = D^{w=0}_{m_1,k_1,m_2,k_2} D^{w=0}_{m_1,-k_1,m_2,-k_2} =
 \frac{4 (k_1+k_2)^2 + (m_1+m_2)^2}{4 (k_1+k_2)^2 + (m_1-m_2)^2}\,.
\eea
Similarly one also has
\bea  \label{remarquable2}
\fl \quad | S^{w=0}_{m,k} |^2 = S^{w=0}_{m,k} S^{w=0}_{m,-k}  =\prod_{q=0}^{m-1} \frac{1}{4 k^2+q^2} .
\eea 

Eq. (\ref{inths}) together with Eqs. (\ref{sw}) and (\ref{dw}) give the complete expression for
the space integral of the Bethe states in the half-space model. Note that at large $w$ one has
\bea
\prod_{i=1}^{n_s} S_w \to \frac{1}{(-2 w)^n}, \qquad D_w \to 1, \qquad  \int^{w}  \Psi_\mu  \to  n! \frac{1}{w^n},
\eea
which is the correct normalization of the sharp edge limit since $e^{w y} \to  \delta(y)/w$.

\subsection{Generating function}

We have now all ingredients to compute the generating function $g(s) =  1 +  \sum_{n_s=1}^\infty \frac{Z(n_s,s)}{n_s!}  $ defined in
(\ref{defg}) where, from (\ref{Zns}) and using (\ref{inths}), we have
\bea
&& \fl 
 Z(n_s,s) = \nn \\
 && \fl \quad \sum_{m_1,\dots m_{n_s}=1}^\infty 
 \prod_{j=1}^{n_s} \left[ \frac{2^{m_j}}{m_j} \int_{k_j} 
S^w_{m_j,k_j}  e^{(m_j^3 - m_j) \frac{t}{12}- m_j k_j^2 t - \lambda m_j s - i x m_j k_j} \right]  \prod_{1 \leq i < j \leq n_s} \tilde D^w_{m_i,k_i,m_j,k_j}, 
\eea
where $S^w$ and $D^w$ are given in (\ref{sw}) and (\ref{dw}) and  we defined
\bea
\tilde D^w_{m_i,k_i,m_j,k_j} =  D^w_{m_i,k_i,m_j,k_j} \Phi_{k_i,m_i,k_j,m_j},
\eea 
where the factor coming from the norm is
\be \label{phi2} 
\Phi_{k_i,m_i,k_j,m_j}=\frac{4(k_i-k_j)^2 +(m_i-m_j)^2}{4(k_i-k_j)^2 +(m_i+m_j)^2}.
\ee
The main difficulty comes from the double products which preclude independent summations over $m_i,k_i$, and 
can be seen as an interaction
between strings. In the $w \to \infty$ limit (sharp wedge) it simplifies since $D^w \to 1$ and the resulting term
can be written as a determinant. This allows to decouple the interaction writing the  determinant 
as a sum over permutations. For arbitrary $w$ it does not seem that such a rewriting exists. In the
limit of $w \to 0$ however a determinantal structure (pfaffian) also arises, as we will now show. 

\subsection{Small $w$ limit: study of the poles}  \label{sec:poles}

Upon inspection we find that the limit of interest, $w=0^+$, is dominated by poles near zero momentum in the $k_i$ integrations. These are
the analog of the states contributing in the large but fixed finite size $L$ at $w=0$ discussed in Section \ref{DIRECT}. Those appeared
as delta functions forcing a set of combinations of sums of $k_j$ to vanish for the contributing states. Here instead they
are poles in the same combinations of $k_j$. The empirical rule which appeared there, i.e. that strings appear to pair up in the
 states with non zero contribution, will appear much more clearly here.

Let us first examine the behavior of the factors $S^w$ and $D^w$ when $w \to 0^+$. From Eq. (\ref{sw}) 
there is a divergence at $k_i=0$, which takes the form
\bea \label{poleS}
S^w_{m_i,k_i} = \frac{(-1)^{m_i}}{\Gamma(m_i) (2 i k_i + 2 w)} + s^w_{m_i,k_i},
\eea
where $s^w$ is the ``regular part" which remains finite for any momentum when $w \to 0$, and is $O(1)$ when $k_i \sim w$. For the
inter-string factor it is clear from Eqs. (\ref{dw2}) and (\ref{remarquable}) that when $m_1 \neq m_2$ $D^w_{m_i,k_i,m_j,k_j}$ remains
finite for all $k_1,k_2$ as $w \to 0$. 
However,  for $m_1=m_2$  there is a divergence when a pair of momenta vanish (i.e. $k_i+k_j=0$), which from Eq. (\ref{dw2}) is
 \bea \label{poleD}
 D^w_{m_i,k_i,m_j,k_j} = \frac{(-1)^{m_i} m_i}{ i (k_i +k_j)+ 2 w} \delta_{m_i,m_j} + d^w_{m_i,k_i,m_j,k_j},
 \eea
  Again we denote $d^w$ the ``regular part" 
 which remains finite for any momentum when $w \to 0$, and is $O(1)$ when $k_i+k_j \sim w$. 
 By contrast the factor $\Phi$ in (\ref{phi2}) never diverges. However
 it does play an important role: since in each pair of strings a factor $\Phi$ occurs with each $D^w$ we should
 put them together in evaluation of the divergence, i.e.
  \bea \label{poleDPhi}
 \tilde D^w_{m_i,k_i,m_j,k_j} =  \frac{(-1)^{m_i} m_i}{ i (k_i +k_j)+ 2 w} \delta_{m_i,m_j} \frac{(k_i-k_j)^2}{(k_i-k_j)^2 + m_i^2} + \cdots.
 \eea
 
We now argue that the only regions in the integral $\int_{k_1,..k_{n_s}}$ which contribute to $Z(n_s,s)$ {\it in the limit of small $w$ and
of large negative $x$} are obtained by splitting the $n_s=2 N+M$ strings into $N$ pairs of strings of (nearly) opposite momenta 
and $M$ single strings of (nearly) zero momentum. More precisely these regions are
\bea \label{config}
\fl \quad k_{i_1} \sim w, \quad \dots \quad k_{i_M} \sim w , \quad k_{i_{M+1}}+k_{i_{M+2}} \sim w , \quad \dots 
 \quad k_{i_{M+2N-1}} + k_{i_{M+2 N}} \sim w\,.
\eea 

To understand how this works, and the respective role of the limits $w \to 0^+$ and $x \to - \infty$ let us examine
the product 
\bea \label{disc} 
\prod_{i} S_i \prod_{i<j} D_{ij} \Phi_{ij} \,,
\eea 
where here 
we denote for simplicity $S_i=S^w_{k_i,m_i}$, $D^w_{ij}=D^w_{m_i,k_i,m_j,k_j}$ and $\Phi_{ij}=\Phi_{m_i,k_i,m_j,k_j}$. 
We now expand each factor $S_i = S_i^{pole} + s_i$ and $D_{ij}=D_{ij}^{pole} + d_{ij}$ in pole plus
regular part according to Eqs. (\ref{poleS},\ref{poleD}). In each of the $2^{n_s(n_s+1)/2}$ terms of this expansion,
each string momentum $k_l$ can thus (i) appear in exactly one pole (ii) appear in more than one pole (overlapping poles)
(iii) appear only in regular parts (unsaturated poles). We now argue that:

- the limit $w \to 0^+$ rules out (ii). 

- the limit $x \to - \infty$ rules out (iii). 

Hence the only configurations which contribute in these limits are of the form (\ref{config}), i.e
they are such that any given string momentum $k_\ell$, $\ell=1,..n_s$, appears in exactly one pole.
The overlapping poles are found to be subdominant for $w \to 0$ because of phase space considerations, and
the unsaturated poles decay to zero {\it exponentially} as $x \to \infty$. 

Consider first overlapping poles, and focus on the Jacobian in $k_j$ integration. Let us first 
check that (\ref{config}) indeed yields $O(1)$ momentum integral. One can rescale each $k_{i_a} \to w \tilde k_{i_a}$, $a=1,...M$
and similarly each pair momentum as $(k_{i_{a-1}},k_{i_a}) \to (- k_{i_a} + \frac{w}{2} \tilde k_{i_a},k_{i_a} + \frac{w}{2} \tilde k_{i_a})$,
$a=M+2,M+4,..M+2N$. This leads to a Jacobian $w^{M + N}$ which is compensated by the same factor $1/w^{M+ N}$ coming from
the poles. 

Suppose now that one of the momenta, say $k_i$, appears both in the pole of a $S_i$ factor and in one pair factor, say $D_{ij}$. One must thus rescale both $k_i$ and $k_j$ by $w$, and then one sees from (\ref{poleDPhi}) that an extra $w^2$ factor arises due to the $\Phi$ factor.
Even if all the 3 poles $S_i S_j D_{ij}$ are considered, the jacobian is $w^2$ while the poles give $w^{-3} w^2=w^{-1}$. Hence this contribution is negligible as $w \to 0$ \footnote{this assumes convergence of the rescaled integrals, but examination shows at worse
logarithmic divergences which do not change the conclusion of being subdominant 
}. The same mechanism is at play for instance to exclude simultaneous
poles in $D_{ij} D_{j\ell} D_{i\ell}$, since the simultaneous conditions $k_i+k_j = O(w), k_i+k_\ell = O(w), k_j+k_\ell = O(w)$ 
imply $k_i \sim k_j \sim k_\ell = O(w)$, hence once again the $\Phi$ (norm) factors in (\ref{poleDPhi}) make this contribution
subdominant. One can extend these arguments to a larger number of overlapping poles with similar conclusions.

This still does not show that only the configurations (\ref{config}) remain. To prove this
we need to show that each string momentum should appear in at least one pole. This is
the role of the $x \to - \infty$ limit. The idea is simple. Let us first integrate over all
momenta which appear in poles. The remaining integral is $O(w^0)$ hence survives
as $w \to 0$, however it now contains momenta which appear in only regular
terms. The factors $\exp(- \sum_j m_j k_j^2 t)$ can be removed using diffusion kernels
as shown in \ref{sec:examples}. Thus one can first examine the $t=0$ integral. 
Thanks to the factor $\exp(- i \sum_j m_j k_j x)$ this integral can be performed
by contour integration in the sector where all ${\rm Im} k_j \geq 0$. Examination of poles in this integral shows 
that it decays exponentially as $x \to - \infty$, and this property is not changed by the
convolution with diffusion kernels. 
The general argument is then rather intuitive, but giving a general proof is difficult, and for this reason we simply examine
some examples in \ref{sec:examples}. 

Finally, a last property emerges which simplifies even further the calculation. For each pole
since we can close the contour in the upper half-plane, we can use the Cauchy residue formula 
which is equivalent to \footnote{note that this corrects a misprint in \cite{we-flat}}
\bea
\lim_{w \to 0^+} \frac{1}{i k + w} \to  2 \pi \delta(k) \,.
\eea 
This replacement is equivalent, for configurations of the type (\ref{config}), to perform first the $w$ rescaling 
and then integrate. Replacing each pole by a $\delta$-function, one then
obtains a $x$-independent result, which can be shown to equal the limit $x \to -\infty$, 
i.e. the flat initial condition for KPZ. This is also discussed in \ref{sec:examples}. 

To summarize, the recipe is as follows: expand the product $\prod_{i} S_i \prod_{i<j} D_{ij}$ 
in poles plus regular part according to
\bea 
&& S^w_{m_i,k_i} \to  \frac{(-1)^{m_i}}{2 \Gamma(m_i)} 2 \pi \delta(k_i) + s^0_{m_i,k_i} \,,\\
&& D^w_{m_i,k_i,m_j,k_j} \to (-1)^{m_i} m_i \delta_{m_i,m_j} ~ 2 \pi \delta(k_i+k_j)  + d^w_{m_i,k_i,m_j,k_j}\,,
\eea
with the constraint that each momentum $k_\ell$ appears only in exactly one pole, 
and perform the remaining integrals. The result is the sum of the residues associated to configurations where
the $n_s$ strings split into $N$ pairs of strings of opposite momenta with same particle
number $m$ and $M$ single strings of zero momentum with all distinct number of particles. 
Of course there are many such partitions, hence many contributions, since $M$ and $N$ 
can vary from $M=1,..n_s$ and $N=0,.. {\rm Int}[n_s/2]$ with the constraint $M+2 N = n_s$. 
Our claim is that this gives the same result as the finite size method of Section \ref{DIRECT} as it is 
directly checked below for all results available from the direct method.

\subsection{Examples of configurations}

To illustrate the method, let us consider the simplest class of terms (i) $M=n_s$ i.e. 
$n_s$ strings with zero momenta (ii) $2 N=n_s$ all strings paired. 

\subsubsection{All unpaired configurations.}

The residue of the pole where all $k_i$ vanish (pole in all $S_i$) reads
\medskip
\bea
&& \fl 
 Z(n_s,s)|_{N=0} =  \\
 && \fl \sum_{m_1 \neq m_2 \neq \dots m_{n_s}=1}^\infty 
 \prod_{j=1}^{n_s} [ \frac{2^{m_j}}{m_j} 
  e^{(m_j^3 - m_j) \frac{t}{12} - \lambda m_j s}  \frac{(-1)^{m_j}}{2 \Gamma(m_j)} ] \prod_{1 \leq i < j \leq n_s} D^w_{m_i,0,m_j,0}  \nn\Phi_{0,m_i,0,m_j} \,,
\eea
where the summation is over all different $m_i$. It contains an additional factor $\frac{1}{2}$ for each integration
$\int \frac{dk_j}{2 \pi}$. Note the absence of $x$ dependence. One now uses that
for $m_1 \neq m_2$
\bea
&& \fl D^0_{m_1,0,m_2,0}  = (-1)^{\min(m_1,m_2)} \frac{m_1+m_2}{|m_2-m_1|}  , \qquad 
\Phi_{m_1,0,m_2,0}  = \frac{(m_1-m_2)^2}{(m_1+m_1)^2}\,,
\eea
which yields
\bea && \fl 
Z(n_s,s)=  \frac{1}{2^{n_s}}\! \sum_{m_1,\dots m_{n_s}=1}^\infty  (-2)^{\sum_j m_j} \prod_j \frac{1}{m_j!} 
e^{m_j^3  \frac{\lambda^3}{3}- \lambda m_j s} \prod_{1 \leq i<j \leq n_s}  (-1)^{\min(m_i,m_j)} \frac{|m_i - m_j|}{m_i+m_j}, \nn \\
&& \label{zunpaired}
\eea
where we extended the sum to all $m_i$, which is possible because of the $(m_i-m_j)^2$ factors in the norm.

\subsubsection{All paired string configurations.}

The other extreme case is for all the strings are paired, i.e. $2 N = n_s$, $M=0$, assuming $n_s$ even. 
We can label their momenta $k_1,-k_1,k_2,-k_2,..k_p,-k_p,..$ with $p=1,..N$ and denote their
sizes as $m_1,m_1,m_2,m_2,..$. We find
\bea
\fl  Z(n_s=2N,s) &=&  \sum_{m_1,\dots m_{N}=1}^\infty 2^{2 \sum_j m_j}
 \prod_{p=1}^{N} \frac{1}{m_p^2} \int_{k_p} 
 \prod_{p=1}^{N}e^{2 (m_p^3-m_p)  \frac{\lambda^3}{3}- 8 m_p k_p^2 \lambda^3 + 2 \lambda m_p s}   
 \nn  \\ \fl &&  \times 
 \prod_{p=1}^N (-1)^{m_p} m_p
 S^0_{m_p,k_p} S^0_{m_p,- k_p}  \Phi_{m_p,k_p,m_p,-k_p} 
  \nn  \\ \fl &&  \times 
  \prod_{1\leq p<q\leq N} 
 \tilde D^0_{m_p,k_p,m_q,k_q} \tilde D^0_{m_p,-k_p,m_q,k_q} \tilde D^0_{m_p,k_p,m_q,-k_q} \tilde D^0_{m_p,-k_p,m_q,-k_q} .
\end{eqnarray}
From Eqs. (\ref{remarquable2}) and (\ref{phi2}) we have
\bea \label{canc1}
S^0_{m,k} S^0_{m,- k}  \Phi_{m,k,m,-k} = \prod_{q=1}^m \frac{1}{4 k^2 + q^2}. 
\eea 
We also see from Eqs. (\ref{remarquable}) and (\ref{phi2}) that
\bea \label{canc2}
D^0_{m_1,k_1,m_2,k_2} D^0_{m_1,- k_1,m_2,- k_2} = \frac{1}{\Phi_{m_1,k_1,m_2,- k_2}} ,
\eea 
and so
\bea
\fl \tilde D^0_{m_p,k_p,m_q,k_q} \tilde D^0_{m_p,-k_p,m_q,k_q} \tilde D^0_{m_p,k_p,m_q,-k_q} \tilde D^0_{m_p,-k_p,m_q,-k_q} 
 = \Phi_{m_p,k_p,m_q,k_q} \Phi_{m_p,k_p,m_q,- k_q}\,,
\eea 
which leads to
\bea
 \fl Z(n_s=2N,s) &=&   \frac{1}{2^N}  \sum_{m_1,\dots m_{N}=1}^\infty  2^{2 \sum_j m_j}  \nn\\ \fl &&
 \prod_{p=1}^{N} \Big[ \int_{k_p}  \frac{(-1)^{m_p} }{m_p}
   \prod_{q=1}^{m_p} \frac{1}{4 k_p^2 + q^2} 
  e^{2 (m_p^3 - m_p) \frac{\lambda^3}{3}- 8 m_p k_p^2 \lambda^3 + 2 \lambda m_p s} \Big] 
  \nn\\ \fl &&  \prod_{1\leq p<q\leq N}
\frac{4 (k_p-k_q)^2 +(m_p-m_q)^2}{4 (k_p-k_q)^2 +(m_p+m_q)^2}
\frac{4 (k_p+k_q)^2 +(m_p-m_q)^2}{4 (k_p+k_q)^2 +(m_p+m_q)^2} .
\end{eqnarray}

\medskip

\subsection{Full generating function}  \label{sec:fullgenerating}

\medskip

We now display the complete result for the generating function. We
label the configurations as follows. There are $N$ strings, $p=1,..N$, $(n_1,..n_p,..n_N)$ which are paired, and 
$M$ strings which are singles denoted $j=1,..M$ with $m_1,..m_j,..m_M$. The 
summation over the number of strings can be replaced by independent summation over $N$ and $M$
as follows
\bea \label{gener}
&& g(s) = 1 + \sum_{\stackrel{N,M \geq 0}{(N,M) \neq (0,0)}} \frac{1}{(2 N)! M!} Z(N,M) .
\eea
The partition sum at fixed $N,M$ is now
\bea \label{gener2}
&& \fl Z(N,M) = \sum_{n_1,\dots n_{N}=1}^\infty  \sum_{m_1,\dots m_{M}=1}^\infty
\frac{1}{2^{M}}   (-2)^{\sum_{j=1}^M m_j+ 2 \sum_{p=1}^N n_p} \prod_{j=1}^M \frac{1}{m_j!} 
e^{(m_j^3-m_j)  \frac{\lambda^3}{3}- \lambda m_j s}  \\
&& \fl \times \!\!\! \prod_{1 \leq i<j \leq M} \!\! (-1)^{\min(m_i,m_j)} \frac{|m_i - m_j|}{m_i+m_j} 
 \prod_{p=1}^{N}  \int_{k_p}  \frac{1 }{n_p}
  \prod_{p=1}^N \prod_{q=1}^{n_p} \frac{1}{4 k_p^2 + q^2} 
    \prod_{p=1}^{N}e^{2 (n_p^3-n_p)  \frac{\lambda^3}{3}- 8 n_p k_p^2 \lambda^3 - 2 \lambda n_p s} \nn \\
    && \fl \times 
   \prod_{1\leq p<q\leq N}
\frac{4 (k_p-k_q)^2 +(n_p-n_q)^2}{4 (k_p-k_q)^2 +(n_p+n_q)^2}
\frac{4 (k_p+k_q)^2 +(n_p-n_q)^2}{4 (k_p+k_q)^2 +(n_p+n_q)^2} \nonumber \\
&& \fl \times  (-1)^{ \sum_{p=1}^N n_p}  (2 N -1)!!  \prod_{1 \le p \le N,1 \le j \le M} \frac{4 k_p^2 + (n_p-m_j)^2}{4 k_p^2 + (n_p+m_j)^2},
 \nn
\end{eqnarray}
where the last line is made of the factors $\tilde D^0_{n_p,k_p,m_j,0} \tilde D^0_{n_p,-k_p,m_j,0} = \Phi_{n_p,k_p,m_j,0}$
from a partial cancellation with the norm factors due to (\ref{canc2}). The combinatorial factor
$(2N-1)!!= \frac{(2 N)!}{ 2^N N!} $ is the number of ways to pair $2N$ strings. 

This formula allows to recover the moments $\overline{Z^n}$ by expansion in powers of $\exp(- \lambda s)$.
We checked it as (i) against the results for the full space obtained in Section \ref{DIRECT} up to $n=4$ 
(ii) at $t=0$ where $\overline{Z^n}=1$ is recovered up to $n=12$. 

\section{Generating function as a Fredholm  Pfaffian} 
\label{Pfsec}

We can now take the formula (\ref{gener},\ref{gener2}) that we just derived as the starting point for $g(s)$. 
We show in this section that it has a determinantal structure, more precisely a Pfaffian structure. Hence we
first recall the definition of the Pfaffian of a matrix: for any antisymmetric matrix $A$ of size $2 n \times 2 n$
\be
{\rm pf} A = \sum_{\sigma \in S_{2n}, \sigma(2j-1)<\sigma(2 j)} \hspace{-2mm} (-1)^\sigma \prod_{i=1}^{n} 
A_{\sigma(2 i-1),\sigma(2 i)}\,,
\ee
with $({\rm pf} A)^2 = {\rm det} A$. For $n=1$ it is ${\rm pf} A=A_{12}$, for $n=2$, ${\rm pf} A=A_{12} A_{34} - A_{13} A_{24} + A_{14} A_{23}$ and so on, the number of terms of the sum being $(2 n-1)!!$. 

\subsection{$n_s$ even}

Let us first concentrate on the case with an even number of string $n_s$ which requires in Eq. (\ref{gener2}) $M$ to 
be an even integer and $N$ arbitrary. 
First we use Schur's Pfaffian identity for $X_1,..X_{2n}$ \cite{japanesePfaffian} 
\be
{\rm pf} \bigg( \frac{X_i - X_j}{X_i + X_j} \bigg)_{2n \times 2n} = \prod_{1 \leq i <  j \leq 2 n} \frac{X_i-X_j}{X_i+X_j}\,,
\ee
which, for $2 n = n_s = M + 2 N$, allows to rewrite the product in Eq. (\ref{gener}) as
\bea
\fl \prod_{1 \leq i<j \leq M}  \frac{m_i - m_j}{m_i+m_j}
   \prod_{1\leq p<q\leq N} 
\frac{4 (k_p-k_q)^2 +(n_p-n_q)^2}{4 (k_p-k_q)^2 +(n_p+n_q)^2}
\frac{4 (k_p+k_q)^2 +(n_p-n_q)^2}{4 (k_p+k_q)^2 +(n_p+n_q)^2} \nn\\ \fl \qquad\qquad \times
  \prod_{1 \le p \le N,1 \le j \le M} \frac{4 k_p^2 + (n_p-m_j)^2}{4 k_p^2 + (n_p+m_j)^2}
=
\left(\prod_{p=1}^{N} \frac{n_p}{2 i k_p}\right) {\rm pf} \left(\frac{X_i-X_j}{X_i+X_j}\right)_{n_s \times n_s}\,,
\label{PFdou}
\eea
with
\bea
X_{2p-1}&=& n_p+2i k_p\,, \qquad p=1,\dots N\,,\\ 
X_{2p}   &=& n_p-2i k_p\,, \qquad p=1,\dots N\,,\\ 
X_{p+2N}&=& m_p\,,\qquad p=1,\dots M\,.
\eea

We will need another important Pfaffian identity (valid for $M$ even), checked
up to large order using mathematica, and which, to our knowledge is new
\be
\fl \prod_{1 \leq i<j \leq M}  (-1)^{\min(m_i,m_j)} {\rm sgn}(m_i-m_j) = {\rm pf}[ (-1)^{\min(m_i,m_j)} {\rm sgn}(m_i-m_j) ]_{M \times M}\,,
\label{PFm}
\ee
valid for any set of integers $m_i$ (we define ${\rm sgn}(0)=0$).

It is now convenient to treat $n_p$ and $m_p$ on the same footing. To this aim we define a new set of $n_s$ integers $g_j$, $j=1,..n_s$ such that 
\bea
g_{2p-1}&=& n_p \,, \qquad p=1,\dots N\,,\\ 
g_{2p}   &=& n_p  \,, \qquad p=1,\dots N\,,\\ 
g_{p+2N}&=& m_p\,,\qquad p=1,\dots M\,,
\eea
and the $n_s$ momentum integration variables $Q_j$, $j=1,..n_s$
\bea
Q_{2p-1}&=&  k_p\,, \qquad p=1,\dots N\,,\\ 
Q_{2p}   &=& - k_p\,, \qquad p=1,\dots N\,,\\ 
Q_{p+2N}&=& 0\,,\qquad p=1,\dots M\,.
\eea
We want now to sum independently over all the integers $g_j$ and integrate over all momenta $Q_j$. 
In order to do so, since $g_j$ and $Q_j$ are not independent variables, we introduce several delta functions.
Putting together the various pieces above and inserting in Eq. (\ref{gener}) we obtain the intermediate expression 
\bea&& \fl 
Z(N,M) = \frac{(2 N -1)!!}{ 2^M}\sum_{g_1, \dots g_{n_s}=1}^\infty
 (-2)^{\sum_{j} g_j}  
 \prod_{j=1}^{n_s}  \int_{Q_j}   \prod_{q=1}^{g_j} \frac{1}{2 i Q_j + q} 
    \prod_{j=1}^{n_s}e^{ (g_j^3-g_j)  \frac{\lambda^3}{3}-  g_j Q_j^2 \lambda^3 -  \lambda g_j s}\nn\\
 && \fl\quad  \times (-1)^{ \sum_{p=1}^N n_p}  {\rm pf}_{(\ref{PFm})} {\rm pf}_{(\ref{PFdou})} 
 \prod_{p=1}^{N}  \delta_{g_{2p},g_{2p-1}} \frac{2\pi}{2 i Q_{2p}} \delta(Q_{2 p}+Q_{2 p-1}) \prod_{j=1}^{M} 2\pi \delta(Q_{2 N+j})\,,
 \end{eqnarray}
where ${\rm pf}_{(\ref{PFm})}$ and  ${\rm pf}_{(\ref{PFdou})}$ are the Pfaffian in Eqs. (\ref{PFm}) and (\ref{PFdou}) 
respectively, with $X_j = g_j + i Q_j$.

Now we can use the Pfaffian identity 
${\rm pf} \big(\lambda_i \lambda_j A_{ij} \big)_{2 n \times 2 n} = (\prod_{i=1}^{2n} \lambda_i) {\rm pf} A$
and rewrite
\bea \label{Pf1}
&& \fl
2^{-M} {\rm pf}_{(\ref{PFm})} \prod_{j=1}^{M} 2\pi \delta(Q_{2 N+j}) =
{\rm pf}  \left(\frac{(2\pi)^2}{4}  \delta(Q_i) \delta(Q_j) (-1)^{\min(g_i,g_j)} {\rm sgn}(g_i-g_j)\right)_{M  \times M}\!\!\!\!\,.
\eea
Next we can use the permutation symmetry of the integration measure $\prod_{j=1}^{2 N} \sum_{g_j} \int_{Q_j}$ which allows to make the
substitution
\bea  
&& \fl (2 N -1)!! (-1)^{ \sum_{p=1}^N n_p} \prod_{p=1}^{N}  \delta_{g_{2p},g_{2p-1}} \frac{2\pi}{2 i Q_{2p}} \delta(Q_{2 p}+Q_{2 p-1}) \nn \\
&& 
~~~~~~~~~~~~ \to {\rm pf}\left(  (-1)^{g_i}\delta_{g_{i},g_{j}} \frac{2\pi}{2 i Q_{i}} \delta(Q_{i}+Q_{j})\right)_{2N\times 2N}, \label{Pf2}
\eea
{\it in the integrand} since the remainder of the integrand is also a symmetric function in the permutations of strings
within the $2 N$ paired strings. Note that the matrix inside the Pfaffian is indeed antisymmetric thanks to the 
$\delta$-function (it can also be written equivalently as an explicit antisymmetric matrix). 

Now we can further use the full symmetry of the integration measure $\prod_{j=1}^{2 N + M=n_s} \sum_{g_j} \int_{Q_j}$ 
with respect to {\it all} $n_s$ variables so that again, inside the integrand one can replace the product of these 
two Pfaffian (\ref{Pf1}) and (\ref{Pf2}) by a single one
\bea && \fl
\frac{1}{(2 N)! M!} {\rm pf}_{(\ref{Pf1})} {\rm pf}_{(\ref{Pf2})} \to  \\
&& \fl \frac{1}{n_s!} 
 {\rm pf} \left(\frac{2 \pi}{2 i Q_i} \delta(Q_i+Q_j) (-1)^{g_i} \delta_{g_i,g_j} + \frac{(2 \pi)^2}{4}  \delta(Q_i) \delta(Q_j) (-1)^{\min(g_i,g_j)} 
 {\rm sgn}(g_i-g_j)\right)_{n_s\times n_s}  \label{PfQ} 
\eea
using again the symmetry of the remainder of the integrand by permutations of all the $n_s$ strings. 

Putting all together, we can now write the generating function for $n_s$ even in a Pfaffian form much more
symmetric with respect to all $n_s$ strings
\be
g(s)=1+\sum_{n_s=1}^\infty \frac1{n_s!} Z(n_s,s)\,,
\ee
where
\bea\fl
 Z(n_s,s) &&= \sum_{g_i \geq 1}  (-2)^{\sum_{j} g_j} 
   \prod_{j=1}^{n_s} \int_{Q_j} \prod_{q=1}^{g_j} \frac{1}{2 i Q_j + q} 
   e^{\frac{\lambda^3}{3} (g_j^3 - g_j) - 4 g_j Q_j^2 \lambda^3 - \lambda g_j s} 
\nonumber \\
\fl && \!\!\!\!\!\!\!\!\!\!\!\!\!\!
\times {\rm pf}\bigg(\frac{2\pi}{2 i Q_i} \delta(Q_i+Q_j) (-1)^{g_i} \delta_{g_i,g_j} + \frac{(2\pi)^2}{4}  \delta(Q_i) \delta(Q_j) (-1)^{\min(g_i,g_j)} {\rm sgn}(g_i-g_j)\bigg)_{n_s  \times n_s} \nn \\
\fl&& \times
     {\rm pf}\bigg( \frac{2 i Q_i + g_i - 2 i Q_j - g_j}{2 i Q_i + g_i + 2 i Q_j + g_j} \bigg)_{n_s  \times n_s}.
\eea
This formula has also been checked explicitly to reproduce the same expansion as Eqs.  
(\ref{gener}-\ref{gener2}) for $n_s=2,4$ and up to replica number $n=10$. 

\subsection{$n_s$ odd and general $n_s$}

Now we first need the generalization to $M$ odd of Eqs. (\ref{PFdou}) and (\ref{PFm}). In order to do this, 
it is useful to think of the case $n_s$ odd as a sum over an addition dummy variable
$g_{n_s+1}=m_{M+1}=0$ and over an additional integration over $Q_{n_s+1}=0$.
One can then check that Eq. (\ref{PFdou}) remains valid for $M$ odd
where the r.h.s. is now a $(n_s+1) \times (n_s+1)$ Pfaffian with $X_{n_s+1}=0$. Equivalently,
Eq. (\ref{PFdou}) remains valid for $M$ odd if one substitutes there
\bea
&& \fl {\rm pf}_{(\ref{PFdou})} \to {\rm pf}\left( \begin{array}{ll} B&V \cr -V^T&0 \end{array} \right)_{n_s+1,n_s+1} 
\quad , \quad  B_{ij} =  \frac{2 i Q_i + g_i - 2 i Q_j - g_j}{2 i Q_i + g_i + 2 i Q_j + g_j} , \quad   V_i = 1
\eea
in terms of the original two sets of $n_s$ variables $g_j$ and $Q_j$, i.e. adding a constant unit
column vector in the Pfaffian. 

A similar reasoning shows that for $M$ odd Eq. (\ref{PFm}) must be modified into
\bea
\fl \prod_{1 \leq i<j \leq M}  \!\!(-1)^{\min(m_i,m_j)} {\rm sgn}(m_i-m_j) = 
{\rm pf}\left(\begin{array}{ll}
 (-1)^{\min(m_i,m_j)} {\rm sgn}(m_i-m_j) & {\mathbf 1}\cr -{\mathbf 1}^T& 0\end{array}\right)_{M+1 \times M+1}\!\!,
\nn\\
\label{PFmodd}
\eea
where again $V={\mathbf 1}$ represents the vector with all elements equal to 1.

The argument of the previous section can then be repeated, introducing delta-functions and 
symmetrizing the integration over 
the odd $n_s$ original variables $g_j,Q_j$, $j=1,..n_s$. By inspection one sees that (\ref{PfQ}) still
holds provided one substitutes
\bea
&& pf_{\ref{PfQ}} \to  {\rm pf}\left( \begin{array}{ll} A&U\cr -U^T&0\end{array} \right)_{(n_s+1) \times (n_s+1)} , \qquad  U_i = \frac{2\pi}{2}  \delta(Q_i), \\
&& \fl A_{ij} = \frac{2\pi}{2 i Q_i} \delta(Q_i+Q_j) (-1)^{g_i} \delta_{g_i,g_j} + \frac{(2\pi)^2}{4}  \delta(Q_i) \delta(Q_j) (-1)^{\min(g_i,g_j)} {\rm sgn}(g_i-g_j).
\eea 

We can now transform, for any $n_s$, the product of two above Pfaffians into a single Pfaffian using the properties of the Pfaffian
of a tensorial product (for $n_s$ even) or a slight modification thereof (for $n_s$ odd).  It reads as follows:
consider a $2 n_s$ by $2 n_s$ antisymmetric matrix of the form
\bea
&& C = \left(\begin{array}{ll} A&D\cr  -D^T&B\end{array}\right), \qquad D_{ij}= U_i V_j \,,
\eea
where $A$ and $B$ are two antisymmetric $n_s$ by $n_s$ matrices and $U$ and $V$ are $n_s$ vectors, then:
\bea
\fl&&  {\rm pf}( C )_{2 n_s,2 n_s} =  {\rm pf}( A )_{n_s,n_s} {\rm pf}( B )_{n_s,n_s}    \, , \quad n_s\; {\rm even}, \\
\fl&& {\rm pf}(C )_{2 n_s,2 n_s} = {\rm pf}\left( \begin{array}{ll} A&U\cr -U^T&0\end{array} \right)_{n_s+1,n_s+1} 
{\rm pf}\left( \begin{array}{ll} B&V \cr -V^T&0 \end{array} \right)_{n_s+1,n_s+1}, \quad n_s\; {\rm odd}.
\eea
Applying it to the matrices $A,B$ and vectors $U,V$ defined above we obtain the final expression valid both for $n_s$ odd and even
{\small
\bea \fl &&
Z(n_s,s) = \sum_{m_i \geq 1}  
   \prod_{j=1}^{n_s} \int_{k_j} \prod_{q=1}^{m_j} \frac{-2}{2 i k_j + q} 
   e^{\frac{\lambda^3}{3} (m_j^3 - m_j) - 4 m_j k_j^2 \lambda^3 - \lambda m_j s} 
\\ && \fl
\times 
{\rm pf}\left(
\begin{array}{cc}
 \frac{2 \pi}{2 i k_i} \delta(k_i+k_j) (-1)^{m_i} \delta_{m_i,m_j} + \frac{(2 \pi)^2}{4}  \delta(k_i) \delta(k_j) (-1)^{\min(m_i,m_j)} {\rm sgn}(m_i-m_j) & \frac{1}{2}  (2 \pi) \delta(k_i)   \\
- \frac{1}{2} (2 \pi)  \delta(k_j)  &  \frac{2 i k_i + m_i - 2 i k_j - m_j}{2 i k_i + m_i + 2 i k_j + m_j} 
\end{array} \right)_{2 n_s  \times 2 n_s} . \nonumber 
\eea
}
which was previously displayed in \cite{we-flat}. Here we restored standard variables $Q_j \to k_j$ and $g_j \to m_j$
to indicate string momenta and number of particle in the string respectively. We recall that here and below
$\int_{k_j} = \int_{-\infty}^{+\infty} \frac{d k_j}{2 \pi}$. This formula has been
checked against intermediate formula by expanding up to $n_s=3,4$ and 
shown to reproduce (\ref{gener2}) up to $n=10$.

Until now we have not performed the usual shift in the free energy, but we perform it from now on.
It was discussed in Section  (\ref{sec:ns1}) to which we refer for details. It amounts to define (\ref{defv0}) and focus from now on the
generating function associated to the variable $f=-\xi_t$ in Eq. (\ref{defxi}). 


\subsection{Airyzation} \label{sec:Airy} 

Using the Airy trick (\ref{airytrick}) and shifting $y_j \to y_j +s$ we get
{\small
\bea \label{fork} \fl
&& Z(n_s,s) = \sum_{m_i \geq 1}  (-2)^{\sum_{j} m_j} 
   \prod_{j=1}^{n_s} \int_{k_j,y_j} Ai(y_j+s) \prod_{q=1}^{m_j} \frac{1}{2 i k_j + q} 
   e^{\lambda m_j y_j - 4 m_j k_j^2 \lambda^3} 
\nonumber \\
\fl && 
\times 
{\rm pf}
\left(\begin{array}{cc}
 \frac{\pi}{ i k_i} \delta(k_i+k_j) (-1)^{m_i} \delta_{m_i,m_j} + \pi^2  \delta(k_i) \delta(k_j) (-1)^{\min(m_i,m_j)} {\rm sgn}(m_i-m_j) & {\pi}  \delta(k_i)   \\
-\pi \delta(k_j)  &  \frac{2 i k_i + m_i - 2 i k_j - m_j}{2 i k_i + m_i + 2 i k_j + m_j} 
\end{array} \right)_{2 n_s  \times 2 n_s} \nonumber 
\eea
}
where here and below $\int_{y_j}=\int_{-\infty}^{\infty} dy_j$. 
To proceed we will make extensive use of the following identity
\bea \fl  \label{id1}
{\rm pf}
\left(\begin{array}{cc}
\lambda_i A_{ij} \lambda_j  & \lambda_i U_i V_j \\
- \lambda_j U_j V_i  & B_{ij} 
\end{array} \right)_{2 n_s  \times 2 n_s} = \lambda_1..\lambda_{n_s} 
{\rm pf}
\left(\begin{array}{cc}
A_{ij} & U_i V_j \\
- U_j V_i  & B_{ij} 
\end{array} \right)_{2 n_s  \times 2 n_s} .
\eea
The same identity exists with the role of $(A,U)$ and $(B,V)$ interchanged. It can be generalized
by formally replacing $\lambda_i \to \int dv_i a_i(v_i)$ or, as below, with summations,
e.g. $\lambda_i \to \sum_{m_i} a_i(m_i)$, leading for instance to (for $2 n_s  \times 2 n_s$ Pfaffians)
\bea \fl \label{id2}
{\rm pf}
\left(\begin{array}{cc}
A_{ij}   & U_i \int dv_j a_j(v_j) V_j(v_j) \\
- U_j \int dv_i a_i(v_i) V_i(v_i)  & \int dv_i dv_j a_i(v_i) a_j(v_j) B_{ij}(v_i,v_j) 
\end{array} \right) = \\  \qquad\qquad=
 [ \prod_{i=1}^{n_s} \int dv_i a_i(v_i) ] 
{\rm pf}
\left(\begin{array}{cc}
A_{ij} & U_i V_j(v_j) \\
- U_j V_i(v_i)  & B_{ij}(v_i,v_j) 
\end{array} \right), \nonumber 
&&
\eea
i.e. it allows to enter integrations inside the Pfaffian. 
The same identity again exists with the role of $(A,U)$ and $(B,V)$ interchanged.

There are several ways to proceed from here. 
In the following we propose one that is particularly suited to extract the infinite time limit, but  
 there may be others to arrive at different but equivalent forms for the kernels.
We start by  rescaling $k_j \to k_j/\lambda$ (using Eq. (\ref{id1})) then shifting $y_j \to y_j + 4 k_j^2$ obtaining
{\small
\bea\fl
&& Z(n_s,s) = \sum_{m_i \geq 1}  (-2)^{\sum_{j} m_j} 
   \prod_{j=1}^{n_s} \int_{k_j,y_j} Ai(y_j+s+4 k_j^2) \prod_{q=1}^{m_j} \frac{1}{2 i k_j/\lambda + q} 
   e^{\lambda m_j y_j} 
\nonumber \\ \fl && 
\times 
{\rm pf}
\left(\begin{array}{cc}
 \frac{\pi}{i k_i} \delta(k_i+k_j) (-1)^{m_i} \delta_{m_i,m_j} + \pi^2  \delta(k_i) \delta(k_j) (-1)^{\min(m_i,m_j)} {\rm sgn}(m_i-m_j) & \pi  \delta(k_i)   \\
- \pi\delta(k_j)  &  \frac{2 i k_i + \lambda m_i - 2 i k_j - \lambda m_j}{2 i k_i + \lambda m_i + 2 i k_j + \lambda m_j} 
\end{array} \right)_{2 n_s  \times 2 n_s} \nonumber 
\eea
}

Now we use
\bea
&& \fl \frac{2 i k_i + \lambda m_i - 2 i k_j - \lambda m_j}{2 i k_i + \lambda m_i + 2 i k_j + \lambda m_j}=
\int_{v_i>0,v_j>0} \!\!\!dv_i dv_j   
\delta(v_i -v_j) (\partial_{v_j}-\partial_{v_i}) e^{- v_i(2 i k_i + \lambda m_i) -v_j(2 i k_j + \lambda m_j)} \nn \\
&& = 
2\int_{v_i>0,v_j>0} dv_i dv_j  \delta'(v_i -v_j) e^{- v_i(2 i k_i + \lambda m_i) -v_j(2 i k_j + \lambda m_j)}\,,
\eea
since the boundary terms cancel. We can now use Eq. (\ref{id2}) with 
$a_i(v_i)=e^{-v_i (2 i k_i + \lambda m_i)}$ and obtain
\bea
&& \fl Z(n_s,s) =  \sum_{m_i \geq 1}  
   \prod_{j=1}^{n_s} \int_{k_j,y_j} Ai(y_j+s+v_j+4 k_j^2) 
  (-2)^{m_j}  e^{-2 i v_j k_j + \lambda m_j y_j} \prod_{q=1}^{m_j} \frac{1}{2 i k_j/\lambda + q}
   \nonumber \times \\ && \fl 
{\rm pf}
\left(\begin{array}{cc}
 \frac{\pi}{ i k_i} \delta(k_i+k_j) (-1)^{m_i} \delta_{m_i,m_j} + \pi^2  \delta(k_i) \delta(k_j) (-1)^{\min(m_i,m_j)} {\rm sgn}(m_i-m_j) & \pi \delta(k_i)  \delta(v_j) \\
- \pi  \delta(k_j)  \delta(v_i) &   2\delta'(v_i- v_j)
\end{array} \right)_{2 n_s  \times 2 n_s} \nn \\
&& 
\eea
where we have further shifted $y_j \to y_j + v_j$. Entering the sums inside the Pfaffian this leads to
\bea
\fl&& Z(n_s,s) =  
    \prod_{j=1}^{n_s} \int_{k_j,y_j,v_j>0} Ai(y_j+s+v_j+4k_j^2) e^{- 2 i v_j k_j }  \nn
    \\ \fl   && \qquad\qquad \times 
{\rm pf}\bigg(
\begin{array}{cc}
\tilde K_{11}(y_i,k_i,v_i;y_j,k_j,v_j)  
& \tilde K_{12}(y_i,k_i,v_i;y_j,k_j,v_j) 
\\
- \tilde K_{12}(y_j,k_j,v_j;y_i,k_i,v_i)   
&  \tilde K_{22}(y_i,k_i,v_i;y_j,k_j,v_j)
\end{array} \bigg)_{2 n_s  \times 2 n_s}, 
\eea
with (we keep the arguments of the kernels implicit)
\bea
\fl&&   \tilde K_{11}
=  \frac{\pi}{ i k_i} \delta(k_i+k_j)  f_{k_i/\lambda}[4 e^{ \lambda (y_i+y_j)}]
 + \pi^2 \delta(k_i) \delta(k_j) F[2 e^{\lambda y_i} , 2 e^{\lambda y_j} ],  \\
\fl&& \tilde K_{12}
= \pi \delta(k_i) (e^{-2 e^{\lambda y_i}}-1) \delta(v_j), \\
\fl&& \tilde K_{22}
= 2\delta'(v_i-v_j),
\eea
where the summations lead to the definition of two functions 
\bea 
\fl f_{k}[z] &=& 
 \sum_{m=1}^\infty \prod_{q=1}^{m} \frac{1}{4 k^2 + q^2} 
 (-z)^m = - z \frac{ \, _1F_2\left(1;2-2 i k,2 i
   k+2;- z\right)}{4 k^2+1}, \label{Kdef4}  \\
\fl F(z_i,z_j)&=&\sum_{m,m' \geq 1} (-1)^{\min(m,m')} {\rm sgn}(m-m') \frac{1}{m!} \frac{1}{m'!}  (-z_i)^m (-z_j)^{m'} \label{Kdef5}  \\
\fl& =&  \sinh(z_2-z_1) +e^{-z_2} -e^{-z_1}  \nn\\ \fl&&+
\int_0^{1} du J_0(2 \sqrt{z_1 z_2 u}) [ z_1 \sinh(z_1(1-u)) - z_2 \sinh(z_2(1-u))].\nonumber
\eea
While the resummation of $f$ is immediate the one for $F$ has been performed from
a Borel transform as explained in the \ref{app:borel}. 

The advantage of these manipulations is that the integrations over $k_j$ and $y_j$ 
can be brought inside the Pfaffian, using
again (\ref{id2}), which leads to our main result \footnote{correcting a minor misprint in \cite{we-flat}}
\bea
&& Z(n_s,s) =  
    \prod_{j=1}^{n_s} \int_{v_j>0} 
{\rm pf}\bigg(
\begin{array}{cc}
K_{11}(v_i;v_j)  
& K_{12}(v_i;v_j) 
\\
-  K_{12}(v_j;v_i)   
&  K_{22}(v_i;v_j)
\end{array} \bigg)_{2 n_s  \times 2 n_s} ,
\label{znsfinal} 
\eea
with
\bea
\fl&&   K_{11}
=\!
 \int \frac{dk}{2 \pi} dy_1 dy_2 Ai(y_1+s+v_i+4k^2)  Ai(y_2+s+v_j+4k^2) 
  \frac{e^{- 2 i (v_i -v_j) k}}{2 i k}  f_{k/\lambda}[4 e^{ \lambda (y_1+y_2)}] \nn
\\
\fl&& \qquad\qquad +  \int dy_1 dy_2  \frac{1}{4} F[2 e^{\lambda y_1} , 2 e^{\lambda y_2} ]  Ai(y_1+s+v_i) Ai(y_2+s+v_j),  \label{Kdef1} \\
\fl&& K_{12}
= \tilde K(v_i) \delta(v_j), \qquad \tilde K(v_i) = \frac{1}{2} \int dy Ai(y+s+v_i) (e^{-2 e^{\lambda y}}-1),  \label{Kdef2}  \\
\fl&& K_{22}
= 2\delta'(v_i-v_j). \label{Kdef3} 
\eea
Note that our definition of the Pfaffian of a matrix Kernel $K_{ab}(v_1,v_2)=K(a,v_1;b,v_2)$ assumes
the order $1,v_1;1,v_2;.. 1,v_{n_s};2,v_1;2,v_2;..2;v_{n_s}$.
Here and below we will use indifferently the symbol $K(v_1,v_2)=K_{11}(v_1,v_2)$. In evaluations below
one should not forget that since $K_{12}(v_1,v_2)$ is of the form $U(v_1) V(v_2)$ it cancels in
the final result for $n_s$ even, and it appears linearly for odd $n_s$. In addition the
result depends on $K_{11}$ and $K_{22}$ only as a function of the product kernel $K_{11} K_{22}$,
see \ref{app:pfaffian}.

To summarize we have shown that the generating function (\ref{defg}) can be written as 
\bea \label{gener85}
\!\!\!\!\!\! && g(s) = 1 + \sum_{n_s=1} \frac{1}{n_s!}  \prod_{j=1}^{n_s} \int_{v_j>0} 
{\rm pf}\bigg(
\begin{array}{cc}
K_{11}(v_i;v_j)  
& K_{12}(v_i;v_j) 
\\
-  K_{12}(v_j;v_i)   
&  K_{22}(v_i;v_j)
\end{array} \bigg)_{2 n_s  \times 2 n_s} \,.
\eea 
Before relating this to a Fredholm Pfaffian let us explicitly evaluate the simplest cases. 

\subsection{Evaluation for $n_s=1,2$}

Let us first show the one-string contribution, i.e. $n_s=1$
\be \label{1}
\fl Z(1,s) =\! \int_{v>0} \!\! K_{12}(v,v) = Tr K_{12} = \tilde K(0) = \frac{1}{2} \int dy \int_0^{+\infty} dv  (e^{-2 e^{\lambda y}}-1) Ai(y+s) ,
\ee
which is identical to the one we obtained in the analysis at fixed $n$ in Section \ref{sec:ns1}.
As shown there, this gives the leading asymptotics of $g(s)$ for large $s>0$. 

Because $K_{12}$ is a rank one operator, hence $K_{12}^2= ({\rm Tr} K_{12}) K_{12}$ and ${\rm Tr} K_{12}^p = \tilde K(0)^p$, the
evaluation of the Pfaffian for $n_s=2$ gives simply (see \ref{app:pfaffian})
\bea
Z(2,s) = - {\rm Tr} K_{11} K_{22} =  - 2 \int_{v>0} K_{10}(v,v) 
\eea
where we have defined the kernel $K_{10}$, important for what follows, as
\bea \label{defK10}
K_{10}(v_1,v_2) = \partial_{v_1} K_{11}(v_1,v_2).
\eea
Notice that since $f_k$ is even in $k$, in (\ref{Kdef1}) 
one can replace $\frac{1}{2 i k} e^{- 2 i (v_i -v_j) k}
\to  - \frac{1}{2  k} \sin(2  (v_i -v_j) k)$ hence
\bea
&& \fl Z(2,s) = 2
  \int \frac{dk}{2 \pi} dy_1 dy_2 ~ K_{Ai}(y_1+s+4 k^2,y_2+s+4 k^2) \\
  && \times 
  \left[  f_{k/\lambda}[4 e^{ \lambda (y_1+y_2)}] + 2 \pi \delta(k) \frac{1}{4} F[2 e^{\lambda y_1} , 2 e^{\lambda y_2} ] \right],
\eea
in terms of the {\it Airy Kernel}
\bea
K_{Ai}(y_1,y_2) = \int_{v>0} Ai(y_1+v) Ai(y_2+v) \,.
\eea

The product operators read
\bea \label{relationK}
&& (K_{11} K_{22})(v_1,v_2) = 2 K_{10}(v_2,v_1) - 2 K_{11}(v_1,0) \delta(v_2)\,, \\
&& (K_{22} K_{11})(v_1,v_2) = 2 K_{10}(v_1,v_2) + 2 \delta(v_1) K_{11}(0,v_2) \,,
\eea
as can be checked by applying to a test function. Hence $K_{22} K_{11}$ is equal to
$2 K_{10}$, up to a rank one operator which furthermore is nilpotent (i.e. of square zero) 
since $K_{11}(0,0)=0$. 

\subsection{Fredholm Pfaffian} \label{sec:FP} 

We now recall the definition of a Fredholm Pfaffian, which we denote as
\bea \label{FPdef}
\fl && g_1(s) =  {\rm Pf}[ {\bf J} + {\bf K} ] := \sum_{n_s=0}^\infty \frac{1}{n_s!}  \prod_{j=1}^{n_s}  
 \int_{v_i>0} \widetilde{\rm pf}[{\bf K} (v_i,v_j)]_{2n_s,2n_s}  , \qquad {\bf J}=
\left(\begin{array}{cc}
0 & I  \\
- I & 0
\end{array} \right),
\eea
where here ${\bf K}$ is the antisymmetric $2$ by $2$ matrix kernel of components $K_{ab} \equiv K_{ab}(v_i,v_j)$ 
defined in (\ref{Kdef1}-\ref{Kdef3}) and (\ref{Kdef4},\ref{Kdef5}) and $I$ is the identity
acting in functional space. Fredholm Pfaffians
are generalizations of Fredholm determinants, for more details
see e.g. \cite{Rainspfaffians,kanzieper}. 

In (\ref{FPdef}) the symbol $\widetilde {\rm pf}$ recalls that the Pfaffian is defined with the order 
$1,v_1;2,v_1;1,v_2;2,v_2;..1,v_{n_s};2,v_{n_s}$. However in our formula (\ref{gener85}) we use the
other ordering $1,v_1;1,v_2;.. 1,v_{n_s};2,v_1;2,v_2;..2;v_{n_s}$, which differs 
by a permutation $\sigma$ of $2 n_s$ elements with signature $(-1)^{\frac{n_s(n_s-1)}{2}}$. The two definitions are
thus related through $\widetilde{\rm pf} = (-1)^{\frac{n_s(n_s-1)}{2}} {\rm pf}$. \footnote{equivalently the
transformation can be written as $\widetilde {\rm pf}(A) = {\rm pf}(B^T A B) = \det(B) {\rm pf}(A)$ where 
the matrix $B_{ij}=\delta_{i,\sigma(j)}$ is a permutation matrix of $2 n_s$ elements
with $\det(B)=(-1)^{\frac{n_s(n_s-1)}{2}}$.} Hence $g(s)$ and $g_1(s)$ are closely
related, but not identical, i.e. one has
\bea \label{compare}
&& g(s) = \sum_{n_s} \frac{1}{n_s!} Z(n_s,s)\,, \\
&& g_1(s) = \sum_{n_s} \frac{1}{n_s!} (-1)^{\frac{n_s(n_s-1)}{2}} Z(n_s,s)\,.
\eea 
Now, one useful property of the Fredholm Pfaffian is that it can be written as the square root of a Fredholm determinant (FD), namely one has
\bea
 \! \! \! \! \! \! \! \! \! \! \! \! && g_1(s)^2 = {\rm Pf} [ {\bf J} + {\bf K} ]^2 = {\rm Det}[{\bf I} -  {\bf J} {\bf K}] = {\rm Det} \left(\begin{array}{cc} I + K_{12}^T & - K_{22}  \\ 
K_{11}  & I + K_{12}
\end{array} \right) ,
\eea
which is the generalization of the following relation which can be checked for \label{beautiful} arbitrary antisymmetric matrices $A,B$ and arbitrary matrix $C$:
\bea \label{vbeautiful} 
\! \! \! \! \! \! \! \! \! \! \! \! && \left( \widetilde{\rm pf} \left(\begin{array}{ll} A&I + C\cr -I -C^T&B\end{array}\right) \right)^2
= \det \left(\begin{array}{ll} I + C^T & - B \cr  A &I + C\end{array}\right) .
\eea
Some basic properties of FD are recalled in \ref{app:pfaffian}.

Now we notice that, although not true in general, due to the fact that $K_{12}$ is a rank one operator
there also exist a Pfaffian representation for $g(s)$. Going back to the way we constructed the
Pfaffian form for $Z(n_s,s)$ in the previous section we note that (i) if $n_s$ is even then $Z(n_s,s) \sim (K_{11} K_{22})^{n_s/2}$
while (ii) if $n_s$ is odd, due to the fact that $K_{12}$ is a projector, then $Z(n_s,s) \sim (K_{11} K_{22})^{(n_s-1)/2} K_{12}$. 
Hence the formal change $K_{22} \to - K_{22}$ provides exactly the factor $(-1)^{n_s(n_s-1)/2}$, i.e. from
formula (\ref{compare})
\bea
g(s) = g_1(s)_{K_{22} \to - K_{22}}.
\eea 
Thus $g(s)$ can be written also as the square root of a Fredholm determinant (FD)
\bea \label{gFP}
&& g(s)^2 = {\rm Det}\left[ \left(\begin{array}{cc} I + K_{12}^T &  K_{22}  \\ 
K_{11}  & I + K_{12}
\end{array} \right) \right],
\eea
in terms of the Kernels defined in (\ref{Kdef1}-\ref{Kdef3}) and (\ref{Kdef4},\ref{Kdef5}).
Hence it is also a Fredholm Pfaffian, although we will from now on use only the relation
to the FD. 

We can perform some further manipulations. Consider the 
general identity for matrices:
\bea
\fl \det\left[ \left(\begin{array}{cc} \delta_{ij} + u_i v_j  &  B_{ij}  \\ 
A_{ij}  &   \delta_{ij} + v_i u_j 
\end{array} \right) \right] = \det\left[ \left(\begin{array}{cc} \delta_{ij}   &  B_{ij}  \\ 
A_{ij}  &  \delta_{ij} 
\end{array} \right)\right]  ( 1 + \langle u | (1- A B)^{-1} | v \rangle)^2  ,
\eea 
where $A$ and $B$ are arbitrary antisymmetric matrices, 
and extend it to FD. From now on we will be using the standard
bra-ket notations. From (\ref{gFP}) using that $K_{12}(v_1,v_2) =\tilde K(v_1)  \delta(v_2)$
is a rank one operator we find
\bea
 g(s)^2 &=& {\rm Det}\left[ \left(\begin{array}{cc} I  &  K_{22}  \\ 
K_{11}  &  I
\end{array} \right) \right] (1 + \langle\delta | (1- K_{11} K_{22} )^{-1} | \tilde K\rangle )^2 .
\eea
This can be rewritten as
\bea
 g(s) & =& \sqrt{{\rm Det}[ I - K_{11} K_{22} ] } (1 + \langle\delta | (1- K_{11} K_{22} )^{-1} | \tilde K\rangle) \\
& = &\sqrt{{\rm Det}[ I - K_{11} K_{22} ] } (1+Tr (1- K_{11} K_{22} )^{-1} K_{12} ) \\
& = &\sqrt{{\rm Det}[ I - K_{22} K_{11} ] } (1+\langle\tilde K | (1- K_{22} K_{11} )^{-1} | \delta\rangle.
\eea
In the \ref{app:Qn}, using (\ref{relationK}) we show that
\bea \label{iddet} 
&& {\rm Det}[ I - K_{11} K_{22} ] = {\rm Det}[ I - K_{22} K_{11} ] = {\rm Det}[ 1- 2 K_{10}] ,
\eea
and we recall that we denote $K_{10}(v_1,v_2)= \partial_{v_1} K(v_1,v_2)$ where $K=K_{11}$.
In the \ref{app:Qn} we also show that
\bea  \label{iddet2} 
\langle\tilde K | (1- K_{22} K_{11} )^{-1} | \delta\rangle = \langle\tilde K | (1- 2 K_{10} )^{-1} | \delta\rangle.
\eea 
Hence we arrive to the main result for the generating function $g(s)$
\bea  \label{mainresult}
g(s) = \sqrt{ {\rm Det}[ 1- 2 K_{10}] } (1 + \langle\tilde K | (1- 2 K_{10} )^{-1} | \delta\rangle). 
\eea 
where we recall the definition of the Kernel $K_{10}$ in (\ref{defK10}) and of the function $\tilde K$ in (\ref{Kdef2}). 

\section{Large time limit} \label{sec:larget} 

We now show that the free energy probability distribution converges to the GOE Tracy Widom distribution
at large time (i.e long polymer length). In the large time limit (i.e. large $\lambda=(\bar c^2 t/4)^{1/3}$) 
one already sees from (\ref{Kdef2}) that
\bea \label{Klarge}
&&  \lim_{\lambda \to \infty} \tilde K(v) = - \frac{1}{2} \int_{y>0} Ai(y+s+v) .
\eea
Hence from (\ref{1}) the one-string ($n_s=1$) partition sum reads
\bea
\lim_{\lambda \to \infty} Z(1,s) = - \int_{y>0} Ai(2 y + s) = - {\rm Tr} B_s,
\eea
where $B_s(x,y)$ is the Kernel
\bea
B_s(x,y) = \theta(x) Ai(x+y+s) \theta(y)\,.
\eea
As shown by Ferrari and Spohn \cite{ferrari2} this is precisely the GOE kernel which enters the 
Fredholm determinant expression for the GOE Tracy Widom distribution
\bea
F_1(s) = \det[I - B_s] = \sum_{n=0}^\infty \frac{1}{n!} z(n,s) 
\eea
We will now show that it extends to all $n_s$ i.e. that
\be \label{lim}
\lim_{\lambda \to + \infty} Z(n_s,s)= z(n_s,s) := (-1)^{n_s} \int_{x_1,..x_{n_s}} \hspace{-3mm} \det[{B}_s(x_i,x_j)]_{n_s \times n_s}, 
\ee 
and equivalently
\bea \label{desired}
\lim_{\lambda \to + \infty} g(s) = \lim_{\lambda \to + \infty} {\rm Prob}(f > -s) = F_1(s) \,,
\eea 
i.e. the free energy probability distribution converges to the GOE Tracy Widom distribution. 

Considerable simplifications occur in the expressions (\ref{Kdef1},\ref{Kdef2},\ref{Kdef3}) for 
the FP Kernel. First, we can rewrite (\ref{Klarge}) in the bra-ket notation:
\bea
 \lim_{\lambda \to \infty} \tilde K(v) = - \frac{1}{2} \langle1|B_s,
\eea 
involving the GOE Kernel, where $\langle1|(v_1)=1$ is a constant vector. Second, we 
notice that the functions $f_{k}$ and $F$ which appear in the finite
time Kernel simplify considerably in the large $\lambda$ limit
\bea \label{limitform1}
&& \lim_{\lambda \to + \infty} f_{k/\lambda}(4 e^{ \lambda y})
= - \theta(y)\,, \\
&& \lim_{\lambda \to + \infty} F(2 e^{\lambda y_1} , 2 e^{\lambda y_2}) ]
= \theta(y_1+y_2) (\theta(y_1)\theta(-y_2) - \theta(y_2)\theta(-y_1)). \label{limitform2}
\eea
The first formula is immediate from (\ref{Kdef4}) and the second is shown in the \ref{app:asymptF}. Third, another important property 
shown in the Appendix is that for $\lambda \to + \infty$:
\bea 
&&  \fl 2 K_{10}(v_1;v_2) =  \int_{y>0} Ai(y + v_1 +s) Ai(y + v_2 +s) - \frac{1}{2} Ai(s+v_1) \int_{y>0} Ai(y+s+v_2) \nn \\
&& = K_{Ai}(s+v_1,s+v_2) - Ai(s+v_1) \tilde K(v_2), \label{asymptK10} 
\eea
i.e. it is exactly {\it the Airy Kernel plus a projector} and can be written as
\bea \label{K10large} 
2 K_{10} = B_s^2 - \frac{1}{2}  |B_s \delta\rangle \langle1 B_s|.
\eea

We need two identities which are known for matrices but we assume extend
readily to operators. The so-called matrix determinant lemma
\bea \label{form1}
{\rm Det}( A + |u \rangle\langle v| ) = (1 + \langle v | A^{-1} | u \rangle) \, {\rm Det} A \,,
\eea 
for any invertible operator $A$, and the so-called Sherman-Morrison formula
\bea \label{form2}
( A + |u \rangle \langle v| )^{-1} = A^{-1} - \frac{A^{-1} |u\rangle\langle v| A^{-1} }{1 + \langle v | A^{-1} | u \rangle} .
\eea 
We use these identities for $A=1-B_s^2$ and $|u\rangle \langle v| =  \frac{1}{2} |B_s \delta \rangle \langle 1 B_s|$. 

Let us now examine the expression (\ref{mainresult}) for $g(s)$ in the large
$\lambda$ limit. From (\ref{K10large}) and using (\ref{form1}) we obtain
\bea
\fl {\rm Det}(I - 2 K_{10}) = {\rm Det}\left[I - B_s^2 + \frac{1}{2} |B_s \delta \rangle \langle 1 B_s|\right] 
=  {\rm Det}\left[(I - B_s^2) (1 + \frac{1}{2} \langle 1 | \frac{B_s^2}{1-B_s^2} |\delta\rangle) \right]. \nn
\eea
Now the formula (\ref{form2}) gives
\bea
(I - 2 K_{10})^{-1} = (1-B_s^2)^{-1} - \frac{1}{2} \frac{(1-B_s^2)^{-1}  |B_s \delta  \rangle \langle 1 B_s| (1-B_s^2)^{-1}}{1 + \frac{1}{2} 
\langle 1 | \frac{B_s^2}{1-B_s^2} |\delta\rangle },
\eea 
which allows to write the second factor in (\ref{mainresult}) as
\bea
&& \fl  1 + \langle\tilde K | (1- 2 K_{10} )^{-1} | \delta\rangle = 1 - \frac{1}{2} \langle1|B_s (1- 2 K_{10} )^{-1} | \delta\rangle \\
&& \fl\qquad = 1 - \frac{1}{2} \langle1 | \frac{B_s}{1-B_s^2} |\delta\rangle \left(1 - \frac{1}{2} \frac{\langle 1 | \frac{B_s^2}{1-B_s^2} |\delta\rangle}{(1+\frac{1}{2} \langle1 | \frac{B_s^2}{1-B_s^2} |\delta\rangle ) } \right) = \frac{1 - \frac{1}{2} \langle 1 | \frac{B_s}{1+B_s} |\delta\rangle}{1+\frac{1}{2} \langle1 | \frac{B_s^2}{1-B_s^2} |\delta\rangle}. \nn
\eea
We can now put all together and write
\bea 
 g(s)^2 &=& {\rm Det}[ 1- 2 K_{10}]  (1 + \langle\tilde K | (1- 2 K_{10} )^{-1} | \delta\rangle)^2 \\
& = & {\rm Det}(I - B_s^2) \frac{(1 - \frac{1}{2} \langle1 | \frac{B_s}{1+B_s} |\delta\rangle)^2}{1+\frac{1}{2} \langle1 | \frac{B_s^2}{1-B_s^2} |\delta\rangle} \\
& =&  {\rm Det}(I - B_s^2) \frac{(\frac{1}{2} + \frac{1}{2} \langle1 | \frac{1}{1+B_s} |\delta\rangle)^2}{\frac{1}{2} + \frac{1}{4}  \langle1 | \frac{1}{1-B_s} |\delta\rangle + \frac{1}{4}  \langle1 | \frac{1}{1+B_s} |\delta\rangle}, \label{lastline}
\eea 
where we have used that $\langle1  |\delta\rangle=1$. 
Now we will use the result of Ferrari and Spohn in Ref. \cite{ferrari2}:
\bea
\frac{{\rm Det}(I-B_s)}{ {\rm Det}( I + B_s)} = \langle1 | \frac{1}{1+B_s} |\delta\rangle.
\eea
Examination of their proof shows that this property holds for any kernel $B_s(x,y)$ function of $x+y$ only, hence it
should also hold for $-B_s(x+y)$ which implies that one also has
\bea
\frac{{\rm Det}(I+B_s)}{ {\rm Det}( I - B_s)} = \langle1 | \frac{1}{1-B_s} |\delta\rangle.
\eea
Using these two identities we can now simplify (\ref{lastline}) into
\bea
g(s)^2 = {\rm Det}(I-B_s)^2,
\eea 
which, upon taking the square-root, proves the desired result (\ref{desired}) and the convergence to GOE. 

\section{Conclusions} \label{conclusion}

To summarize we have obtained, for arbitrary time, the generating function $g(s)$ for the distribution of the free energy 
of the DP with one free end, equivalently of the height of the continuum KPZ interface with a flat initial condition.
It takes the form of a Fredholm Pfaffian, i.e the square root of a Fredholm determinant. 
At large time we have shown that the distribution of free energy crosses over to the GOE Tracy Widom distribution
$F_1(s)$. These results should be of importance and testable in experiments. It is rewarding that
they have been obtained using the replica method, which thus has allowed to solve the three main
classes (droplet, flat, stationary). 

The method of solution that we found is somewhat indirect, involving a half-space wedge model and taking several limits. 
We have described why this is so, but it would be very interesting to find a more direct method to derive the result. It is
not clear whether this is a purely technical, or more fundamental difficulty. 
On the other hand the finite time Kernel that we have found is not very simple. There too, it would be
interesting to explore whether simpler equivalent expressions could be found. 
We also note that the (half-space and symmetric) wedge model is of interest in itself and deserve further studies. 
Further properties of the solution at finite time, including extracting $P(f)$ and numerics are of high interest
and left for future studies \cite{uslong}.

\medskip

{\it Acknowledgements}

We thank A. Rosso for numerous discussions and for helpful numerical checks of (i) low integer moments of $Z$ 
at small $t$ (ii) the variance of $\ln Z$ at large $t$, which will be reported elsewhere. 
PC thanks LPTENS, and PLD thanks KITP for hospitality and partial support through NSF grant
PHY05-51164, as well as KITPC. This work was supported by ANR grant 09-BLAN-0097-01/2 (PLD) 
and by the ERC under the Starting Grant n. 279391 EDEQS (PC).


\appendix

\section{Statistical tilt symmetry (STS)} \label{sts}

It is easy to show on the replicated Hamiltonian that the moments
\be
{\cal Z}_n(\{x_i\},t; \{y_i\},0) := \overline{Z(x_1 t|y_1,0)..Z(x_n t|y_n,0)}\,,
\ee 
satisfy the STS symmetry property
\bea
\fl\quad {\cal Z}_n(\{x_i + a + b\},t; \{y_i + a\},0) = e^{- \frac{n}{4 t} b^2 - \frac{1}{2 t} b \sum_i (x_i - y_i) } {\cal Z}_n(\{x_i\},t; \{y_i\},0)\,.
\eea
Choosing $a=0$, $b=x$ and $x_i=0$ STS implies
\bea
\fl\quad {\cal Z}_n(\{x\} ,t; \{y_i\},0) = 
e^{- \frac{n}{4 t} x^2 + \frac{1}{2 t} x \sum_i  y_i } {\cal Z}_n(\{0\},t; \{y_i\},0)\,,
\eea
and so
\bea
\fl \int_{y_i<0} dy_i e^{w \sum_i y_i} {\cal Z}_n(\{x\} ,t; \{y_i\},0) = 
e^{-\frac{n x^2}{4 t}} \int_{y_i<0} e^{ (w +\frac{x}{2 t}) \sum_i y_i  } {\cal Z}_n(\{0\} ,t; \{y_i\},0).
\eea

\section{Direct approach $n=4$} \label{app:n4} 

Let us give the sum of all contributions, with all integrals displayed explicitly, and then discuss each of them separately. 
We find
\begin{eqnarray}
&& \fl \overline{Z^4} =   8 e^{5 c^2 t} - 8 e^{2 c^2 t} + 24 \bar c^3 \! \int_{k}  \frac{1}{(c^2 + 4 k^2)(c^2+k^2)} e^{ - 4 k^2 t + c^2 t}
- 48 \! \int_{k} 
 \frac{\bar c}{(9 c^2 + 4 k^2)} e^{ - 2 k^2 t + \frac{c^2}{2} t} \nn \\
 && \fl +  48  \int_{k_1,k_3}  
\frac{c^2  (k_1^2 - k_3^2)^2}{(c^2 + 4 k_1^2) (c^2 + 4 k_3^2) (c^4 + (k_1^2 - k_3^2)^2 + 
   2 c^2 (k_1^2 + k_3^2))} e^{ - 2 (k_1^2 + k_3^2) t}\,, \nn
\end{eqnarray}
where everywhere we denote momenta integrals $\int_k = \int \frac{dk}{2 \pi}$. 
Upon explicit momenta integration it yields the final result given in the text.
We now discuss each term as they appear and for each give the spatial integral $\int \Psi_\mu$ and the norm $\Phi$.

(i) First term $n_s=1$ and $m_1=4$ (one 4-string of zero momentum):
\begin{eqnarray}
&& \int \Psi_\mu =  \frac{128}{\bar c^3} L \quad , \quad \Phi = 1.
\end{eqnarray}

(ii) Second term, discussed below.

(iii) Third term $n_s=2$ and $m_1=m_2=m=2$ (2 2-strings, $(k+i c/2, k-i c/2, - k+ic/2,-k-i c/2)$):
\begin{eqnarray}
&& \int \Psi_\mu = \frac{384 \bar c }{k^2 (c^2 + 4 k^2) } L \quad , \quad \Phi = \frac{k^2}{k^2 + c^2}\,,
\end{eqnarray}
and we have used $\frac1{m_1^2 m_2^2} \sum_{k_1,k_2} \to \frac1{m^4} m L \int_k$ and
a symmetry factor of unity. 

(iv) Fourth term $n_s=3$ (2 1-strings and one 2-string  $(k_1+i \frac{c}{2}, k_1-i \frac{c}{2},k_2,-2 k_1 -k_2)$.
The BE imply $k_1=0$. This leads to
\begin{eqnarray}
&& \int \Psi_\mu = - 
\frac{96  (9 c^2 + 4 k_2^2) }{k_2^2 (c^2 + 4 k_2^2)} L^2  , \qquad  \Phi = \frac{4 k_2^2 (c^2 + 4 k_2^2)}{(9 c^2 + 4 k_2^2)^2}\,,
\end{eqnarray}
and there is a factor of $3$ from the 3 choices $(m_1,m_2,m_3)=(2,1,1),(1,2,1),(1,1,2)$.

(v) $n_s=4$, four 1-strings. One checks that BE leave only the possibility $k_2=-k_1$ and $k_4=-k_2$ which yields
\begin{eqnarray}
&& \int \Psi_\mu = 
\frac{24 c^2 (c^4 + (k_1^2 - k_3^2)^2 + 2 c^2 (k_1^2 + k_3^2))}{k_1^2 k_3^2 (k_1^2 - k_3^2)^2}  L^2\,,
\end{eqnarray}
while we have
\begin{eqnarray}
&& \Phi = 
\frac{16 k_1^2 k_3^2 (k_1^2 - k_3^2)^4}{(c^2 + 4 k_1^2) (c^2 + (k_1 - k_3)^2)^2 (c^2 + 4 k_3^2) 
(c^2 + (k_1 + k_3)^2)^2} \,,
\end{eqnarray}
with a symmetry factor of $3$ for three ways to pair the momenta.

We now come back to the second term. We found this term by considering the wedge $w>0$ on the
infinite space $L=+\infty$ in the limit $w \to 0^+$. It correspond to $n_s=2$ and $m_1=3$, $m_2=1$.
As discussed in the text, it is hard to guess from consideration of fixed $L$. In the wedge calculation
it corresponds to the contribution
 \bea
&& \fl \overline{Z^4}|_{missing} = \int_{k_1,k_2} \frac{64 w \left(8 w^2+6 w+1\right) \left((k_1-k_2)^2+1\right)}{\left((k_1-k_2)^2+4\right)
   \left(k_1^2+w^2\right) \left(k_1^2+(w+1)^2\right)} \\
&& \times   \frac{e^{-4 \lambda^3
   \left(3 k_1^2+k_2^2-2\right)} \left(w \left((k_1+k_2)^2-2\right)+3
   k_1 k_2+4 w^3-5 w^2\right)}{ \left(4 k_1^2+(2 w+1)^2\right) \left(k_2^2+w^2\right) \left((k_1+k_2)^2+(2 w+1)^2\right)}\,. \nn
 \eea
 Performing the change of variable $k_i \to w k_i$ which amounts to focus on the pole $k_1=k_2=0$
 one finds at small $w$
 \begin{eqnarray}
&& \fl \overline{Z^4}|_{missing} = - 32 e^{8 \lambda^3} \int_{k_1,k_2}  \frac{1}{(1 + k_1^2) (1 + k_2^2)} + O(w)
= - 8 e^{8 \lambda^3} + O(w)\,.
\end{eqnarray}

\section{$n_s=2$ for direct method} \label{app:ns2} 

Consider first 2 $m$-strings ($m_1=m_2=m$, $n=2 m$) with $k_1=-k_2=k$. To compute $\int \Psi$ we use formula (\ref{laplaceL}) for the Laplace transform (LT) of the $B_n(\lambda)$ factors. 
After inspection we find that we must extract the $1/s^2$ pole of the LT, other poles gives subdominant contributions after using the BE. In particular 
all $1/s^p$ poles with $p>2$, i.e. 2 or more zeroes in the set $\mu^P$ (associated to the set of $\{ \lambda_{P_i} \}$), have a zero $A_P$ factor.
Since all $\mu_n^P=0$ the expression simplifies into
\begin{eqnarray}
&& \int \Psi = L  \sum_{P,A_P \neq 0} A_P  \prod_{i=1}^{n-1} \frac{1}{ - i \mu^P_i} \,.
\end{eqnarray}
This allows to compute more easily $\int \Psi$ with mathematica up to $n=8$ and guess the general form given in the text. This form
is later established more systematically from the wedge initial condition. 

Consider now different strings $m_1 \neq m_2$. We have checked that $k_1=k_2=0$ leading to $\int \Psi  \sim L^2$ is the only possibility, the
others give subdominant contributions in $L$ after using BE. Here we will consider $m_1$ and $m_2$ to be of opposite parity, otherwise $\lambda_j=0$ occurs twice which is excluded. The other possibilities are discussed in the text.  We must thus look for the $1/s^3$ in the Laplace transform (\ref{laplaceL}), i.e.
two of the $\mu_i$ must vanish. It is then easy to see that only two configurations contribute (i.e. have a non-zero $A_P$) (i) string $m_1$ on the left with decreasing imaginary parts and string $m_2$ on the right with decreasing imaginary parts (ii) the reverse. We thus consider
\bea
&& \lambda_i = \frac{i c}{2} (m_1+1-2 i) \quad , \quad 1 \leq i \leq m_1, \\
&& \lambda_{m_1+j} = \frac{i c}{2} (m_2+1-2 j) \quad , \quad 1 \leq j \leq m_2, 
\eea
which has a $A_P=A$ factor
\bea
&& \fl A= m_1! m_2! \prod_{i=1,m_1} \prod_{j=1,m_2} (1 + \frac{2}{m_1 - m_2 - 2 i + 2j })  
= m_1! m_2! \frac{ (m_1+m_2)  }{ |m_1-m_2|} (-1)^{\min(m_1,m_2)},  \nn
\eea
where the $m_1!$ and $m_2!$ factors are the intra-string (or single string) contributions using the identity 
$\prod_{1 \leq i < j \leq n} (1 + \frac{1}{j-i}) = n!$. 
Note that this formula makes sense only for strings of opposite parity (it diverges otherwise). 

After inverse LT the final result is $\int \Psi  \sim L^2 = n! A L^2/C$ with
\bea
&& \fl C=  (-i)^{m_1+m_2-2}  \prod_{j=2}^{m_2}  \sum_{p=j}^{m_2} \lambda_{m_1+ p}  \prod_{i=2}^{m_1}  \sum_{p=i}^{m_1} \lambda_p =  (-i)^{m_1+m_2-2} h(m_1) h(m_2), \\
&& \fl h(m)= \prod_{i=2}^{m}  \sum_{p=i}^{m} \frac{i c}{2} (m +1-2 p) =  (-1)^m i^{m+1} 2^{1 - m} c^{-1 + m} \Gamma(m)^2.
\eea
Putting all together it gives the result displayed in the text, using $c=-\bar c$.

\section{The wedge initial condition in the full space}
\label{FSsec}

From section \ref{sec:replica} we want to calculate
\be
  \overline{Z_w(x,t)^n} =   
 \sum_\mu \frac{ \Psi_\mu^*(x,..x)}{||\mu ||^2} e^{-t E_\mu}  \int^w \Psi_\mu\,,
\ee
with
\be
 \int^w \Psi_\mu := \int_{-\infty}^\infty  \Big(\prod_{j=1}^n dy_j\Big) e^{-w \sum_{i=1}^n |y_i|}  \Psi_\mu(y_1,..y_n)\,.
\ee
We introduce the auxiliary integrals 
\be
\fl B_{n,p}(\lambda_1,..\lambda_n) = \int_{-\infty< x_1<x_2<..< x_p < 0 < x_{p+1} <..x_n<+\infty } \hspace{-3cm}
 dx_1.. dx_n e^{i \lambda_1 x_1 + .. + i \lambda_n x_n + w (x_1 +..+ x_p) - w (x_{p+1} + .. + x_{n})} \,,
\ee
and
\be
B_{n}(\lambda_1,..\lambda_n)= \sum_{p=0}^n B_{n,p}(\lambda_1,..\lambda_n)\,,
\ee
so that 
\be
  \int d^n y  \Psi_n(y)  = n! \sum_P B_{n}(\lambda_{P_1},..\lambda_{P_n}) 
\prod_{n \geq \ell > k \geq 1} (1- \frac{i c}{\lambda_{P_\ell} - \lambda_{P_k}})\,.
\ee
We have
\begin{eqnarray}
\fl  B_{n,p}(\lambda_1,..\lambda_n) &=& G_{p}(\lambda_1,..\lambda_p) G_{n-p}(-\lambda_{n},..-\lambda_{p+1}) , \\
\fl  G_{p}(\lambda_1,..\lambda_p) &=& \int_{-\infty< x_1<x_2< ..<x_p < 0} \hspace{-2cm}
dx_1.. dx_n e^{(w+ i \lambda_1) x_1 + .. + (w+ i \lambda_p) x_p }\nn \\ &=& 
\frac{1}{(w+ i \lambda_1)(2 w+i \lambda_1+i \lambda_2)..(p w + i \lambda_1 + ..+ i \lambda_p)}\,.  \nonumber 
\end{eqnarray}
It should be already evident that the main technical difficulty in this approach is that the symmetry of the problem 
introduces double pieces in the calculations that instead will be absent in the half-space model. However,
it is a useful model to recover the results for the uniform initial condition by simply taking from the start $L=+\infty$,
and considering $w \to 0^+$ at the end. Like this we have obtained the moments $\overline{Z^n}$, and
up to $n=4$ we have recovered the results of the direct approach. As mentioned there, it allowed to
identify and obtain the contributions of the deformed strings. We extended it also to larger values of $n$. 
Since these calculations are not really illuminating, we only present here calculations at fixed $n_s$. We
present $n_s=1$ in details and show how the various limits are recovered. We will not show details of
$n_s=2$ but we did perform it and checked that it reproduces the results displayed in Section \ref{sec:ns2}. 

\subsection{One string partition sum, $n_s=1$.}

For simplicity we will set $\bar c=-c=1$ with no loss of generality (upon rescaling space). 
For a single string the rapidities are $\lambda_a = - \frac{i}{2} (m+1-2 a)+ k$ and so
\bea
 B_{n,p}(w,\lambda_1,..\lambda_n) = G_{p}(w+ i k, \lambda_1,..\lambda_p) G_{n-p}(w-i k, -\lambda_{n},..-\lambda_{p+1})\,. \nn
\eea
This gives simply
\begin{eqnarray}
\fl  B_{m,p}&=& \frac{(-2)^m \Gamma \left(1-m-2 (w-i k) \right) \Gamma
   \left(1-m- 2 (i k+w) \right)}{\Gamma (p+1) \Gamma (m-p+1) \Gamma
   \left(1-p-2 (w-i k) \right) \Gamma \left(1-m+p-2 (i
   k+w)\right)}\,, \\
 \fl   B_m &=& \frac{(-2)^m \Gamma \left(1-4 w \right) \Gamma \left(1-m + 2 i k-2 w) \right) \Gamma \left(1-m - 2 i k-2 w) \right) }{\Gamma (m+1) \Gamma \left(1-m- 4 w \right) \Gamma
   \left(1 + 2 i k- 2 w \right) \Gamma \left(1 - 2 i k- 2 w \right)}\,,
\end{eqnarray}
from which we can write the needed integral
\be
\fl \int^w \Psi =
\frac{(-2)^m  \Gamma (m+1) \Gamma \left(1-4 w \right) \Gamma
   \left(1- m + 2 i k- 2 w \right) \Gamma \left(1-m - 2 i k- 2 w \right)}{\Gamma \left(1-m-4 w\right) 
   \Gamma\left(1+ 2 i k-2 w \right) \Gamma \left(1-2 i k-2 w \right)}\,.
\ee

Inserting the integral in the expressions for $\overline{Z^n}$ we have 
\be
 \overline{Z^n}  =  \frac{1}{n^2} \frac{1}{L} \sum_k  \int^w \Psi e^{- n k^2 t} e^{(n^3-n) \frac{c^2 t}{12}} .
\ee
Let us first compute for finite $w$ in which case we can replace $\sum_k \to n L \int_k$, leading to
\bea
\fl \overline{Z^n}  &= & \frac{ (-2)^{n} \Gamma (n) \Gamma \left(1- 4 w \right)}{\Gamma \left(1-n- 4 w\right)} e^{(n^3-n) \frac{ t}{12}} 
\! \int_k  e^{- n k^2  t}
\frac{\Gamma \left(1-2 i k-n- 2 w\right) \Gamma \left(1+2 i k-n-2 w \right)}{
\Gamma \left(1-2 i k- 2   w\right) \Gamma \left(1+2 i k- 2 w\right)}\nn \\
\fl &  = & \frac{(-2)^{n} \Gamma (n) \Gamma \left(1- 4 w\right)}{\Gamma \left(1-n- 4 w \right)} e^{(n^3-n) \frac{ t}{12}}   \int_k  e^{- n k^2  t} 
\prod_{j=1}^n \frac{1}{4 k^2 + (j-1+ 2 w)^2} .
 \eea
Using $\Gamma(-n+x) \approx \frac{(-1)^n}{n!} \frac{1}{x}$, we see that for $w \to 0$ the term $j=1$
dominates and acts like a $\delta(k)$ function and we get
\be
 \overline{Z^n}  =  2^{n-1} \Gamma (n) (n-1)! e^{(n^3-n) \frac{c^2 t}{12}} \prod_{j=2}^n \frac{1}{ (j-1)^2} 
= 2^{n-1} e^{(n^3-n) \frac{c^2 t}{12}}\,,
\ee
recovering Eq. (\ref{moments1}) obtained for the uniform initial condition (direct method). 

However this approach allows also to study the wedge problem for arbitrary $w$ and not only the flat initial condition.
In fact for finite $w$ we have (performing the usual shift of the free energy)
   \begin{eqnarray}
&& \fl  Z(1,s) = \sum_{m=1}^\infty \frac{(-1)^m}{m!} e^{- \lambda m s} \overline{Z^m} \\
&& \fl = \sum_{m=1}^\infty 
     \frac{2^m \Gamma \left(1- 4 w \right)e^{m^3 \frac{\lambda^3}{3} - \lambda m s} }{m \Gamma \left(1-m- 4 w \right)} 
     \!  \int_k  e^{- 4 m k^2 \lambda^3}
\frac{\Gamma \left(1-2 i k-m-2 w\right) \Gamma \left(1+2 i   k-m-2 w \right)}{
\Gamma \left(1-2 i k- 2 w \right) \Gamma \left(1+2 i k- 2 w \right)} \nn \\
&&  \fl  = \frac{1}{\lambda} \Gamma \Big(1- 4 w \Big) \int_{k,y} Ai(y+ 4 k^2 +s ) \nn \\
&& \fl \times \sum_{m=1}^\infty 
     \frac{2^m e^{m \lambda y } }{m \Gamma \left(1-m- 4 w \right)}    
\frac{\Gamma \left(1-2 i \frac{k}{\lambda}-m- 2 w \right) \Gamma \left(1+2 i
   \frac{k}{\lambda}-m- 2 w \right)}{\Gamma \left(1-2 i \frac{k}{\lambda}- 2
   w \right) \Gamma \left(1+2 i \frac{k}{\lambda}- 2 w \right)}, \nonumber 
   \end{eqnarray} 
where the rescaling $k \to k/\lambda$ has been performed followed by the shift $y \to y + 4 k^2 + s$. Quite surprising the sum can obtained in terms of an hypergeometric function. One finds the one-string partition sum
\be
 Z(1,x) =   \int dy \frac{dk}{2 \pi}  Ai(y+ 4 k^2 - x) F(\lambda,k,w,y)\,,
\ee
in terms of the function
\bea
&& \fl F(\lambda,k,w,y)=  - \frac{2 w}{\lambda} \frac{1}{(\frac{k}{\lambda})^2 + w^2}  \\
&&  \times e^{\lambda y}
{}_3F_3(\{1, 1, 1 + 4 w \}, \{2, 1 - 2 i \frac{k}{\lambda} + 2 w, 1 + 2 i \frac{k}{\lambda} + 2 w\}, -2 e^{\lambda y}). \nn 
\eea

Although the function $F$ seems quite complicated,  for fixed $\lambda$, $k$ and $w$ the large $y>0$ behavior is very simple
\be
 F(\lambda,k, w,y) \sim - y + C + O(e^{- 4  w \lambda y} ) + O(e^{- e^{\lambda y}}), \nonumber 
\ee
as well as $F(\lambda,k, w,y) \sim e^{\lambda y}$ for large $y<0$. 

We can now discuss the various limits:

(i) very long polymer $\lambda \to \infty$ with fixed wedge $w$. Then we obtain
\begin{eqnarray}
&& \fl Z(1,x) \approx - \int_{y>0} dy \frac{dk}{2 \pi} y Ai(y+ 4 k^2 + s)  = - \int_{w>0} \frac{dw}{3 \pi} w^{3/2} Ai(w + s),  
\label{oldres}
\end{eqnarray} 
 since the hypergeometric function behaves as a $\ln (e^{\lambda y})$ for large argument and
 as $e^{\lambda y}$ for small ones. This is exactly the result for the fixed endpoint case (KPZ droplet initial condition)
 which appears as the attractor at large time for fixed $w$ 
 (note that the reduced free energy shift in $x$, $\ln(2/w)/\lambda$ is subdominant). 
 
 (ii) flat initial condition $w \to 0$ at fixed $\lambda$. Then using that $\frac{w}{w^2 + (k/\lambda)^2} \to \pi \delta(k/\lambda)$ we 
 find 
 \bea
 F(\lambda,k, w,y)  \to 2 \pi \delta(k) \times \frac{1}{2} (e^{- 2 e^{\lambda y}}-1), 
 \eea 
 hence one recovers the result (\ref{firstla}) from the direct approach. 
 
 Besides these two limits, the above
 expression however allows to identify different regimes when scaling both
 $w$ to zero and $\lambda$ to infinity, although this is left for future study. 
%


\section{More on pole structure of half-space model}
\label{sec:examples} 

Here we illustrate some important properties discussed in the text for the half-space
model for $w \to 0$, namely:

(i) to compute $\overline{Z(x,t)^n}$ at large but finite $x,t$ in the limit $w \to 0$ 
one can use the substitution $\frac{1}{i k + w} \to  \pi \delta(k) + P \frac{1}{i k}$
in the integrals (as described in the text). Neglecting terms with finite poles give only exponentially decaying corrections 
in $x$. This should be useful to describe the universality class $Ai_2 \to Ai_1$, i.e. the flat to wedge class.

(ii) if furthermore $x \to - \infty$ (to describe the flat initial condition class) one
can use $\frac{1}{i k + w} \to  2 \pi \delta(k)$ as discussed in the text. 

We just give some simple illustrative examples with no attempt to construct a proof.

\subsection{No disorder, i.e. $n=1$}



It is already instructive to discuss the trivial case without disorder. Equivalently this is the case $n=1$ in presence of
disorder. Consider the half-space model
\bea
&& \overline{Z(x,t)} = \int \frac{dk}{2 \pi} \int_{y<0} dy e^{i k y+w y} e^{-k^2 t - i x k} =
\int \frac{dk}{2 \pi} \frac{1}{i k + w} e^{-k^2 t - i x k} \\
&& = \int \frac{da}{\sqrt{4 \pi t}} e^{-\frac{a^2}{4 t}} \int \frac{dk}{2 \pi} \frac{1}{i k + w} e^{- i (x+a) k},
\eea 
performing the $k$ integration, there is a pole at $k= i w$. For $x+a >0$ the contour can be
closed in lower half plane and it gives zero. For $x+a<0$ the contour can be closed in upper
half plane and it gives the full residue via the Cauchy formula and so
\bea \label{trivial} 
&& \overline{Z(x,t)} = \int \frac{da}{\sqrt{4 \pi t}} e^{-\frac{a^2}{4 t}} \frac{1}{2} ( 1 - {\rm sgn}(x+a) ) e^{w (x+a)} \\
&& \to_{w \to \pm 0^+} = \frac{1}{2} {\rm erfc}(\frac{x}{2 t^{1/2}}) \to_{x \to - \infty} = 1. \nn
\eea 
The two terms correspond to the substitution $\frac{1}{i k + w} = \pi \delta(k) + P \frac{1}{i k}$. The 
$\pi \delta(k)$ gives the $1/2$ while the principal value gives the (odd) sign function. However in the
limit $x \to - \infty$ one sees that for any fixed $a$ the replacement $\frac{1}{i k + w} = 2 \pi \delta(k)$, which
corresponds to the Cauchy formula, holds. The conclusion is that for $x \to - \infty$ at fixed $t$ one can
always close the integration contour on the upper half plane, hence use the Cauchy formula. In other
words the pieces $\pi \delta(k)$ and $P \frac{1}{i k}$ contribute equally in that limit. Note
that to justify this one needs to use convolutions via diffusion kernels. This is made more
systematic below.

\subsection{An example with $n_s=2$}

Here we show on an example that the contributions neglected in the text (case called (iii)
in the discussion below Eq. (\ref{disc}) indeed decay exponentially. 

Consider $n_s=2$ and for simplicity the contribution of $m_1=1$ and $m_2=2$ which is
\bea
&& \fl 
 Z(n_s,s)|_{m_1=1,m_2=2} = \frac{2^2}{2} \int_{k_1,k_2} 
S^w_{1,k_1}  S^w_{2,k_1} D^w_{1,k_1,2,k_2} \Phi^w_{1,k_1,2,k_2}
e^{(m_j^3 - m_j) \frac{t}{12}- m_j k_j^2 t - \lambda m_j s - i x m_j k_j} \,,  \nn
\eea
where we have
\bea
&& S^{pole}_{1,k_1} = - \frac{1}{2( i k_1 + w)} , \qquad s_{1,k_1}=0, \\
&& S^{pole}_{2,k_2} = \frac{1}{2( i k_2 + w)} , \qquad s_{2,k_2} = - \frac{1}{1 + 2 i k_2 + 2 w}, \\
&& D^{pole}_{1,k_1,2,k_2} = 0  , \qquad d_{1,k_1,2,k_2}= \frac{3 i + 2 k_1 + 2 k_2 - 4 i w}{-i + 2 k_1 + 2 k_2 - 4 i w}. 
\eea
The term $S^{pole}_1 S^{pole}_2 d_{12}$ is the one included in the text. The one left out is
\bea
&& \fl \frac{2^2}{2} \int_{k_1,k_2} 
S^{pole}_{1,k_1}  s_{2,k_1} d_{1,k_1,2,k_2} \Phi^w_{1,k_1,2,k_2}
e^{(m_j^3 - m_j) \frac{t}{12}- m_j k_j^2 t - \lambda m_j s - i x m_j k_j}   \nn \\
&&  \fl = \frac{2^2}{2} e^{(m_j^3 - m_j) \frac{t}{12} - \lambda m_j s }  \int_{k_1,k_2}
\frac{3 i + 2 k_1 + 2 k_2 - 4 i w}{2 (k_1 - i w) (1 + 2 i k_2 + 2 w) (1 + 2 i k_1 + 2 i k_2 + 4 w)} 
 \\ &&  \qquad\qquad\qquad\qquad\qquad\qquad\times
\frac{4 (k_1-k_2)^2 +1}{4 (k_1-k_2)^2+9} e^{- k_1^2 t - 2 k_2^2 t - i x (k_1+ 2 k_2)}. \nn
\eea 
Replacing $\frac{1}{ i k_1 + w} \to \frac{1}{2} (2 \pi) \delta(k_1)$ and then setting $w=0$, it simplifies to
\bea
&&  \fl = \frac{2^2}{2} \frac{i}{2} e^{(m_j^3 - m_j) \frac{t}{12} - \lambda m_j s }  \int_{k_2}
\frac{-(i + 2 k_2)}{2 (- i + 2 k_2) (-3 i + 2 i k_2)}  e^{- 2 k_2^2 t - i x (2 k_2)} \nn \\
&& = \frac{2^2}{2} \frac{i}{2} e^{(m_j^3 - m_j) \frac{t}{12} - \lambda m_j s }  \int_{k_2} 
[ \frac{1}{2(- i+ 2 k_2)} - \frac{1}{- 3 i + 2 k_2} ] e^{- 2 k_2^2 t - i x (2 k_2)}.
\eea 
The integrals at $t=0$ are exponentially decaying in $x$
\bea
&& \int_{k_2} \frac{1}{2(- i+ 2 k_2)} e^{- 2 i k_2 x} = \frac{i}{8} e^{-|x|} (1- {\rm sgn}(x)), \\
&& \int_{k_2}  \frac{1}{- 3 i + 2 k_2} e^{- 2 i k_2 x} = \frac{i}{4} e^{-3 |x|} (1- {\rm sgn}(x)).
\eea
At finite $t$ they are ``broadened" by diffusion but they still decay exponentially:
\bea
&& \fl \int_{k_2} \frac{1}{2(- i+ 2 k_2)}  e^{- 2 k_2^2 t - i x (2 k_2)} = \int \frac{da}{\sqrt{8 \pi t}} e^{- \frac{a^2}{8 t}} \int_{k_2} \frac{1}{2(- i+ 2 k_2)}
e^{- 2 i k_2 (x + \frac{a}{2})} \\
&&\fl = \frac{i}{4} \int_{a< - 2 x} \frac{da}{\sqrt{8 \pi t}} e^{- \frac{a^2}{8 t}} e^{(\frac{a}{2}+x)}  = \frac{1}{8} i e^{x+ \frac{t}{2}} {\rm erfc}\left(\frac{x+t}{ \sqrt{2} \sqrt{t}}\right) \sim \frac{1}{4} i e^{x+ \frac{t}{2}}  , \qquad x \to - \infty. \nn
\eea
On the side $x \to + \infty$ the decay is even faster, as $e^{-x^2/(2 t)}$. 

%

\subsection{A more general argument}

One can make the argument about the limit $x \to - \infty$ slightly more general by systematic use of diffusion kernel which allows
to close the integration contour in the $k$ complex plane. We can write
\bea
&& \fl 
 Z(n_s,s) = \sum_{m_1,\dots m_{n_s}=1}^\infty 
 \prod_{j=1}^{n_s} \frac{2^{m_j}}{m_j} e^{(m_j^3 - m_j) \frac{t}{12} - \lambda m_j s } G_{{\bf m}}({\bf x},t)|_{x_j=x} ,
\eea
where ${\bf m}=(m_1,..m_{n_s})$ and ${\bf x}=(x_1,..x_{n_s})$ and we have introduced the
multi-point integrals
\bea
&& \fl G_{{\bf m}}({\bf x},t) = \prod_{j=1}^{n_s} \int_{k_j} 
S^w_{m_j,k_j}  e^{- m_j k_j^2 t  - i x_j m_j k_j}   \prod_{1 \leq i < j \leq n_s} \tilde D^w_{m_i,k_i,m_j,k_j}. 
\eea
Writing $G_{{\bf m}}({\bf x},t)$ as a function of its initial value
\bea
&& G_{{\bf m}}({\bf x},t) = \prod_{j=1}^{n_s} \int \frac{da_j}{\sqrt{4 \pi t/m_j}} e^{- \frac{m_j a_j^2}{4 t}}, 
G_{{\bf m}}({\bf x}+{\bf a} ,t=0),
\eea 
we have to study simply
\bea
&&  G_{{\bf m}}({\bf x},0) = \prod_{j=1}^{n_s} \int_{k_j} 
S^w_{m_j,k_j}  e^{ - i x_j m_j k_j}   \prod_{1 \leq i < j \leq n_s} \tilde D^w_{m_i,k_i,m_j,k_j} ,
\eea
and in particular its limit when all $x_j \to x + a_j \to - \infty$. In that limit
this integral can be computed by contour integrals closing in ${\rm Im} k_j >0$, since
all factors $S,\tilde D$ are rational fractions of the $k_j$ and there is an exponential
convergence factor $e^{ - i x_j m_j k_j}$. We now examine the structure of poles and zeros
\bea
&& \fl \Phi_{ij}  \quad {\rm poles:} \quad k_i-k_j = \pm \frac{i}{2} (m_i + m_j), \quad  {\rm zeroes:} \quad k_i-k_j = \pm \frac{i}{2} (m_i - m_j), \\
&& \fl S^w_i    \quad {\rm poles:} \quad  k_j = i w + \frac{i}{2} q_j , \quad q_j=0,1,..m_j-1  \quad , \quad {\rm no ~~ zeroes}, \\
&& \fl D^w_{ij} \quad {\rm poles:} \quad k_i + k_j = 2 i w + i q_{ij} , \quad  q_{ij}=\frac{1}{2} |m_i-m_j|,... \frac{1}{2} (m_i+m_j) -1.
\eea 
These are all simple poles and zeroes. They can become multiple however in the
products, e.g. $S^0_i D^0_{ij} S^0_j$ has a double pole for $q_{ij}=q_i+q_j$. 
The remaining zeroes of $D^w_{ij}$ (which do not cancel the poles) correspond to
negative $q_{ij}= - \frac{1}{2} |m_i-m_j|-1,.. - \frac{1}{2} (m_i+m_j)$.

The poles which have been discussed in the text correspond to $q_j=0$ and $q_{ij}=0$
and their residue lead to the dominant contribution as $w \to 0$, analyzed in the text. 
The other poles lead to decaying exponentials as $x \to - \infty$. Note that it
is crucial to work at fixed $t$, i.e. the $x \to - \infty$ limit should be taken before the
large time limit.

\subsection{Half-space model, $n_s=1$}

It is instructive to study the one string partition sum $n_s=1$ of the half-space model at fixed $w$. 

First, using (\ref{inths}) and (\ref{sw}) one has for the single string state with $n$ particles 
\bea
 \int^w \Psi = n! (-2)^n \frac{\Gamma(1-n- 2 i k - 2 w)}{\Gamma(1-2 i k - 2 w)}=n! 2^n \frac{\Gamma(2 i k +2 w)}{\Gamma(n+2 i k + 2 w)}.
\end{eqnarray}
For large $w$ this gives $\int^w \Psi = (\frac{1}{w})^n n!$ consistent with the result for fixed endpoints.
Using (\ref{zn}) this leads to
\begin{eqnarray} \label{form11}
&& \fl \overline{Z(x,t)^n}|_{n_s=1} =  2^n \Gamma(n) e^{(n^3-n) \frac{\lambda^3}{3}} 
\int \frac{dk}{2 \pi} e^{- 4 n k^2 \lambda^3} \frac{\Gamma(2 i k +2 w)}{\Gamma(n+2 i k + 2 w)} e^{- i n k x },
\end{eqnarray}
with $\lambda^3=t/4$ since here we work in units where $\bar c=1$. Note that using the diffusion
propagator of the previous section, one can perform the $k$ integral with poles at $k=i (w + p/2)$,
$p=0,1,2..$ and rewrite this as
\bea
&& \fl \overline{Z(x,t)^n}|_{n_s=1} =  2^{n-1} e^{(n^3-n) \frac{t}{12}} 
\int \frac{da}{\sqrt{4 \pi t/n}} e^{- \frac{n a^2}{4 t} } \theta(-(x+a)) e^{n w(a+x)} (1+e^{\frac{n}{2} (a+x)})^{n-1}, \nn
\eea
where one recovers (\ref{trivial}) for $n=1$. Now it is clear on this expression that
for $w \to 0^+$ and $x \to - \infty$ at fixed $t$ one recovers the result from the direct
method (\ref{moments1}) up to exponentially decaying terms. 

Using (\ref{form11}) we obtain the partition sum, performing the usual manipulations ($k \to k/\lambda$, Airy trick, shift of $y \to y+s+4 k^2$), as
 \begin{eqnarray}
&& Z(1,s) = \sum_{m=1}^\infty \frac{(-1)^m}{m!} e^{- \lambda m s} \overline{Z(x,t)^m} \\
 && = \frac{1}{\lambda} \int dy \frac{dk}{2 \pi} Ai(y + 4 k^2 + s)  \sum_{m=1}^\infty  \frac{(-2)^m}{m} 
\frac{\Gamma(2 i \frac{k}{\lambda} +2 w)}{\Gamma(m +2 i \frac{k}{\lambda} + 2 w)} e^{- i m \frac{k}{\lambda} x} .
 \eea
Performing the sum we have
 \bea
&& \fl Z(1,s)  =  \int dy \frac{dk}{2 \pi} Ai(y + 4 k^2 + s) F(\lambda,k,w,y) \\
&& \fl F(\lambda,k,w,y) = - \frac{e^{- i \frac{k}{\lambda} x + \lambda y} }{i k + \lambda w}
{}_2F_2(\{1, 1\}, \{2, 1 + 2 i \frac{k}{\lambda} + 2 w \}, -2  e^{- i x k/\lambda + \lambda y} ). \nn
 \eea

For large $\lambda$ at fixed $w$ and $x$ one finds $F(\lambda,k,w,y)  \sim - y \theta(y)+O(\frac{1}{\lambda})$ hence one recovers
the fixed endpoint result
  \begin{eqnarray}
&& Z(1,s) \to  -  \int_{y>0} \frac{dk}{2 \pi} y Ai(y+ 4 k^2 - x) = - \int_{t>0} \frac{dt}{3 \pi} t^{3/2} Ai(t-x) .
 \end{eqnarray} 

For small $w=0^+$ and fixed $\lambda$ we can replace $1/(i k + \lambda w) = \pi \delta(k) + \frac{1}{i k}$. The delta function piece gives
\begin{eqnarray} \label{half}
&& Z(1,s)|_{\delta piece} =  \frac{1}{4} \int_{y>0} (e^{-2 e^{\lambda y}}-1) Ai(y+s), 
 \end{eqnarray} 
 independently of $x$, which is half of the result for the flat space (\ref{firstla}). The second piece comes
 from the principal value
\begin{eqnarray}
&& \fl Z_{PV}(1,s) = - \! \int dy \frac{dk}{2 \pi} \frac{Ai(y+ 4 k^2 + s)}{i k}  e^{\lambda y} e^{- i x k/\lambda }
{}_2F_2(\{1, 1\}, \{2, 1 + 2 i \frac{k}{\lambda} \}, -2 e^{\lambda y} e^{- i x k/\lambda }), \nn
 \end{eqnarray} 
 and can be calculated by antisymetrizing the integral in $k$. From the above considerations
 it should converge for $x\to -\infty$ to the same value (\ref{half}) and add up to the
 full result (\ref{firstla}) as if one had substituted from the start $1/(i k + \lambda w) = 2 \pi \delta(k)$. 
 This statement is equivalent to  
 \bea
\fl  \lim_{x \to - \infty}  -  \int \frac{dk}{2 \pi} \frac{1}{i k}  e^{- i x k - t k^2}
{}_2F_2(\{1, 1\}, \{2, 1 + 2 i k \}, -2 z e^{- i x k - t k^2 }) = \frac{1}{4 z} (e^{-2 z}-1), \nn
 \eea 
which we have checked via a series expansion. One can see also that at large $|x|$ the
piece $Z_{PV}(1,s)$ becomes an antisymmetric function of $x$.

\section{Summation for the function $F$} \label{app:borel} 

To compute the function $F(z_1,z_2)$ defined by the double sum in (\ref{Kdef5}), we note that its 
Borel integral transform satisfies
\bea
\int_0^\infty dt_1 dt_2 e^{-t_1 -t_2} F(z_1 t_1 ,z_2 t_2)= \frac{z_1}{1+z_1} \frac{z_2}{1+z_2} \frac{z_1-z_2}{1+z_1 z_2} .
\eea
Denoting $s_1=1/z_1$ and $s_2=1/z_2$, upon the change $t_i \to t_i/z_i$ it can be brought in the form
of a double Laplace transform, hence we have
\bea
&& \fl F(z_1 ,z_2) = LT^{-1}_{s_1 \to z_1,s_2 \to z_2} \frac{1}{s_1 s_2} \frac{1}{1+s_1} \frac{1}{1+s_2} \frac{s_2-s_1}{1+s_1 s_2}  \\
&& \fl = LT^{-1}_{s_1 \to z_1} [ \frac{(s_1+1) e^{-z_2}+(s_1-1) s_1}{s_1-s_1^3} ]
+ LT^{-1}_{s_1 \to z_1} [  \frac{\left(s_1^2+1\right) e^{-\frac{z_2}{s_1}}}{s_1
   \left(s_1^2-1\right)} ] \\
   && \fl = -e^{z_1-z_2}-e^{-z_1}+e^{-z_2} + J_0(2 \sqrt{z_1 z_2}) +   \int_0^{z_1} dt J_0(2 \sqrt{z_2 t}) (e^{z_1- t} - e^{t-z_1}) ,
\eea
we have used the (formal) result that the LT of $e^{\mp z} \int_0^z dt e^{\pm t} g(t)$ is $g(s)/(s \pm 1)$. This function is antisymmetric in
$z_1,z_2$ and to make it more explicit we make some change of variable which brings it in the form (\ref{Kdef5}) given in the text.
One can then check explicitly that it admits the double Taylor expansion given in the text. Note that although it looks like
divergent at large $z_1,z_2$ there are massive cancellations and the function $F$ is in fact bounded and well behaved.
This is further discussed in \ref{app:asymptF}.


\section{Fredholm determinants and Pfaffian} \label{app:pfaffian}

\subsection{Series expansion of a Fredholm determinant.} 

It is useful to recall the series expansion of a Fredholm determinant of Kernel $M(x,y)$ 
where $I$ is the identity:
\bea \label{expandFD} 
&& \fl {\rm Det}(I + M) = e^{  \Tr \ln(I +M) } = 1 + \Tr M + \frac{1}{2} ( (\Tr M)^2 - \Tr M^2)  \\
&& \fl + \frac{1}{6} ((\Tr M)^3 - 3 \Tr M  \Tr M^2 + 2 \Tr M^3)\nn \\
&& \fl + \frac{1}{24} ((\Tr M)^4 - 6 (\Tr M)^2 \Tr M^2 + 3 (\Tr M^2)^2 + 8 \Tr M \Tr M^3 - 6 \Tr M^4) + \dots\,.\nn
\eea
which can also be used unambiguously for a matrix operator $M_{ab}(x,y)=M(a,x;b,y)$, $a=1,2$, $b=1,2$, e.g. it gives 
\bea
&& \fl {\rm Det}(I + M) = \Tr M_{11} + \Tr M_{22} \\
&& + \frac{1}{2} ( ( \Tr M_{11} + \Tr M_{22} )^2 - (\Tr M_{11}^2 + \Tr M_{22}^2 + 2 \Tr M_{12} M_{21} ) ) + O(M^3).  \nn
\eea 

%


\subsection{Generating function $g(s)$ in terms of pfaffians: series expansion.} 

Consider now the generating function studied in the text
\bea
g(s) = \sum_{n_s=0}^\infty \frac{1}{n_s!} Z(n_s), \quad  Z(n_s) = \prod_{j=1}^{n_s} \int_{v_j>0} {\rm pf}[{\bf K} (v_i,v_j)]_{2n_s,2n_s}\,, \nn
\eea 
with the definition of the pfaffian of a matrix Kernel $K_{ab}(v_1,v_2)=K(a,v_1;b,v_2)$ 
with the order $1,v_1;1,v_2;.. 1,v_{n_s};2,v_1;2,v_2;..2;v_{n_s}$, namely
\bea
&& {\rm Pf}[ {\bf J} + {\bf K} ] := 1+  \int_{v_1>0} pf \left(\begin{array}{cc} 0&K_{12}(v_1,v_1) \cr
                                          - K_{12}(v_1,v_1)  & 0 \cr
                                          \end{array}\right) \\
&& \fl + \frac{1}{2!}  \int_{v_1>0,v_2>0}
{\rm pf}  \left(\begin{array}{cccc} 0&K_{11}(v_1,v_2)&K_{12}(v_1,v_1) &K_{12}(v_1,v_2) \cr  
                                          K_{11}(v_2,v_1) & 0 & K_{12}(v_2,v_1) & K_{12}(v_2,v_2) \cr 
                                          - K_{12}(v_1,v_1) & - K_{12}(v_2,v_1)  & 0 & K_{22}(v_1,v_2) \cr
                                          - K_{12}(v_1,v_2) & - K_{12}(v_2,v_2)  & K_{22}(v_2,v_1) & 0 \cr
                                          \end{array}\right) + O(K^3) \nn\\
&& = 1 + Tr K_{12} - \frac{1}{2} ( (Tr K_{12})^2 - Tr K_{12}^2 + Tr K_{11} K_{22}) + O(K^3), \label{firstone}
\eea 
and, more systematically
\bea  
&& \fl Z(1)= \Tr K_{12} ,\\ &&\fl 
 Z(2) = \Tr K_{12}^2 - (\Tr K_{12})^2 - \Tr K_{11} K_{22},\nn \\
&& \fl Z(3) =  - (\Tr K_{12})^3 + 3\Tr K_{12} \Tr K_{12}^2 - 2 \Tr K_{12}^3 -  3 \Tr K_{12}\Tr K_{11} K_{22} 
\nn \\
&& \fl\qquad\qquad+ 6\Tr K_{11} K_{22} K_{12} ,\nn \\
&& \fl Z(4) = \Tr[K_{12}]^4 - 6 \Tr[K_{12}]^2 \Tr[K_{12}^2] + 3 \Tr[K_{12}^2]^2 + 
 8 \Tr[K_{12}] \Tr[K_{12}^3] - 6 \Tr[K_{12}^4] \nn  \\
 && \fl  + 6 \Tr[K_{12}]^2 \Tr[K_{11} K_{22}]   - 
 6 \Tr[K_{12}^2] \Tr[K_{11} K_{22}] + 3 \Tr[K_{11} K_{22}]^2  - 6 \Tr[(K_{11} K_{22})^2]\nn \\
 && \fl- 
 24 \Tr[K_{12}] \Tr[K_{11}  K_{22} K_{12}]   + 24 \Tr[K_{11} K_{22} K_{12}^2 ]   + 
 12 \Tr[K_{11} K_{12}^T K_{22} K_{12}], \nn
\eea
which simplifies drastically when $K_{12}$ is of rank one, as it  is the case in this paper, i.e. 
$K_{12}^2 = Tr(K_{12}) K_{12}$ or $K_{12}(v_1,v_2)=U(v_1) V(v_2)$. Using
also the antisymmetry of $K_{11},K_{12}$ we have
\bea \label{expandFP} 
&&  Z(1)= \Tr K_{12}   , \\ && Z(2) = - \Tr K_{11} K_{22},\nn \\
&&  Z(3) =  -  3 \Tr K_{12} \Tr K_{11} K_{22} + 6 \Tr K_{11} K_{22} K_{12}   \nn ,\\ &&
Z(4) =  3 \Tr[K_{11} K_{22}]^2 - 6 \Tr[(K_{11} K_{22})^2], \nn
\eea
which was used in the text.

\subsubsection{Series expansion of a Fredholm determinant, $g_1(s)$.}

Now consider the matrix operator
\bea
M = - {\bf J} {\bf K} = \left(\begin{array}{cc}
0 & - I  \\
 I & 0
\end{array} \right) \left(\begin{array}{cc}
K_{11} & K_{12}  \\
- K_{21}^T  & K_{22}
\end{array} \right)  =  \left(\begin{array}{cc}
K_{12}^T & - K_{22}  \\
K_{11}  & K_{12}
\end{array} \right),
\eea
and let us consider the square root of the following FD
considered in the text, which is a Fredholm Pfaffian.
We will write its series expansion to second order. Using Eq.  (\ref{expandFD}) one finds
\bea
&& \fl {\rm Pf}[ {\bf J} + {\bf K}] = \sqrt{ {\rm Det}[{\bf I} -  {\bf J} {\bf K}] } = \sqrt{{\rm Det}(I + M)} \\
&& = e^{ \frac{1}{2} \Tr \ln(1+M) } 
= 1 + \frac{1}{2} \Tr M + \frac{1}{8} ( (\Tr M)^2 - 2 \Tr M^2) + O(M^3)  \nn \\
&& =  1 +  \Tr K_{12} + \frac{1}{2} ( (\Tr K_{12})^2 - \Tr K_{12}^2 + \Tr K_{11} K_{12}) + O(K^3). 
\eea
Now we see that this result differs by a sign from the above expression (\ref{firstone}) for the term $n_s=2$. In general we expect that it
will differ by $(-1)^{n_s(n_s-1)/2}$. This is because the connection from FD to FP relies on
the order $1,v_1;2,v_1;1,v_2;2,v_2;..1,v_{n_s};2,v_{n_s}$, as we now explain. Note that
in our case, since $ (\Tr K_{12})^2 - \Tr K_{12}^2=0$ we can simply change $K_{22} \to - K_{22}$
to switch from one formula to the other one, as explained in the text.

\subsection{More pfaffians and Grassmann path integrals}

It is useful to use Grassmann integrals to manipulate pfaffian and
determinant, and Grassmann path integrals for Fredholm determinants and
Fredholm Pfaffian. Consider Grassman integral with $\eta_i^2=0$, $\int d\eta=0$ and $\int d\eta \eta=1$.
We will use convention $\quad \int d\eta_1.. d\eta_n \eta_1 .. \eta_n = 1$, the other convention
$\quad \int d\eta_1.. d\eta_n \eta_n .. \eta_1 = 1$ differs by $(-1)^{n(n-1)/2}$ since $\eta_n .. \eta_1 = (-1)^{n(n-1)/2} \eta_1 .. \eta_n$
(the same identities below apply usually with $e^{X} \to e^{-X}$). 
Note that the linear change of variables gives a jacobian $=\det^{-1}$. 

One has for an antisymmetric matrix $A$ of size $n=2 p$ even, and an arbitrary matrix $B$ one has
\bea
&& \fl  \int \prod_{i=1}^{n} d\eta_i e^{\frac{1}{2} \eta_i A_{ij} \eta_j} = {\rm pf}[A]   , \quad 
 \int d \bar \eta_1 d\eta_1  d \bar \eta_2 d\eta_2 ..  d \bar \eta_n d\eta_n e^{ \bar \eta_i B_{ij} \eta_j} = \det B. 
\eea
This allows to show easily that if $A$ is antisymmetric of even size $n=2p$, writing $\eta_i = \frac{1}{\sqrt{2}} (\alpha_i + i \beta_i)$ and $\bar \eta_i = \frac{1}{\sqrt{2}} (\alpha_i - i \beta_i)$ one has
\bea
 \fl \det[A] &=&  \int d \bar \eta_1 d\eta_1  d \bar \eta_2 d\eta_2 ..  d \bar \eta_n d\eta_n e^{\bar \eta_i A_{ij} \eta_j}   \\
 \fl &=& \frac{1}{i^n}
\int d \alpha_1 d\beta_1 d \alpha_2 d\beta_2 .. d \alpha_n d\beta_n  e^{\frac{1}{2} \alpha_i A_{ij} \alpha_j + \frac{1}{2}  \beta_i A_{ij} \beta_j }
= {\rm pf}[A]^2, \nn
\eea
since the cross term $i (\alpha_i \beta_j - \beta_i \alpha_j) A_{ij}=(\alpha_i \beta_j + \alpha_j \beta_i) A_{ij}=0$.
From $\int d\bar \eta_1 d\eta_1 =1 $ one gets that $d\bar \eta_1 d\eta_1 = \frac{d\alpha_1 d\beta_1}{i}$. The 
sign of the permutation $\alpha_1 \beta_1..\alpha_n \beta_n \to \alpha_1 \alpha_2.. \alpha_n \beta_1 \beta_2..\beta_n$
is $(-1)^{n(n-1)/2}$. For $n=2 p$ this is $(-1)^p=i^{n}$. Note that the determinant is indeed zero if $A$ has odd dimension. 

Now let us consider the formula (\ref{vbeautiful}) given in the text {\it for block matrices} made of 
arbitrary antisymmetric matrices $A,B$ and an arbitrary matrix $C$. It corresponds to the
other definition of the pfaffian, noted $\widetilde {\rm pf}$. 
To clarify let us consider the grassman integral representation
\bea
\fl && \det  \left(\begin{array}{cc}
I + C^T & - B  \\
A  & I + C
\end{array} \right) = \int \prod_i  d\bar \eta_1(i) d \eta_1(i) d\bar \eta_2(i) d \eta_2(i) e^{S} \\
\fl &&S = \sum_{a=1,2} \sum_j \bar \eta_a(i) \eta_a(i) \\
\fl && \quad + \sum_{ij} [ C(i,j) \bar \eta_2(i) \eta_2(j) +
C(j,i) \bar \eta_1(i) \eta_1(j) - B(i,j)  \bar \eta_1(i) \eta_2(j) +  A(i,j) \bar \eta_2(i) \eta_1(j). \nn
\eea 
It turns out that the transformation
\bea
&& \left(\begin{array}{c}
\bar \eta_1(j)   \\
\bar \eta_2(j) 
\end{array} \right) = \left(\begin{array}{cc}
1 & i  \\
-i  & 1
\end{array} \right) \left(\begin{array}{c}
\alpha_1(j)  - i \beta_1(j) \\
\alpha_2(j) - i \beta_2(j) 
\end{array} \right),  \\
&& \left(\begin{array}{c}  \eta_1(j)   \\
\eta_2(j) 
\end{array} \right) = \left(\begin{array}{cc}
1 & i  \\
-i  & 1
\end{array} \right) \left(\begin{array}{c}
\alpha_1(j)  + i \beta_1(j) \\
\alpha_2(j) + i \beta_2(j) 
\end{array} \right) ,
\eea 
decouples the $\alpha_a(i)$ from the $\beta_a(i)$. One can check that it 
leads to the square of the pfaffian $\widetilde {\rm pf}$. This is because there
are two possible Grassmann measure
\bea
\prod_{i,a} d\eta_{a}(i) = d\eta_1(1) d\eta_2(1)  d\eta_1(2) d\eta_2(2) ... d\eta_1(n_s) d\eta_2(n_s) \quad (c), \\
\prod_{i,a} d\eta_{a}(i) = d\eta_1(1) d\eta_1(2) ....  d\eta_1(n_s) d\eta_2(1) d\eta_2(2) ... d\eta_2(n_s) \quad (d) ,
\eea 
the first one allows for the above change of variable and corresponds to $\widetilde {\rm pf}$ and the
second to ${\rm pf}$. 

Finally a Fredholm determinant can be seen as a one dimensional path integral
\bea
\int \prod_{x} d \bar \eta(x) d\eta(x) e^{\int dx dy \bar \eta(x) (\delta(x-y) + K(x,y) ) \eta(y)} = {\rm Det}[I + K] .
\eea
It can be expanded around the free theory, which allows an easy derivation of its expansion in
terms of partial determinants
\bea
\fl&& \int \prod_{x} d \bar \eta(x) d\eta(x) e^{\int dx  \bar \eta(x) \eta(x)} \sum_{n=0}^\infty \frac{1}{n!}
(\int dx dy \bar \eta(x) K(x,y) ) \eta(y) )^n \\
\fl&&\qquad = \sum_{n=0}^\infty \frac{1}{n!}
\int dx_1..dx_n dy_1..dy_n K(x_1,y_1)..K(x_n,y_n) \langle \bar \eta(x_1) \eta(y_1) .. \bar \eta(x_n) \eta(y_n) \rangle \nn \\
\fl&&\qquad = \sum_{n=0}^\infty \frac{1}{n!} \int dx_1..dx_n \sum_\sigma (-1)^\sigma K(x_1,x_{\sigma(1)}) .. K(x_n,y_{\sigma(n)}) \\
\fl &&\qquad = \sum_{n=0}^\infty \frac{1}{n!} \int dx_1..dx_n \det[ K(x_i,x_j) ]_{n \times n}.
\eea
We have used $\bar \eta_{x_1} \eta_{\sigma(x_1)}... \bar \eta_{x_p} \eta_{\sigma(x_p)} = (-1)^\sigma
\bar \eta_{x_1} \eta_{x_1}... \bar \eta_{x_p} \eta_{x_p}$ which, up to multiplication by $\prod_{x} (1+ \bar \eta_{x} \eta_x)$  yields the desired result. 

Similarly one can define from the Grasmann path integral the FD of a matrix operator and Fredholm Pfaffians. 
We will not pursue further
but indicate again that there are the two possible choices for the Grassmann measure $\prod_{x,a} d\eta_{xa}(x)$ 
which singles out the correct definition of the FP. 

\section{Manipulations on $g(s)$: some details} \label{app:Qn} 

Here we show the formulas (\ref{iddet}) and (\ref{iddet2}) in the main text. Let us use the matrix determinant Lemma (\ref{form1}) applied to operators, with $A = 1-  2 K_{10}$, $|u\rangle =2 |\delta\rangle$ and $|v\rangle=|K(0,.)\rangle$. 
To prove (\ref{iddet}) using (\ref{relationK}) we need to show
\bea \label{vanish}
\langle v| A^{-1} |u\rangle = 2 \langle K(0,.) | (1-2 K_{10})^{-1} |\delta\rangle = 0.
\eea 
Since $\langle K(0,.)|\delta\rangle=K(0,0)=0$, it is thus sufficient to show that for all $n \geq 1$
\bea
Q_n := \int_{v_1,v_2>0} K(0,v_1) (K_{10})^{n}(v_1,0)  = 0 ,
\eea 
which we now prove. 

For  $n=1$ we have 
\bea
Q_1 = \int_{v_1 \geq 0} K(0,v_1) K^{(1,0)}(v_{1},0) = - [\frac{1}{2} K(v_1,0)^2]^{v_1=+\infty}_{v_1=0} = 0,
\eea 
since $K(0,0)=0$ and $K(v_1,v_2)$ vanishes when any of the argument goes to $+\infty$. Here we
denote $K^{(i,j)}(v_1,v_2) = \partial_{v_1}^i \partial_{v_2}^j K(v_1,v_2)$. 
For $n=2$ we have
\bea
  Q_2 &=& \int_{v_1,v_2>0} K(0,v_1) K^{(1,0)}(v_1,v_2)  K^{(1,0)}(v_{2},0) \nn \\
 &  =& - \int_{v_1,v_2>0} K(0,v_1) K^{(1,1)}(v_1,v_2)  K(v_{2},0)  \nn \\
 &=& \int_{v_1,v_2>0} K^{(0,1)}(0,v_1) K^{(0,1)}(v_1,v_2)  K(v_{2},0)  = - Q_2,
\eea
where in the second line we have integrated by parts first over $v_2$ then over $v_1$. The integration on either $v_2$ 
or $v_1$ has no boundary term since $K(v,0)=-K(0,v)$ vanishes both at $v=0$ and $v=\infty$. In the last line we have used antisymmetry of $K$ and 
that $K^{(0,1)}(a,b)=-K^{(1,0)}(b,a)$. Hence $Q_2=0$. 

The same manipulations can be pursued for $n \geq 3$, although now one has to keep
track of the boundary terms. One writes
\bea
\fl &&  Q_n  = \int_{v_1,..v_n>0} K(0,v_1) K^{(1,0)}(v_1,v_2) K^{(1,0)}(v_2,v_3)  .. K^{(1,0)}(v_{n-1},v_n) K^{(1,0)}(v_{n},0), \nn
\eea
and we can now integrate by part all variables starting from $v_n$
\bea
\fl  Q_n & =& (-1)^n \int_{v_1,..v_n>0} \Big\{ K^{(0,1)}(0,v_1) K^{(0,1)}(v_1,v_2) .. .. K^{(0,1)}(v_{n-1},v_n) K(v_{n},0) \nn \\
\fl &&   - [K(0,v_1) K^{(0,1)}(v_1,v_2)]_{v_1=0}^{v_1=\infty} \prod_{l=2}^{n-1} K^{(0,1)}(v_l,v_{l+1}) 
K(v_{n},0) \nn \\
\fl &&   + \sum_{p=2}^{n-1} (-1)^p
K(0,v_1) \prod_{j=1}^{p-2} K^{(1,0)}(v_j,v_{j+1}) [K^{(1,0)}(v_{p-1},v_{p}) K^{(0,1)}(v_{p},v_{p+1})]_{v_p=0}^{v_p=\infty} \nn \\
\fl && \times \prod_{l=p+1}^{n-1} K^{(0,1)}(v_l,v_{l+1}) K(v_{n},0) \nn \\
\fl && + (-1)^n K(0,v_1) \prod_{j=1}^{n-2} K^{(1,0)}(v_j,v_{j+1}) [K^{(1,0)}(v_{n-1},v_{n}) K(v_{n},0)]_{v_n=0}^{v_n=\infty} \Big\},
\eea
where we have collected all boundary terms. Now using the antisymmetry of $K$, the property $K^{(0,1)}(a,b)=-K^{(1,0)}(b,a)$
and relabeling   the integration variables, the above sum can be rewritten as (each term in the order they appear)
\bea
Q_n = - Q_n + 0 + \sum_{p=2}^{n-1} Q_{p-1} Q_{n-p} + 0 \,,
\eea
where the second and last line above vanish due to $K(0,0)=0$ and the third can be expressed from
$Q_k$ with $k < n$. Hence we have proved by recurrence that $Q_n=0$ for all $n \geq 1$. 

To prove (\ref{iddet2}) we now use the Sherman-Morrison formula (\ref{form2}) on the form (\ref{relationK})
\bea
\fl (1- K_{22} K_{11})^{-1} = (1-2 K_{10})^{-1} - 2 \frac{ (1-2 K_{10})^{-1} |\delta\rangle \langle K(0,.)| (1-2 K_{10})^{-1}}{1+ 
2\langle K(0,.)| (1-2 K_{10})^{-1}|\delta\rangle },
\eea 
multiplying left by $\langle \tilde K|$ and right by $|\delta\rangle$ we immediately obtain (\ref{iddet2}) since we just proved that the
second term vanishes from (\ref{vanish}). 

\section{Simplification of $F[2 e^{\lambda y_1},2 e^{\lambda y_2}]$ in the large $\lambda$ limit}
\label{app:asymptF} 

Here we analyze the large $\lambda$ behaviour of
\bea
&& \fl h_\lambda(y_1,y_2) = F[2 e^{\lambda y_1},2 e^{\lambda y_2}] = \sinh(2 e^{\lambda y_2}-2 e^{\lambda y_1}) +e^{-2 e^{\lambda y_2}} -e^{-2 e^{\lambda y_1}} \\
&& \fl + 
\int_0^{1} du J_0(4 e^{\frac{1}{2} \lambda (y_1+y_2) } \sqrt{u}) [ 2 e^{\lambda y_1} \sinh(2 (1-u)e^{\lambda y_1} ) - 2 e^{\lambda y_2} \sinh(2 (1-u) e^{\lambda y_2})].
\eea
We first manipulate this formula to make more apparent the fact that this function is bounded. For that one
can use the identity
\bea
\int_0^{\infty} du J_0(4 e^{\frac{1}{2} \lambda (y_1+y_2) } \sqrt{u})
e^{\lambda y_2} e^{2 (1-u) e^{\lambda y_2}} = \frac{1}{2} e^{2 e^{\lambda y_2}} e^{-2 e^{\lambda y_1}},
\eea
to rewrite
\bea
 \fl h_\lambda(y_1,y_2) &=&e^{-2 e^{\lambda y_2}} -e^{-2 e^{\lambda y_1}} \\
\fl && 
+ \int_1^{+\infty} du J_0(4 e^{\frac{1}{2} \lambda (y_1+y_2) } \sqrt{u}) [ e^{\lambda y_2} e^{2 (1-u)e^{\lambda y_2}}  - 
e^{\lambda y_1} e^{2 (1-u)e^{\lambda y_1}} ] \nn
\\
\fl &&  +
\int_0^{1} du J_0(4 e^{\frac{1}{2} \lambda (y_1+y_2) } \sqrt{u}) [ e^{\lambda y_2} e^{- 2 (1-u)e^{\lambda y_2}} 
- e^{\lambda y_1} e^{- 2 (1-u)e^{\lambda y_1}} )].
\eea
Let us write it as
\bea
&& h_\lambda(y_1,y_2) = e^{-2 e^{\lambda y_2}} -e^{-2 e^{\lambda y_1}}  + I_2 - I_1 + J_2 - J_1, \\
&& I_j = \int_1^{+\infty} du J_0(4 e^{\frac{1}{2} \lambda (y_1+y_2) } \sqrt{u}) e^{\lambda y_j} e^{2 (1-u)e^{\lambda y_j}}, \\
&& J_j = \int_0^{1} du J_0(4 e^{\frac{1}{2} \lambda (y_1+y_2) } \sqrt{u})  e^{\lambda y_j} e^{- 2 (1-u)e^{\lambda y_j}} .
\eea
For $j=1,2$. Note that $|J_0(z)|<1$ for $z>0$ hence we have the exact bounds
\bea
|I_{2,1}| < \frac{1}{2} \quad , \quad |J_j| < \frac{1}{2} (1- e^{-2 e^{\lambda y_j}}). 
\eea 
Hence the function $h_\lambda$ is bounded. 

Let us now examine the limit $\lambda \to \infty$. With no restriction we can set $y_1 < y_2$. 
There are four distinct regions:

\medskip

{\bf (i) $y_1<0$ and $y_2<0$ } then $h_\infty(y_1,y_2) = 0$.

\medskip

{\bf (ii) $y_1<0$ and $y_2> 0$ and $y_1+y_2<0$  }. 

For $y_1+y_2<0$ the argument of the Bessel function in $J_{1,2}$ can be set to zero. The integral can then
be done exactly, which leads to
\bea
J_2 = \frac{1}{2} (1- e^{-2 e^{\lambda y_2}}) \approx \frac{1}{2}  , \quad J_1 = \frac{1}{2} (1- e^{-2 e^{\lambda y_1}}) \approx 0
\eea
Next we can rewrite
\bea
&& I_2 = \int_0^{+\infty} dv J_0(4 e^{\frac{1}{2} \lambda (y_1+y_2) } \sqrt{1+v e^{- \lambda y_2}}) e^{-2 v} \approx \frac{1}{2}, \\
&& I_1= \int_0^{+\infty} dw J_0(4 e^{\frac{1}{2} \lambda (y_1+y_2) } \sqrt{1+w}) .
e^{\lambda y_1} e^{-2 w e^{\lambda y_1}}  \approx 0  
\eea
$I_2$ being clearly is dominated by
the argument of $J_0(z)$ near $z=0$ and converges to $1/2$. Evaluation of $I_1$ with mathematica 
indicates that it vanishes. The final result in that region is then
\bea
&& h_\infty(y_1,y_2) = 0 - 1  + \frac{1}{2} - 0 + \frac{1}{2} - 0 = 0 . 
\eea

\medskip

{\bf (iii) $y_1<0$ and $y_2> 0$ and $y_1+y_2>0$  } 

In that region one finds $J_1 \approx 0$, $I_1 \approx 0$ and one can write
\bea
 \fl J_2 &=&  \int_0^{1} du J_0(4 e^{\frac{1}{2} \lambda (y_1+y_2) } \sqrt{1-u})  e^{\lambda y_2} e^{- 2 u e^{\lambda y_2}} 
 \\
 \fl& =& \int_0^{1} du J_0(4 e^{\frac{1}{2} \lambda (y_1+y_2) } \sqrt{1-u e^{-\lambda y_2}})  e^{- 2 u} \approx 
 \frac{1}{2} J_0(4 e^{\frac{1}{2} \lambda (y_1+y_2) }) \approx 0, \nonumber  
\eea
and
\bea
\fl  I_2 &=& \int_1^{+\infty} du J_0(4 e^{\frac{1}{2} \lambda (y_1+y_2) } \sqrt{u}) e^{\lambda y_2} e^{2 (1-u)e^{\lambda y_2}}\nn\\
\fl & =&  \int_0^{+\infty} du J_0(4 e^{\frac{1}{2} \lambda (y_1+y_2) } \sqrt{1+u}) e^{\lambda y_2} e^{- 2 u e^{\lambda y_2}} \nn
\\
 \fl &=&  \int_0^{+\infty} du J_0(4 e^{\frac{1}{2} \lambda (y_1+y_2) } \sqrt{1+u e^{-\lambda y_2}}) e^{- 2 u } \approx \frac{1}{2} J_0(4 e^{\frac{1}{2} \lambda (y_1+y_2) }) \approx 0.
\eea
The final result in that region is then
\bea
&& h_\infty(y_1,y_2) = 0 - 1  + 0 + 0 + 0 + 0 = -1 .
\eea

\medskip

{\bf (iv) $y_1> 0$ and $y_2> 0$   } 
there we can safely find that all integrals are zero hence $h_\infty(y_1,y_2)=0$. 

Using the antisymmetry of $h_\lambda$, the final conclusion is that
\bea
h_\infty(y_1,y_2) = \theta(y_1+y_2) ( \theta(y_1) \theta(-y_2) -  \theta(y_2) \theta(-y_1)  ) ,
\eea 
as indicated in the text. 

\section{Useful Airy identities} \label{app:Airy}

A useful identity involving Airy functions is \cite{Vallee1997} 
\bea
Ai(u) Ai(v) = \frac{1}{2^{1/3} \pi} \int d\xi Ai( 2^{2/3} (\xi^2 + \frac{u+v}{2})) e^{i(u-v) \xi} ,
\eea 
which allows to prove  the ``duality" identity
\bea
 && \fl \int \frac{dQ}{2 \pi} Ai(a + 4 Q^2) Ai(b+ 4 Q^2) e^{2 i c Q}= \nn \\ &=& \int \frac{d \tilde Q}{4 \pi} d\xi 
\frac{1}{2^{1/3} \pi} Ai(2^{2/3} (\xi^2 + \tilde Q^2 + \frac{a+b}{2} )) e^{i (a-b) \xi + i c \tilde Q}  \nn \\
& =&
\int \frac{d\xi}{2 \pi} Ai(\frac{a+b+c}{2} + 4 \xi^2) Ai(\frac{a+b-c}{2}+ 4 \xi^2) e^{2 i (a-b) \xi}.
\eea

\section{Simplifications of the Kernel $K_{10}$ in the large $\lambda$ limit}

Consider the definition (\ref{Kdef1}) of the Kernel $K_{11}$ in the large $\lambda$ limit (i.e large time) and insert the limit forms (\ref{limitform1},\ref{limitform2}).
The result can be written as the sum $K_{11}=K_{11}^a + K_{11}^b$ with
\bea
\fl    K_{11}^a(v_i;v_j)&=& - \int_{y_1,y_2,Q} Ai(y_1+s+v_i+4Q^2)  Ai(y_2+s+v_j+4Q^2) \nn\\ \fl &&\qquad\qquad \times 
  \frac{e^{- 2 i (v_i -v_j) Q} }{2 i Q} \theta(y_1+y_2), \nn
\\
\fl   K_{11}^b(v_i;v_j)&=&  \frac{1}{4} \left[  \int_{0}^\infty dy_1 \int_{-y_1}^0 dy_2 Ai(y_1+s+v_i) Ai(y_2+s+v_j) - (v_i \leftrightarrow v_j) \right].
\eea

Rewriting $K_{11}^a$ and using the duality identity of \ref{app:Airy} we obtain
\bea
 \fl  K_{11}^a(v_i;v_j)&=& \frac{1}{2} \int_{y_1,y_2,Q} \int_{-(v_i-v_j)}^{v_i-v_j} dc ~ Ai(y_1+s+v_i+4Q^2)  Ai(y_2+s+v_j+4Q^2) 
  \nn \\ \fl && \qquad \times
e^{- 2 i c Q} \theta(y_1+y_2) \nn \\
 \fl & =& \frac{1}{2} \int_{-(v_i-v_j)}^{v_i-v_j} dc \int \frac{d\xi}{2 \pi}  \int_{y_1,y_2} 
Ai(\frac{y_1+y_2+v_i+v_j+c}{2}+s+4\xi^2) 
\nn\\&& \times
Ai(\frac{y_1+y_2+v_i+v_j-c}{2}+s+4\xi^2)  e^{2 i (v_i-v_j+y_1-y_2) \xi} \theta(y_1+y_2) \nn \\
& = & \frac{1}{4} \int_{-(v_i-v_j)}^{v_i-v_j} dc  \int_{y>0} 
Ai(y + \frac{v_i+v_j+c}{2}+s) Ai(y + \frac{v_i+v_j-c}{2}+s),\nn
\eea
the integration over $y_1-y_2$ produces a factor $\pi \delta(\xi)$. Computing the derivative, using integration by parts one finds:
\bea \label{der1}
&&\fl    \partial_{v_i} K_{11}^a(v_i;v_j) =  \frac{1}{2} \int_{y>0} Ai(y + v_i +s) Ai(y + v_j +s) \\
&& 
- \frac{1}{8} \int_{-(v_i-v_j)}^{v_i-v_j} dc Ai( \frac{v_i+v_j+c}{2}+s) Ai(\frac{v_i+v_j-c}{2}+s), \nn
\eea
and note that $(\partial_{v_i} - \partial_{v_j}) K_{11} (v_i;v_j) = K_{Ai}(v_i,v_j)$. The derivative of the second part of $K_{11}$ reads
\bea \label{der3}
&& \fl \partial_{v_i} K_{11}^b(v_i,v_j) \! =\! - \frac{1}{4}\! \int_{y>0}\!\! [ Ai(y+s+v_i) Ai(- y+s+v_j) - Ai(- y+s+v_i) Ai(y+s+v_j) ] \nn
\\
&& - \frac{1}{4} Ai(s+v_i) \int_{y>0} Ai(y+s+v_j) .
\eea
Adding (\ref{der1}) and (\ref{der3}) one obtains $K_{10}$. One notes that then the second line in (\ref{der1})
exactly cancels the first line in (\ref{der3}). To see it we note that the second line in (\ref{der1})
can be rewritten as
\bea
 \fl - \frac{1}{4} \int d\alpha Ai(\alpha+s) Ai(v_i+v_j - \alpha + s) [\theta(v_j < \alpha < v_i) - \theta(v_i < \alpha < v_j) ]  ,
 \eea
 while the first line in (\ref{der3}) can be rewritten as
 \bea
 \fl   - \frac{1}{4} \int d\alpha Ai(\alpha+s) Ai(v_i+v_j - \alpha + s) [\theta(v_i < \alpha < v_j) - \theta(v_j < \alpha < v_i) ] , 
\eea
and the second piece in (\ref{der1}) exactly cancels the first line in (\ref{der3}). 
This proves the formula (\ref{asymptK10}) and (\ref{K10large}) for $K_{10}$ in the large $\lambda$ limit.

\section{Explicit check for $n_s=2,3,4$ of convergence to the GOE}

Here we check the property (\ref{lim}) explicitly up to $n_s=4$. 
The general expansion of a Fredholm determinant
is given in (\ref{expandFD}) with $M=- B_s$ which yields
\bea
 && z(1,s) = - {\rm Tr} B_s ,\\
 && z(2,s) = ({\rm Tr} B_s)^2 - {\rm Tr} B_s^2, \nn \\
&& z(3,s) = - ({\rm Tr} B_s)^3 + 3 {\rm Tr} B_s  {\rm Tr} B_s^2 - 2 {\rm Tr} B_s^3 ,\nn \\
&& z(4,s) = ({\rm Tr} B_s)^4 - 6 ({\rm Tr} B_s)^2 {\rm Tr} B_s^2 + 3 ({\rm Tr} B_s^2)^2 + 8 {\rm Tr} B_s {\rm Tr} B_s^3 - 6 {\rm Tr} B_s^4. \nn
\eea 
On the other hand, calculation of the Pfaffian is given in (\ref{expandFP}) and reads
\bea \label{Zexpr} 
&& \fl Z(1,s) = {\rm Tr} K_{12} \\&&\fl  Z(2,s) =  - {\rm Tr} K_{11} K_{22}, \nn \\
&& \fl Z(3,s) =   -  3 {\rm Tr} K_{12} {\rm Tr} K_{11} K_{22} + 6 {\rm Tr} K_{11} K_{22} K_{12} = 
3 Z(1,s) Z(2,s) +  6 {\rm Tr} K_{11} K_{22} K_{12},  \nn \\
&& \fl Z(4,s) =   3 ({\rm Tr} K_{11} K_{22})^2 - 6 {\rm Tr} (K_{11} K_{22})^2 = 3 Z(2,s)^2 - 6 {\rm Tr} (K_{11} K_{22})^2. \nn
\eea
For $n_s=2$ we have
\bea
&& \fl Z(2,s) =  - 2 \int_{v>0} K_{10}(v,v) =  \int_{v>0, y>0} ( - Ai(y + v +s)^2 + \frac{1}{2} Ai(s+v) Ai(y + v +s) ). \nn \\
&& \label{z2s}
\eea
On the other hand
\bea
&& {\rm Tr} B_s = \int_{x>0} Ai(2 x + s), \quad  {\rm Tr} B_s^2 =  \int_{x_1>0,x_2>0} Ai(x_1+x_2 + s)^2, 
\eea
while the first term in (\ref{z2s}) is clearly $- {\rm Tr} B_s^2$, checking that the second is $( {\rm Tr} B_s)^2$ requires some
combinatorics. Note that it does not require any property of the Airy function, and, as for
all values of $n_s$ below, replacing $Ai(z)$ by any function would give the same result. Hence
to check it the simplest method is to write formally $Ai(z_i) = \int_{t_i} f(t_i) e^{-t_i z_i}$,
or equivalently to replace each $Ai(z_i) \to e^{-t_i z_i}$, integrate and symmetrize over the $t_i$.
Being true for any set of $t_i$ it is true for any function. We find $({\rm Tr} B_s)^2 \to e^{-s(t_1+t_2)}/(4 t_1 t_2)$
while the second term in (\ref{z2s}) yields $e^{-s(t_1+t_2)}/(2 t_2( t_1+ t_2))$ which under symmetrization
is identical. Hence we find $Z(2,s)=z(2,s)$.

To check $n_s=3$, from (\ref{Zexpr}) we simply need to show that
\bea
6 {\rm Tr} K_{11} K_{22} K_{12} = 2 ({\rm Tr} B_s)^3 - 2 {\rm Tr} B_s^3,
\eea 
with
\bea
&&  \fl {\Tr} B_s^3 =  \int_{x_1>0,x_2>0,x_3>0} Ai(x_1+x_2 + s) Ai(x_2+x_3+s) Ai(x_3+x_1+s).
\eea
We have, from (\ref{relationK}) 
\bea
&& \fl 6 \Tr K_{11} K_{22} K_{12} = 6 \int_{v_1>0,v_2>0} (K_{11} K_{22})(v_2,v_1) \tilde K(v_1) \delta(v_2) =
12  \int_{v_1>0} K_{10}(v_1,0) \tilde K(v_1) \nn \\
&& \fl  = - 3 \int_{v_1>0,y_1>0,y_2>0} \!\!\!\!\! Ai(y_1+s+v_1) 
 [ Ai(y_2 + v_1 +s) Ai(y_2  +s) - \frac{1}{2} Ai(s+v_1) Ai(y_2+s) ]. \nn
\eea
Let us again use the exponential trick replacing each $Ai(z_i) \to  e^{- z_i t_i}$, after integration we find
\bea
&& \fl 12 \int_{v_1>0} \tilde K(v_1) K_{10}(v_1,0) \equiv {\rm sym}_{t_1,t_2,t_3} \frac{3(t_2-t_3)}{2 t_1 t_3 (t_1+t_2)(t_2+t_3)} 
 \\
   && \equiv \frac{1}{4 t_1 t_2 t_3} - \frac{2}{(t_1 + t_2) (t_2 + t_3) (t_1 + t_3) } \equiv  2 (\Tr B_s)^3 - 2 \Tr B_s^3, 
\eea
upon symmetrization (we are not showing the trivial $e^{-s(t_1+t_2+t_3)}$ factor). Let us note that
\bea
\fl (\Tr B_s)^p \equiv \frac{1}{2^p t_1..t_p}  , \quad \Tr B_s^p \equiv \frac{1}{(t_1+t_2)(t_2+t_3)..(t_{n-1}+t_n)(t_n+t_1)} .
\eea

To check $n_s=4$, we see from (\ref{Zexpr}) that we only need to show that
\bea \label{ns4}
- 6 \Tr (K_{11} K_{22})^2 = - 2 (\Tr B_s)^4 + 8 \Tr B_s \Tr B_s^3 - 6 \Tr B_s^4. 
\eea 
We have, from (\ref{relationK})
\bea\fl 
\Tr (K_{11} K_{22})^2 = 4 \int_{v_1>0,v_2>0} K_{10}(v_1,v_2) K_{10}(v_2,v_1) - 8 \int_{v>0} K(v,0) K_{10}(v,0). \nn
\eea
Note that the second term can be integrated by part and vanishes. 

We also have
\bea
2 K_{10}(v_1,v_2) \equiv \frac{t_2-t_1}{2 t_2 (t_1+t_2)} e^{- t_1 v_1 - t_2 v_2} ,
\eea
hence after multiplying with the same with exchange of $v_1,v_2$ and varlables $t_3,t_4$ and integration
over $v_1,v_2$ we find
\bea
4 Tr K_{10}^2 \equiv {\rm sym}_{t_1,t_2,t_3,t_4} \frac{(t_1 - t_2) (t_3 - t_4)}{4 t_2 t_4 (t_1 + t_2) (t_2 + t_3)  (t_1 + t_4) (t_3 + t_4)}.
\eea
Multiplying by $-6$ this should be compared with the r.h.s of (\ref{ns4}) 
\bea
&& \fl \frac{-2}{16 t_1 t_2 t_3 t_4} - \frac{6}{(t_1 + t_2) (t_2 + t_3)  (t_3 + t_4) (t_4 + t_1)} +
8 {\rm sym} \frac{1}{2 t_1} \frac{1}{(t_2 + t_3) (t_3 + t_4) (t_4 + t_2)}, \nn
\eea
and we find again that it is equal, i.e. $Z(4,s)=z(4,s)$.


\end{document}